\documentclass[review]{elsarticle}

\usepackage{lineno,hyperref}
\journal{XXXX}

\usepackage[numbers]{natbib}
\usepackage{amsmath,amssymb,amsfonts}
\usepackage{algorithmic}
\usepackage{graphicx}
\usepackage{float}
\usepackage{textcomp}
\usepackage{xcolor}
\usepackage{pifont}
\usepackage{amsthm}
\usepackage{hyperref}
\usepackage{mathrsfs}
\usepackage{booktabs}
\usepackage{hyperref}
\allowdisplaybreaks

\usepackage{url}  %Required
\usepackage{graphicx}  %Required
\usepackage[ruled,vlined,linesnumbered]{algorithm2e}
\usepackage{setspace}
\usepackage{floatpag} 
\floatpagestyle{empty}

\newtheorem{defn}{Definition}

\newtheorem{lem}{Lemma}
\newtheorem{thm}{Theorem}

\newtheorem{rem}{Remark}

\newcommand{\bp}[1]{{\mathbb{P}}\left[{#1}\right]}

\newcommand{\qeda}{\hfill\ensuremath{\blacksquare}}

\DeclareMathOperator*{\erfc}{erfc}

\begin{document}

\begin{frontmatter}

\title{Locally Differentially Private Data Collection and Analysis}

\author[add1,add2]{Teng Wang\corref{mycorrespondingauthor}}
\ead{wangteng0610@stu.xjtu.edu.cn, N1805892E@e.ntu.edu.sg}
\author[add2]{Jun Zhao}
\ead{junzhao@ntu.edu.sg}
\author[add1]{Xinyu Yang}
\ead{yxyphd@mail.xjtu.edu.cn}
\author[add1]{Xuebin Ren}
\ead{xuebinren@mail.xjtu.edu.cn}

% \address[1]{School of Electronic and Information Engineering, Xi'an Jiaotong University, Shaanxi, China}
\address[add1]{Xi'an Jiaotong University, Shaanxi, China}
\address[add2]{Nanyang Technological University, Singapore}
\cortext[mycorrespondingauthor]{Corresponding author}
% \address[2]{School of Computer Science and Engineering, Nanyang Technological University, Singapore}

% \author[mymainaddress,mysecondaryaddress]{Elsevier Inc}

% \author[mysecondaryaddress]{Global Customer Service\corref{mycorrespondingauthor}}
% \cortext[mycorrespondingauthor]{Corresponding author}

% \address[mymainaddress]{1600 John F Kennedy Boulevard, Philadelphia}
% \address[mysecondaryaddress]{360 Park Avenue South, New York}

\begin{abstract}
Local differential privacy (LDP) can provide each user with strong privacy guarantees under untrusted data curators while ensuring accurate statistics derived from privatized data. Due to its powerfulness, LDP has been widely adopted to protect privacy in various tasks (e.g., heavy hitters discovery, probability estimation) and systems (e.g., Google Chrome, Apple iOS). Although \mbox{$\epsilon$-LDP} has been proposed for many years, the more general notion of \mbox{$(\epsilon,\delta)$-LDP} has only been studied in very few papers, which mainly consider mean estimation for numeric data. Besides, prior solutions achieve \mbox{$(\epsilon,\delta)$-LDP} by leveraging Gaussian mechanism, which leads to low accuracy of the aggregated results. In this paper, we propose novel mechanisms that achieve \mbox{$(\epsilon,\delta)$-LDP} with high utility in data analytics and machine learning. Specifically, we first design \mbox{$(\epsilon,\delta)$-LDP} algorithms for collecting multi-dimensional numeric data, which can ensure higher accuracy than the optimal Gaussian mechanism while guaranteeing strong privacy for each user. Then, we investigate different local protocols for categorical attributes under \mbox{$(\epsilon,\delta)$-LDP}. Furthermore, we conduct theoretical analysis on the error bound and variance of the proposed algorithms. Experimental results on real and synthetic datasets demonstrate the high data utility of our proposed algorithms on both simple data statistics and complex machine learning models.
% enables each user to perturb her/his data and then the aggregator can compute quite accurate statistics after collecting randomized data from a large number of users. LDP provides strong privacy guarantees for each participating user when facing untrusted data curators. Due to its powerfulness, LDP has been adopted in Google Chrome, Apple iOS, and Microsoft Windows Insider.
% The majority of existing local models focus on achieving \mbox{$\epsilon$-LDP} for various tasks, e.g., heavy hitters discovery, probability distribution estimation, empirical risk minimization. 
\end{abstract}


\end{frontmatter}

%\linenumbers

\section{Introduction}\label{sec-introduction}

With the rapid development of sensing technology \cite{guo2015mobile}, smart devices, such as mobile phones, smart vehicles, wearable devices, and sensor networks, have increasingly developed into data sources of the era of big data and generated gigantic data continuously \cite{han2015mobile,merlino2016mobile}. Various and massive user data are collected and analyzed to provide invaluable knowledge for different organizations or service providers, which significantly benefits human's daily lives. However, privacy concerns related to user's personal information have been serious challenges when collecting and analyzing user's sensing data under untrusted data curators (such as in untrusted crowdsourcing systems) \cite{yang2015security,jin2018incentive,feng2018survey,tang2019privacy}. 

% Crowdsourcing \cite{guo2015mobile} has promoted the collection and aggregation of user data generated from smart devices, such as mobile phones, smart vehicles, wearable devices, and sensor networks \cite{han2015mobile,merlino2016mobile}. However, privacy concerns related to user's personal information have become serious challenges when collecting users' data in untrusted crowdsourcing systems \cite{yang2015security,jin2018incentive,feng2018survey,tang2019privacy}. 

As a formal privacy protection technique, differential privacy (DP) \cite{dwork06Calibrating,dwork2014algorithmic}, which provides rigorous guarantees for the privacy of each user by adding randomized noise, has been extensively studied in the literature.
% It provides probabilistic guarantees for the privacy of each user by adding randomized noise regardless of adversary's prior knowledge. 
Specifically, a mechanism $\mathcal{M}$ achieves $\epsilon$-DP if for any pair of neighboring datasets $D$ and $D'$ (which differ in one record), it holds that $\bp{\mathcal{M}(D)\in \mathcal{Y}}\leq e^\epsilon\bp{\mathcal{M}(D')\in\mathcal{Y}}$, where $\mathbb{P}$ denotes the probability and $\mathcal{Y}$ is any possible subset of outputs. As a relaxed version of $\epsilon$-DP (also referred to \textit{pure} DP), \mbox{$(\epsilon,\delta)$-DP} \cite{dwork2006our} (also referred to \textit{approximate} DP) has the following meaning (loosely speaking, not exactly speaking): given a typically small probability $\delta$, a mechanism $\mathcal{M}$ achieves $\epsilon$-DP with probability at least $1-\delta$. Formally speaking, a mechanism $\mathcal{M}$ achieves \mbox{$(\epsilon,\delta)$-DP} if $\bp{\mathcal{M}(D)\in \mathcal{Y}}\leq e^\epsilon\bp{\mathcal{M}(D')\in\mathcal{Y}} + \delta$ holds for any pair of neighboring datasets $D$ and $D'$. $(\epsilon,\delta)$-DP can also be understood as being more general than $\epsilon$-DP since the former in the special case of $\delta=0$ becomes the latter.

Since the introduction of differential privacy (DP), a large number of \mbox{DP-based} mechanisms \cite{han2019differentially,gong2018protecting, yang2017survey} have been proposed and applied to numerous scenarios, such as data statistics \cite{xu2013differentially,zhu2015correlated,chen2015differentially}, learning models \cite{abadi2016deep,phan2016differential,zhang2017dynamic,mohassel2017secureml}, and systems \cite{hu2015differential,bittau2017prochlo}. Nonetheless, the traditional differential privacy paradigm under centralized setting \cite{dwork06Calibrating} requires a trustworthy data curator and can not guarantee the privacy of each participate locally when collecting data, thus limiting its applications when facing untrusted data curators.

Given the above discussions, local differential privacy (LDP) \cite{kasiviswanathan2011can,duchi2013local} has been proposed to provide stronger privacy guarantees locally for each user, which no longer relies on a trustworthy data curator. Formally, for any neighboring input tuples $x$ and $x'$ of one user, the mechanism $\mathcal{M}$ satisfies \mbox{$\epsilon$-LDP} if $\mathbb{P}[\mathcal{M}(x) \in \mathcal{Y}] \leq e^\epsilon \cdot \mathbb{P}[\mathcal{M}(x') \in \mathcal{Y}]$, for any possible subset of outputs $\mathcal{Y}$. That is, each user utilizes a LDP-achieving mechanism to perturb her/his data and then sends the noisy information to the \textit{aggregator}. Then, the aggregator combines the perturbed data of all users to estimate the desired statistics. Thus, LDP achieves stronger privacy guarantees than centralized DP for protecting users' data and also protects the aggregator from data breaches since the aggregator does not hold users' true data. Besides, LDP model also ensures that the data of each participating user is invisible to any other users except the participating user.

LDP has attracted much attention in both academia and industry. A large number of studies have designed  mechanisms under \mbox{$\epsilon$-LDP} for various tasks including heavy hitters discovery, probability distribution estimation, empirical risk minimization \cite{qin2016heavy,yang2017copula,cormode2018marginal,Wang19Local,bassily2015local,wang2017locally,wang2019collecting}. Google's system called RAPPOR \cite{erlingsson2014rappor} under \mbox{$\epsilon$-LDP} has been used in Chrome to collect information about users' preferred homepages. Apple \cite{apple2017local,thakurta2017emoji} has implemented LDP in recent iOS and MacOS versions. Microsoft \cite{ding2017collecting} has deployed an LDP-enabled data collection mechanism in Windows Insiders program to collect application usage statistics.  

Although LDP has drawn much attention from the research community in recent years, almost existing mechanisms are proposed under \mbox{$\epsilon$-LDP}. The fundamental research on \mbox{$(\epsilon,\delta)$-LDP} (the relaxed version of \mbox{$\epsilon$-LDP}) has not been addressed sufficiently. Moreover, existing solutions mainly \cite{gaboardi2018locally,joseph2018locally,bun2018heavy,bassily2018linear} leverage the basic Gaussian mechanism \cite{dwork2006our} to achieve \mbox{$(\epsilon,\delta)$-LDP}, which yields a low data utility. We will also demonstrate later that the data utility still remains low even the optimal Gaussian mechanism \cite{balle2018improving} is used. Besides, existing local protocols under \mbox{$(\epsilon,\delta)$-LDP} \cite{gaboardi2018locally,joseph2018locally,bassily2018linear} mainly focus on the task of mean estimation for numeric attributes, without considering the frequency estimation of categorical attributes. 

The purpose of this paper is to propose mechanisms that can achieve \mbox{$(\epsilon,\delta)$-LDP} with high accuracy on various estimation tasks.
In particular, we focus on applying \mbox{$(\epsilon,\delta)$-LDP} to complex multi-dimensional data collection and analysis for   numeric attributes and categorical attributes. Our main contributions are summarized as follows. 
\begin{itemize}
    \item First, we propose novel mechanisms for collecting and analyzing multi-dimensional numeric data under \mbox{$(\epsilon,\delta)$-LDP}, which ensures much higher accuracies than Gaussian mechanism. Besides, we also give the theoretical analysis on the error bound of our proposed mechanisms. 
    \item Second, as for categorical attributes, we investigate several different randomized response protocols which achieve \mbox{$(\epsilon,\delta)$-LDP} and also compare the variance of different protocols. Furthermore, we introduce an optimized local hash mechanism under \mbox{$(\epsilon,\delta)$-LDP} which achieves lower communication overhead and higher accuracy than other mechanisms. 
    \item Third, we conduct extensive experiments on both real-world datasets and synthetic datasets to evaluate the performance of our proposed mechanisms. All the experimental results have demonstrated the high accuracy of our proposed mechanisms on both mean/frequency estimations and machine learning models.
\end{itemize}

This paper is organized as follows. Section~\ref{sec-related} reviews the related work. Section~\ref{sec-preliminaries} formalizes the research problem and introduces local differential privacy as preliminaries. In Sections~\ref{sec-numeric} and~\ref{sec-categorical}, we elaborate our proposed algorithms for numeric attributes and categorical attributes, respectively. Section~\ref{sec-experiments} presents our extensive experimental results. Finally, Section~\ref{sec-conclution} concludes the paper.

% Many researches on differentially private empirical risk minimization for machine learning \cite{dwork2009differential,chaudhuri2011differentially} have been studied with a wide variety of techniques, but aiming for a class of convex optimization problems. For deep neural networks, the loss function is usually convex and difficult to minimize \cite{abadi2016deep}. Thus, stochastic gradient descent (SGD) algorithm plays an important role to solve the empirical risk minimization problem.

\section{Related Work}\label{sec-related}

Differential privacy (DP) \cite{dwork06Calibrating,dwork2014algorithmic}, a classical privacy protection technique with rigorous mathematical proofs, has been studied in the literature for more than a decade. It provides formal privacy guarantees for each record in the dataset \cite{xu2013differentially,zhu2015correlated,chen2015differentially,yang2017survey}. One of the many topics in DP research is differentially private empirical risk minimization for machine learning \cite{dwork2009differential,chaudhuri2011differentially}, especially for deep neural networks \cite{abadi2016deep,phan2016differential,zhang2017dynamic,acs2018differentially,xu2019ganobfuscator}. Also, novel privacy notions related to $(\epsilon,\delta)$-differential privacy such as concentrated differential privacy have also been studied recently \cite{bun2016concentrated,mironov2017renyi,bun2018composable}. However, traditional DP in the centralized setting requires a trusted data curator, thereby limiting the application scenarios.

Therefore, local differential privacy (LDP) \cite{kasiviswanathan2011can,duchi2013local} has received considerable attention recently since it no longer assumes a trusted data curator. Specifically, each user applies LDP to protect her/his local information and reports only the noisy data to an aggregator. This is in the same spirit as the classical randomized response technique \cite{warner1965randomized}. LDP not only provides strong privacy guarantees for each user, but also protects the aggregator from data breaches since the aggregator does not collect users' true data in  the first place. Kasiviswanathan \textit{et al.} \cite{kasiviswanathan2011can} have precisely addressed the powerful characterization of the local private learning algorithms. Google has developed RAPPOR \cite{erlingsson2014rappor} to collect user statistics for Chrome under \mbox{$\epsilon$-LDP} with strong privacy protections and high analysis accuracy on the collected data. Afterward, Fanti \textit{et al.} \cite{fanti2016building} extended RAPPOR to conduct complex joint distribution estimations.

Current researches focus on many related problems under the LDP model, such as mean/frequency estimation \cite{duchi2013local,nguyen2016collecting,wangtt2017locally}, probability distribution estimation \cite{fanti2016building,yang2017copula,Wang19Local}, heavy hitter identification \cite{bassily2015local,qin2016heavy,bun2018heavy}, itemset mining \cite{wang2018locally}, marginal distribution release \cite{cormode2018marginal,zhang2018calm}, and empirical risk minimization \cite{wang2019collecting,wang2018empirical}. Besides, Ye \textit{et al.} \cite{yeprivkv} proposed PrivKV which investigates the frequency and mean estimation on key-value data. And they also proposed PrivKVM which can improve the estimation accuracy further through multiple iterations. By deploying LDP to the recommended system, Shin \textit{et al.} \cite{shin2018privacy} proposed an enhanced matrix factorization mechanism which leverages random projection-based dimension reduction technique to improve the recommendation accuracy while guaranteeing per-user privacy. 

In the setting of \mbox{$(\epsilon,\delta)$-LDP}, Gaboardi \textit{et al.} \cite{gaboardi2018locally} have investigated the upper and lower error bounds of mean estimation when protecting privacy by adding Gaussian noise. Afterward, Joseph \textit{et al.} \cite{joseph2018locally} further provided a smaller lower bound of mean estimation than Gaboardi \textit{et al.} \cite{gaboardi2018locally}. As for heavy hitter discovery problem, Bun \textit{et al.} \cite{bun2018heavy} have focused on the transformation of approximate local private protocol (\mbox{$(\epsilon,\delta)$-LDP}) into a pure local private protocol (\mbox{$\epsilon$-LDP}). Moreover, under the constraint of \mbox{$(\epsilon,\delta)$-LDP}, Bassily \cite{bassily2018linear} proposed algorithms for estimating a set of linear queries in both offline setting and adaptive setting and analyzed the accuracy bound of the proposed algorithms. So far, the above mechanisms under \mbox{$(\epsilon,\delta)$-LDP} are all achieved by the classical Gaussian mechanism \cite{dwork2006our}, which yields low accuracies of the estimation results. Thus, the goal of this paper is to investigate the mechanisms which can achieve \mbox{$(\epsilon,\delta)$-LDP} with higher accuracies on estimation results.

\section{Preliminaries}\label{sec-preliminaries}

This paper considers the local setting that the server collects data from a large number of users under an untrusted data curator. Then, the collected data will be used to compute statistical models or conduct machine learning. Our goal is to design the mechanism which can not only achieve $(\epsilon, \delta)$-local differential privacy (LDP) \cite{kasiviswanathan2011can,duchi2013local}, but also maximize the accuracies on both statistical models and machine learning models.

Formally, let $x=\{x(1),x(2),\cdots,x(N)\}$ be the data of all users, where $N$ is the user population. Each tuple $x(i)= \langle x_1(i),x_2(i),\cdots,x_d(i) \rangle$ $(i\in[1,N])$\footnote{For simplicity, in this paper, we use $i\in[1,N]$ and $j\in[1,d]$ to denote the sets $i\in\{1,2,\cdots,N\}$ and $j\in\{1,2,\cdots,d\}$, respectively.} denotes the data of the $i$-th user, which consists of $d$ attributes $A_1,A_2,\cdots,A_d$. Each $x_j(i)$ $(j\in[1,d])$ denotes the value of the $j$-th attribute of the $i$-th user. These attributes are either numeric or categorical. Without loss of generality, we assume that each numeric attribute holds a domain $[-1,1]$, and each categorical has $k$ distinct values, holding a discrete domain $\{1,2,\cdots,k\}$.

While collecting users' multi-dimensional data under an untrusted data curator, each user $u_i$ adopts a randomized perturbation mechanism $\mathcal{M}$ to perturb her tuple $x(i)$. Then, the perturbed data $\mathcal{M}(x(i))$ instead of raw data will be sent to the aggregator in order to protect privacy information locally against an untrusted aggregator. This paper follows the local differential privacy model and focuses on two types of analytic tasks under \mbox{$(\epsilon,\delta)$-LDP}:
\begin{enumerate}
    \item Basic statistics: mean estimation and frequency estimation. For numeric attribute, we focus on estimating the mean value of each attribute $A_j(j\in[1,d])$ over all $N$ users, that is $\frac{1}{N}\sum_{i=1}^{N}x_j(i)$. As for categorical attribute, the frequency $f_j(k)$ of each possible value $k(k\in[1,k])$ in attribute $A'_j$ will be computed.
    \item Advanced statistics: machine learning models analysis under empirical risk minimization.
\end{enumerate}

Next, we briefly review some conceptions related to $(\epsilon, \delta)$-local differential privacy and machine learning. In the following, we simplify $x(i)$ as $x$ to denote the data tuple of one user by omitting notation $i$.

\subsection{Local Differential Privacy}
Local differential privacy \cite{kasiviswanathan2011can,duchi2013local} has been used to provide strong privacy protection for each user locally,
% It allows that each user $u_i$ can randomized report her data $x(i)$ and then send the perturbed data to the server aggregator. Therefore, the aggregator will never access to the true data of each user, thus providing a strong protection. 
which is defined as follows.

\begin{defn}[$\epsilon$-Local Differential Privacy \cite{kasiviswanathan2011can}] \label{defn-eps-ldp}
A randomized mechanism $\mathcal{M}$ satisfies $\epsilon$-local differential privacy if and only if for any pairs of adjacent input tuples $x$ and $x'$ in the domain of $\mathcal{M}$, and for any possible subset of outputs $\mathcal{Y}$, it always holds
\begin{align}  \label{eqn-eps-ldp}\mathbb{P}[\mathcal{M}(x) \in \mathcal{Y}] \leq e^\epsilon \cdot \mathbb{P}[\mathcal{M}(x') \in \mathcal{Y}],
\end{align}
where the notation $\mathbb{P}[\cdot]$ denotes probability.
\end{defn}

Similar to the case that $(\epsilon, \delta)$-differential privacy \cite{dwork2006our} is a relaxation   of $\epsilon$-differential privacy \cite{dwork06Calibrating},   $(\epsilon, \delta)$-local differential privacy (also called \textit{approximate} LDP) is a relaxation   of $\epsilon$-local differential privacy (also called \textit{pure} LDP). 
% And $(\epsilon, \delta)$-local differential privacy can achieve better accuracy than $\epsilon$-local differential privacy when applied to deep learning algorithms.

\begin{defn}[$(\epsilon, \delta)$-Local Differential Privacy \cite{bassily2018linear}] \label{defn-eps-delta-ldp}
A randomized mechanism $\mathcal{M}$ satisfies $(\epsilon, \delta)$-local differential privacy if and only if for any pairs of adjacent input tuples $x$ and $x'$ in the domain of $\mathcal{M}$, and for any possible subset of outputs $\mathcal{Y}$, it always holds
\begin{align}\label{eqn-eps-delta-ldp}
    \mathbb{P}[\mathcal{M}(x) \in \mathcal{Y}] \leq e^\epsilon \cdot \mathbb{P}[\mathcal{M}(x') \in \mathcal{Y}] + \delta,
\end{align}
where $\delta$ is typically small. Loosely speaking (not exactly speaking), \mbox{$(\epsilon,\delta)$-LDP} means that a mechanism $\mathcal{M}$ achieves \mbox{$\epsilon$-LDP} with probability at least $1-\delta$. By relaxing $\epsilon$-LDP,    $(\epsilon, \delta)$-LDP  is more general since the latter in the special case of $\delta=0$ becomes the former.
\end{defn}

\subsection{Machine Learning based on Empirical Risk Minimization}
Machine learning models, which can be expressed as empirical risk minimization essentially, have been applied to many fields recent years. As for a machine learning task with $N$ training samples $x=\{x(1),x(2),\cdots,x(N)\}$, the loss function $\mathcal{L}(\theta)$ is used to capture how ``bad'' is the predictor when predicting the label of the $i$-th data point, which is parameterized by a $d$-dimensional parameter vector $\theta$ and computed as the average loss of all samples. That is, $\mathcal{L}(\theta)=\frac{1}{N}\sum _i\mathcal{L}(\theta,x(i))$, where $\mathcal{L}(\theta,x(i))$ is the loss of sample $x(i)$. Generally, the training target is to find a $\theta$ that obtains an acceptably small loss. In practice, the stochastic gradient descent (SGD) algorithm is often used to compute the target $\theta$ where we have the minimum (or hopefully) loss. At each iteration $t+1$, the parameter vector is computed as
% \begin{align}\label{defn-grad-iter}
    $\theta_{t+1} = \theta_{t}-\eta \cdot \nabla\mathcal{L}(\theta_{t})$,
% \end{align}
where $\eta$ is the learning rate, $\nabla\mathcal{L}(\theta_{t})$ is the gradient of loss function $\mathcal{L}(\theta_{t})$ at $\theta_{t}$.

When in private settings, each user will submit a noisy gradient $\nabla\mathcal{L}^*(i)$ to the aggregator. In this paper, we assume that each iteration involves a batch $G$ of users. Then, the parameter will be updated as
\begin{align}\label{eqn-sgd}
    \theta_{t+1} = \theta_{t}-\eta \cdot \frac{1}{|G|}\sum\nolimits_{i \in G} \nabla\mathcal{L}^*(i),
\end{align}
where $|G|$ is the batch size.

% The gradient of loss function $\mathcal{L}(\theta)$ at $\theta$ is defined as
% \begin{align}\label{defn-grad}
%     \nabla\mathcal{L}(\theta)=\frac{1}{N}\sum_i\nabla\mathcal{L}(\theta, x(i)).
% \end{align}

% In this paper, we consider three common learning tasks: linear regression, logistic regression and support vector machines (SVM) classification. We adopt three specific loss function definitions as suggested in \cite{wang2019collecting}.

\subsection{Existing Solutions to Achieve \mbox{$(\epsilon,\delta)$-LDP}}

The Gaussian mechanism is a classical solution for achieving $(\epsilon, \delta)$-differential privacy \cite{dwork2006our}, which can also be applied to achieve $(\epsilon, \delta)$-local differential privacy. Most existing studies on \mbox{$(\epsilon,\delta)$-LDP} are based on the Gaussian mechanism \cite{gaboardi2018locally,joseph2018locally,bun2018heavy,bassily2018linear}. Balle and Wang \cite{balle2018improving} have shown that the two classical Gaussian mechanisms of Dwork and Roth~\cite{dwork2014algorithmic} and of Dwork~\textit{et~al.}~\cite{dwork2006our} for $(\epsilon,\delta)$-differential privacy are not optimal. Moreover, they also developed the optimal Gaussian mechanism. Hence, we will discuss the optimal Gaussian mechanism in this paper and its application to $(\epsilon,\delta)$-LDP.

\begin{thm}[\textbf{Optimal Gaussian mechanism (\mbox{Opt-GM}) for $(\epsilon,\delta)$-differential
privacy} \cite{balle2018improving}] \label{thm-DP-OPT} 
The optimal Gaussian mechanism for $(\epsilon,\delta)$-differential
privacy adds Gaussian noise with standard deviation $\sigma$ to each dimension of a query with $\ell_2$-sensitivity $\Delta$, for $\sigma$ given by
\begin{align}\label{eqn-DP-OPT}
\sigma = \frac{\left(\xi+\sqrt{\xi^2+\epsilon}\right)  \cdot \Delta }{\epsilon\sqrt{2}},
\end{align}
where $\ell_2$-sensitivity of a query is the maximal $\ell_2$-norm difference of the true query results on neighboring datasets which differ in just one record, $\xi$ is the solution of $\erfc\left(\xi \right)- e^{\epsilon} \erfc\left( \sqrt{\xi^2 + \epsilon} \right)  =  2 \delta$ and {\rm erfc()} is the complementary error function.
\end{thm}

\begin{figure}
  \centering
  \begin{tabular}{cc} 
    \hspace{-10pt}\includegraphics[height=3.3cm]{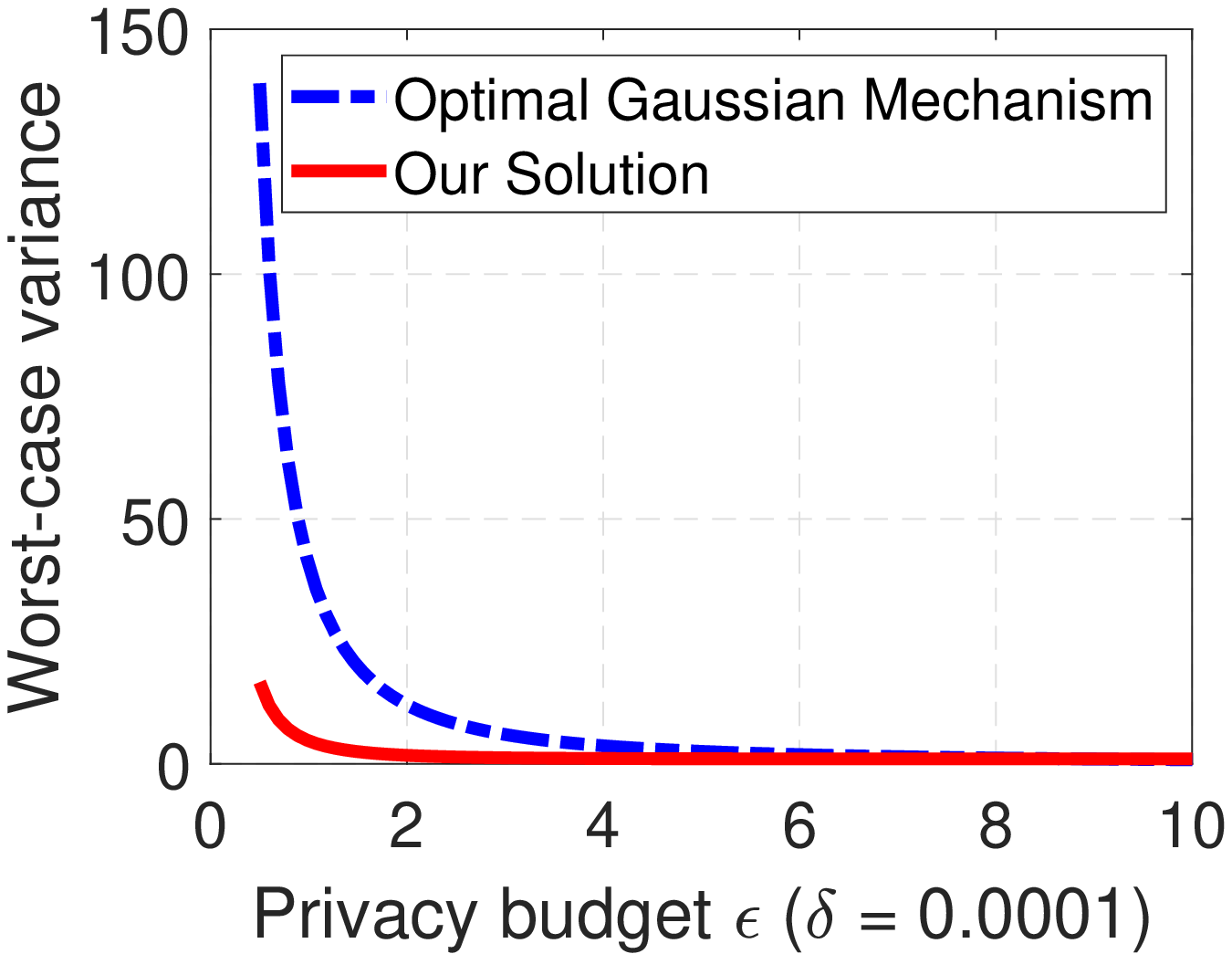}&\hspace{-15pt}
    \includegraphics[height=3.3cm]{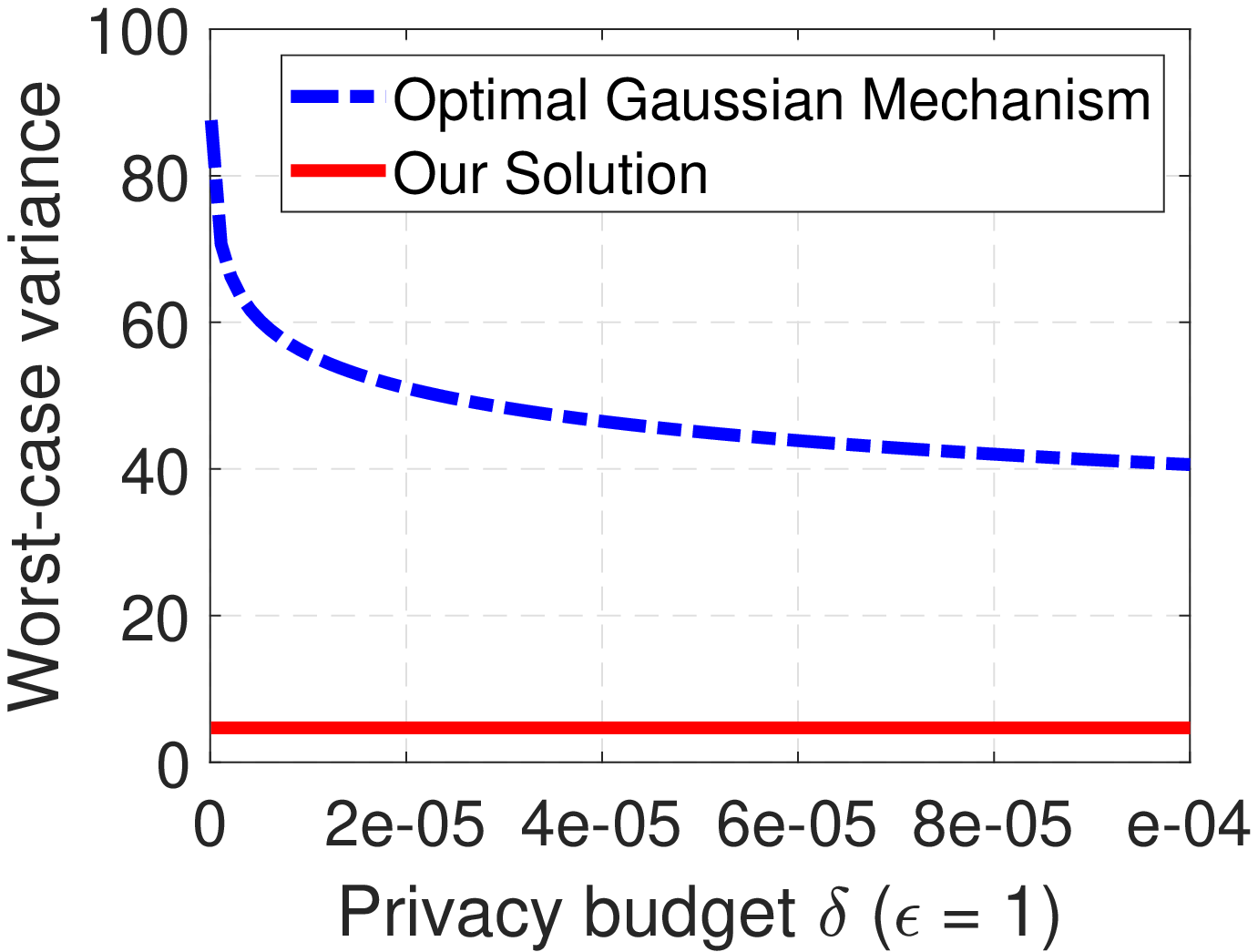}\\[-1pt]
   \scriptsize (a) Variance vs. $\epsilon$ ($\delta=10^{-4}$) & \scriptsize (b) Variance vs. $\delta$ ($\epsilon=1$)
   \end{tabular} 
 	\vspace{-1mm}
    \caption{The worst-case noise variances for one-dimensional numeric data.} \label{comp-var}
	\vspace{-3mm} 
\end{figure}

Then, each user's data will be perturbed by adding randomized Gaussian noise, that is, $x^*(i)=x(i) + \langle\mathcal{N}(0, \sigma^2)\rangle^d$, where $\mathcal{N}(0, \sigma^2)$ denotes a random variable  following a Gaussian distribution with mean $0$ and variance $\sigma^2$. Since we assume each user's data lies in range $[-1,1]$, thus $\ell_2$-sensitivity is $\Delta=2$. Clearly, the estimation for $x^*(i)$ is unbiased since the injected Gaussian noises have zero mean. Besides, the worst-case variance is $\sigma^2$. As shown in Fig.~\ref{comp-var}, we plot the worst-case noise variances of the optimal Gaussian mechanism and our solution (will be introduced later) for one-dimensional numeric data versus different privacy parameters. It can be observed our solution has much smaller variances than the optimal Gaussian mechanism especially when $\epsilon$ is small (i.e., the degree of privacy protection is high). This demonstrates that our solution can ensure high accuracy in reality while providing strong privacy guarantees.

% \begin{figure}\centering
% 	\includegraphics[height=3.5cm]{figures/Var_Nu.eps}
% 	\caption{The worst-case noise variances for one-dimensional numeric data versus $\epsilon$ when $\delta=0.0001$}
% 	\label{comp-var}
% 	\vspace{-5mm}
% \end{figure}

\section{Mean Estimation for Numeric Attributes under \mbox{$(\epsilon, \delta)$-LDP}}\label{sec-numeric}

This section introduces the solutions to achieve $(\epsilon, \delta)$-local differential privacy on multi-dimensional numeric attributes for mean estimation.

\subsection{Our First Solution for Multiple Numeric Attributes under \mbox{$(\epsilon, \delta)$-LDP}}\label{sec-numeric-first}

Under \mbox{$\epsilon$-LDP}, Duchi~\textit{et~at.} \cite{duchi2018minimax} have proposed a classical randomized mechanism for numeric data which has been extended to many scenarios. However, Nguy{\^e}n~\textit{et al.} \cite{nguyen2016collecting} have pointed that Duchi~\textit{et al.}'s solution doesn't achieve \mbox{$\epsilon$-LDP} when $d$ is even. But they don't give the specific proofs. We show Duchi~\textit{et~at.}'s solution in \ref{appen-Duchi-solution} (e.g., Algorithm~\ref{algorithm-duchi}) and give the proofs. 
Moreover, this paper has also fixed this problem by re-defining the probability of sampling a Bernoulli variable $u$ and shown the proofs in Appendix~A.2
of the online full version~\cite{fullversion} due to space limitation.

In what follows, inspired by Duchi~\textit{et al.}'s work, we propose a randomized mechanism on multiple numeric attributes for achieving \mbox{$(\epsilon, \delta)$-LDP}. Firstly, we present and prove the Lemma~\ref{condition-approx-ldp} that will be used later to ensure \mbox{$(\epsilon, \delta)$-LDP}.

\begin{lem}\label{condition-approx-ldp} 
For a randomized mechanism $\mathcal{M}$ whose outputs are discrete, $\mathcal{M}$ satisfies $(\epsilon, \delta)$-local differential privacy if and only if for any pairs of adjacent input tuples $x$ and $x'$ in the domain of $\mathcal{M}$, and for any possible output $x^*$, it always holds
\begin{align}\label{eqn-eps-delta-ldp-single}
    \mathbb{P}[\mathcal{M}(x) = x^*] \leq e^\epsilon \cdot \mathbb{P}[\mathcal{M}(x')=x^*] + \delta.
\end{align}
And Eq.~(\ref{eqn-eps-delta-ldp}) and Eq.~(\ref{eqn-eps-delta-ldp-single}) is equivalent to each other.
\end{lem}

\noindent\textbf{Proof.} For Eq.~(\ref{eqn-eps-delta-ldp}) $\Rightarrow$ Eq.~(\ref{eqn-eps-delta-ldp-single}), this can be easily achieved by letting $\mathcal{Y}=\{x^*\}$. For Eq.~(\ref{eqn-eps-delta-ldp-single}) $\Rightarrow$ Eq.~(\ref{eqn-eps-delta-ldp}), we have
\begin{small}
\begin{align}
    \bp{\mathcal{M}(x)\in\mathcal{Y}}
    &=\sum_{x^*\in\mathcal{Y}}\bp{\mathcal{M}(x)=x^*}\nonumber\\
    &\leq \sum_{x^*\in\mathcal{Y}}\bigg(e^\epsilon\bp{\mathcal{M}(x')=x^*}+\delta\bigg)\nonumber\\
    &=\bigg( \sum_{x^*\in\mathcal{Y}}e^\epsilon\bp{\mathcal{M}(x')=x^*} \bigg)+\left | \mathcal{Y} \right |\cdot\delta \nonumber\\
    &\leq \bp{\mathcal{M}(x')\in\mathcal{Y}} + \delta.
\end{align}
\end{small}
Thus, it has proved Eq.~(\ref{eqn-eps-delta-ldp}) $\Leftrightarrow$ Eq.~(\ref{eqn-eps-delta-ldp-single}). \qeda

Followed the definition before, each user's $d$-dimensional data is denoted as $x=(x_1,x_2,\cdots,x_d)$ (We will omit the notation $i$ in the analysis for simplicity since we focus on one arbitrary user $i$ here). And each $x_j\in[-1,1]$ is the value of the $j$-th attribute $A_j$, where $j\in[1,d]$. Under \mbox{$(\epsilon, \delta)$-LDP}, each user's data $x\in[-1,1]^d$ will be perturbed into $x^* \in \{-B, B\}^d$, where $B$ is a constant decided by $d$, $\epsilon$ and $\delta$. Before chosen $B$, we first compute $C_d$ as
\begin{align}
    C_d=
    \begin{cases}
    2^{d-1},&\text{~if~}d\text{~is odd},\\
    2^{d-1}-\frac{1}{2}\binom{d}{d/2},&\text{~otherwise}.
    \end{cases}
\end{align}
Then, $B$ is calculated by
\begin{align} \label{B-value}
    B=
    \begin{cases}
    \frac{2^d+C_d\cdot(e^\epsilon-1)}{\binom{d-1}{(d-1)/2}\cdot(e^\epsilon +2^d\cdot \delta-1)},&\text{~if~}d\text{~is odd},\\[3pt]
    \frac{2^d+C_d\cdot(e^\epsilon-1)}{\binom{d-1}{d/2}\cdot(e^\epsilon +2^d\cdot \delta -1)},&\text{~otherwise}.
    \end{cases}
\end{align}

\begin{algorithm}[tb]\label{ldp-algo-our-multi}
\small
\setstretch{1} 
 \caption{Mechanism for Multi-dimensional Numeric Data under \mbox{$(\epsilon, \delta)$-LDP} (\mbox{Mechanism-1})}
  \KwIn{tuple $x\in [-1, 1]^d$ and privacy parameters $\epsilon$ and $\delta$} 
  \KwOut{perturbed tuple $x^*\in \{-B, B\}^d$}
    Generate a random vector $V : = [V_1, V_2, \ldots, V_d] \in \{-1,1\}^d$ by sampling each $V_j$ independently from the following distribution:
    \begin{align}
    \mathbb{P}[V_j=v_j]=\begin{cases}
    \frac{1}{2}+\frac{1}{2}x_j,~~\text{if}~~v_j=1\\
    \frac{1}{2}-\frac{1}{2}x_j,~~\text{if}~~v_j=-1
    \end{cases}\nonumber
    \end{align}\\
    {In the case of $V$ is sampled as $v$, let $T^+(v)$ (resp. $T^-(v)$) be the set of all tuples $x^*\in\{-B,B\}^d$ such that $x^*\cdot v > 0$ (resp. $x^*\cdot v\leq 0$)\;}
    {Sample a Bernoulli variable $u=1$ with probability $\alpha$, for $\alpha$ given by Eq.~(\ref{alpha-val-our}), i.e., $\alpha:=
    \begin{cases}
    \frac{e^\epsilon+C_d\cdot\delta}{e^\epsilon+1},&\text{~if~}d\text{~is odd,} \\
    \frac{e^\epsilon \cdot C_d + \delta \cdot C_d(2^d-C_d)}{(e^\epsilon-1)C_d+2^d},&\text{~if~}d\text{~is even.}
    \end{cases}$ for $C_d:=
    \begin{cases}
    2^{d-1},&\text{~if~}d\text{~is odd},\\
    2^{d-1}-\frac{1}{2}\binom{d}{d/2},&\text{~otherwise}.
    \end{cases}$\;}   
    \eIf{$u=1$}
    {\textbf{return} a tuple $x^*$ uniformly from $T^+(v)$\;}
    {\textbf{return} a tuple $x^*$ uniformly from $T^-(v)$\;}
\end{algorithm}

Algorithm~\ref{ldp-algo-our-multi} shows the pseudo-code of our mechanism. It firstly discretizes the $d$-dimensional data into $V\in\{-1,1\}^d$ which will be used to sample $T^+(v)$ and $T^-(v)$. Then, a noisy tuple will be returned based on the value of a Bernoulli variable $u$, where the probability of $u=1$ is $\alpha$.

In what follows, we will show the computation of $\alpha$ while achieving \mbox{$(\epsilon, \delta)$-LDP}. Firstly, we analyze the size of $T^+(v)$ and $T^-(v)$. Recall that $T^+(v)$ (resp. $T^-(v)$) is the set of all tuples $x^*\in\{-B,B\}^d$ such that $x^*\cdot v > 0$ (resp. $x^*\cdot v\leq 0$). The analysis includes two cases, e.g., $d$ is odd and $d$ is even.

\textbf{Case 1: $d$ is odd}. Since $v\in\{-1,1\}^d$ and $x^*\in\{-B,B\}^d$, suppose there are $k$ positions that vectors $x^*$ and $v$  have the same sign (i.e., $d-k$ positions have the different sign). Therefore, once $v$ is sampled based on the input $x$, it can easily know that $x^*\cdot v > 0$ will be guaranteed if and only if $k>d-k$ (i.e., $k \geq (d+1)/2$ since $d$ is odd), and $x^*\cdot v \leq 0$ will be guaranteed if and only if $k\leq d-k$ (i.e., $k \leq (d-1)/2$). 
% Thus, we can get $\left | T^+(v) \right |=\sum_{j\geq(d+1)/2}\binom{d}{j}$. Similarly, . Thus, we get $\left | T^-(v) \right |=\sum_{j\leq(d-1)/2}\binom{d}{j}$.
%For simplicity, we omit $x$ and $x'$ when denoting the size of $T_x^+$ and $T_x^-$ in the following.
Therefore, when $d$ is odd, it holds
\begin{align}\label{eqn-T-odd}
    % \begin{cases}
    \left | T^+(v) \right |=\sum_{j\geq\frac{d+1}{2}}\binom{d}{j},~\left | T^-(v) \right |=\sum_{j\leq\frac{d-1}{2}}\binom{d}{j}.
    % \end{cases}
\end{align}
From Eq.~(\ref{eqn-T-odd}), it can be observed that $\left | T^+(v) \right |=\left | T^-(v) \right |$ since $d$ is odd. Recall that $\left | T^+(v) \right |+\left | T^-(v) \right |=2^d$, thus we can obtain
\begin{align}\label{eqn-T-odd-val}
    \left | T^+(v) \right |=\left | T^-(v) \right |=2^{d-1},\text{~~if~}d\text{~is odd}.
\end{align}
As we can seen, the size of both $\left | T^+(v) \right |$ and $\left | T^-(v) \right |$ is independent of $v$. Thus, when given input $x'$ and sampled $v'$, it will hold $\left | T^+(v') \right |=\left | T^+(v) \right |$ and $\left | T^-(v') \right |=\left | T^-(v) \right |$.

\textbf{Case 2: $d$ is even}. Same as \textbf{Case 1}, assume there are $k$ positions that vectors $x^*$ and $v$ have the same sign (i.e., $d-k$ positions have the different sign). Therefore, we can know that $x^*\cdot v > 0$ will be guaranteed if and only if $k>d-k$ (i.e., $k \geq (d+2)/2$ since $d$ is even), and $x^*\cdot v \leq 0$ will be guaranteed if and only if $k\leq d-k$ (i.e., $k \leq d/2$).
% Thus, it gets $\left | T^+(v) \right |=\sum_{j\geq (d+2)/2}\binom{d}{j}$. Similarly, we can know that $x^*\cdot v \leq 0$ will be guaranteed if and only if $k\leq d-k$, that is $k \leq d/2$. Thus, it gets $\left | T^-(v) \right |=\sum_{j\leq d/2}\binom{d}{j}$. 
Hence, when $d$ is even, it holds
\begin{align}\label{eqn-T-even}
    % \begin{cases}
    \left | T^+(v) \right |=\sum_{j\geq\frac{d+2}{2}}\binom{d}{j},~
    \left | T^-(v) \right |=\sum_{j\leq\frac{d}{2}}\binom{d}{j}.
    % \end{cases}
\end{align}
Base on Eq.~(\ref{eqn-T-even}), it holds that $\left | T^+(v) \right | + \left | T^-(v) \right |=2^d$ and $\left | T^-(v) \right | - \left | T^+(v) \right |=\binom{d}{d/2}$. Thus, we can get
\begin{align}\label{eqn-T-even-val}
    \begin{cases}
    \left | T^+(v) \right |=2^{d-1}-\frac{1}{2}\binom{d}{d/2},\\[3pt]
    \left | T^-(v) \right |=2^{d-1}+\frac{1}{2}\binom{d}{d/2}. 
    \end{cases}
\end{align}

Assume that we sample a Bernoulli variable $u=1$ with probability $\alpha$ (note that $\alpha>1/2$) in our mechanism. Thus, given a perturbed output $x^*$ of input $x$, it holds
\begin{small}
\begin{align} \label{eqn-prob-left}
 &   \mathbb{P}[\mathcal{M}(x)=x^*]= \alpha \mathbb{P}[\mathcal{M}(x)=x^*~|~u=1] + (1-\alpha) \mathbb{P}[\mathcal{M}(x)=x^*~|~u=0] \nonumber\\
    & =  \Bigg\{ \sum_{v \in \{-1,1\}^d} \bigg[ \alpha \mathbb{P}[ x^*\in T^+(v)] + (1-\alpha) \mathbb{P}[ x^*\in T^-(v)] \bigg] \times \mathbb{P}[v~|~x] \Bigg\} \nonumber\\
    & =  \Bigg\{ \sum_{v \in \{-1,1\}^d} \bigg[ \alpha \mathbb{P}[ x^*\in T^+(v)] + (1-\alpha) \mathbb{P}[ x^*\in T^-(v)] \bigg]  \times \prod_{j=1}^d \left(  \frac{1}{2}+\frac{1}{2} x_j \cdot v_j \right) \Bigg\}  \nonumber\\
    & =  \Bigg\{ \sum_{v \in \{-1,1\}^d} \bigg[   \frac{\alpha}{|T^+(v)|} \times  \boldsymbol{1}[ x^*\in T^+(v)] +  \frac{1-\alpha}{|T^-(v)|}\times  \boldsymbol{1}[ x^*\in T^-(v)] \bigg]  \times \prod_{j=1}^d \left(  \frac{1}{2}+\frac{1}{2} x_j \cdot v_j \right) \Bigg\}  \nonumber\\
    & =  \Bigg\{ \sum_{v \in \{-1,1\}^d} \bigg[   \frac{\alpha}{|T^+(v)|} \times  \boldsymbol{1}[ x^*\cdot v>0]  +  \frac{1-\alpha}{|T^-(v)|}\times  \boldsymbol{1}[ x^*\cdot v\leq 0] \bigg]  \times \prod_{j=1}^d \left(  \frac{1}{2}+\frac{1}{2} x_j \cdot v_j \right) \Bigg\}   \nonumber\\
    & =  \Bigg\{ \sum_{ _{x^*\cdot v>0}^{v \in \{-1,1\}^d:}} \bigg[   \frac{\alpha}{|T^+(v)|}   \times \prod_{j=1}^d \left(  \frac{1}{2}+\frac{1}{2} x_j \cdot v_j \right) \bigg]\Bigg\}  + \Bigg\{ \sum_{ _{x^*\cdot v \leq 0}^{v \in \{-1,1\}^d:}} \bigg[   \frac{1-\alpha}{|T^-(v)|}\times    \prod_{j=1}^d \left(  \frac{1}{2}+\frac{1}{2} x_j \cdot v_j \right) \bigg] \Bigg\} .
\end{align}
\end{small}

In the same way, given a perturbed output $x^*$ of input $x'$, it can also get
\begin{small}
\begin{align} \label{eqn-prob-right}
 &   \mathbb{P}[\mathcal{M}(x')=x^*]
     = \alpha \mathbb{P}[\mathcal{M}(x')=x^*~|~u=1] + (1-\alpha) \mathbb{P}[\mathcal{M}(x')=x^*~|~u=0] \nonumber\\
    & =  \Bigg\{ \sum_{v' \in \{-1,1\}^d} \bigg[ \alpha \mathbb{P}[ x^*\in T^+(v')]  + (1-\alpha) \mathbb{P}[ x^*\in T^-(v')] \bigg] \times \mathbb{P}[v'~|~x] \Bigg\} \nonumber\\
    & =  \Bigg\{ \sum_{ _{x^*\cdot v'>0}^{v' \in \{-1,1\}^d:}} \bigg[   \frac{\alpha}{|T^+(v')|}   \times \prod_{j=1}^d \left(  \frac{1}{2}+\frac{1}{2} x'_j \cdot v'_j \right) \bigg]\Bigg\}  + \Bigg\{ \sum_{ _{x^*\cdot v' \leq 0}^{v' \in \{-1,1\}^d:}} \bigg[   \frac{1-\alpha}{|T^-(v')|}\times    \prod_{j=1}^d \left(  \frac{1}{2}+\frac{1}{2} x'_j \cdot v'_j \right) \bigg] \Bigg\} .
\end{align}
\end{small}

In order to satisfy $(\epsilon, \delta)$-local differential privacy, it needs to ensure for any $x\in [-1, 1]^d$ and $x'\in [-1, 1]^d$ that Eq.~(\ref{eqn-eps-delta-ldp-single}) is always satisfied for any output $x^*\in\mathcal{Y}$. Thus, it can be seen that as long as Eq.~(\ref{eqn-eps-delta-ldp-single}) is satisfied when $\mathbb{P}[\mathcal{M}(x)=x^*]$ takes the maximum value and $\mathbb{P}[\mathcal{M}(x')=x^*]$ takes the minimum value, then mechanism $\mathcal{M}(\cdot)$ will satisfy $(\epsilon, \delta)$-local differential privacy. Here and in the following, we may omit $v$ in the $|T^+(v)|$ and $|T^-(v)|$ for simplicity since the size of them is independent of $v$.

\begin{lem}\label{max-min-value} 
The Eq.~(\ref{eqn-prob-left}) will take the maximum value when 
\begin{align}
    x\in \{ v:v\in\{-1,1\}^d, x^*\cdot v>0 \},
\end{align}
and the maximum value is
\begin{align}
    \max{\mathbb{P}[\mathcal{M}(x)=x^*]}=\frac{\alpha}{|T^+(v)|}.
\end{align}
And, the Eq.~(\ref{eqn-prob-right}) will take the minimum value when
\begin{align}
    x'\in \{ v':v'\in\{-1,1\}^d, x^*\cdot v'\leq 0 \},
\end{align}
and the minimum value is
\begin{align} 
    \min{\mathbb{P}[\mathcal{M}(x')=x^*]}=\frac{1-\alpha}{|T^-(v')|}.
\end{align} 
\end{lem}

\noindent\textbf{Proof.} Eq.~(\ref{eqn-prob-left}) can be induced as 
\begin{small}
\begin{align} \label{eqn-prob-left-1}
 &   \mathbb{P}[\mathcal{M}(x)=x^*]=
 \nonumber\\
    & \Bigg\{ \sum_{ _{x^*\cdot v>0}^{v \in \{-1,1\}^d:}} \bigg[   \frac{\alpha}{|T^+(v)|}   \times \prod_{j=1}^d \left(  \frac{1}{2}+\frac{1}{2} x_j \cdot v_j \right) \bigg]\Bigg\}  \nonumber\\
    & \quad + \Bigg\{ \sum_{ _{x^*\cdot v \leq 0}^{v \in \{-1,1\}^d:}} \bigg[   \frac{1-\alpha}{|T^-(v)|}\times    \prod_{j=1}^d \left(  \frac{1}{2}+\frac{1}{2} x_j \cdot v_j \right) \bigg] \Bigg\} \nonumber\\
    &= \Bigg\{ \frac{\alpha}{|T^+|} \sum_{ _{x^*\cdot v>0}^{v \in \{-1,1\}^d:}} \prod_{j=1}^d \left(  \frac{1}{2}+\frac{1}{2} x_j \cdot v_j \right) \Bigg\} \nonumber\\
    & \quad+\Bigg\{ \frac{1-\alpha}{|T^-|} \sum_{ _{x^*\cdot v \leq 0}^{v \in \{-1,1\}^d:}} \prod_{j=1}^d \left(  \frac{1}{2}+\frac{1}{2} x_j \cdot v_j \right) \Bigg\}.
\end{align}
\end{small}

It can be seen that
\begin{small}
\begin{align} \label{eqn-prob-left-2}
 &   \sum_{ _{x^*\cdot v>0}^{v \in \{-1,1\}^d:}} \prod_{j=1}^d \left(  \frac{1}{2}+\frac{1}{2} x_j \cdot v_j \right) + \sum_{ _{x^*\cdot v \leq 0}^{v \in \{-1,1\}^d:}} \prod_{j=1}^d \left(  \frac{1}{2}+\frac{1}{2} x_j \cdot v_j \right)   \nonumber\\
    &  = \sum_{v \in \{-1,1\}^d} \prod_{j=1}^d \left(  \frac{1}{2}+\frac{1}{2} x_j \cdot v_j \right) \nonumber\\
    &  =   \prod_{j=1}^d  \left[ \left(  \frac{1}{2}+\frac{1}{2} x_j   \right) + \left(  \frac{1}{2} - \frac{1}{2} x_j   \right) \right]    =  \prod_{j=1}^d 1  =  1  .
\end{align}
\end{small}
We define $A$ as $ \sum_{ _{x^*\cdot v>0}^{v \in \{-1,1\}^d:}} \prod_{j=1}^d \left(  \frac{1}{2}+\frac{1}{2} x_j \cdot v_j \right) $. Then $\sum_{ _{x^*\cdot v \leq 0}^{v \in \{-1,1\}^d:}} \prod_{j=1}^d \left(  \frac{1}{2}+\frac{1}{2} x_j \cdot v_j \right) $ equals $1-A$. Thus, Eq.~(\ref{eqn-prob-left}) can be deduced as
\begin{align} \label{eqn-prob-left-3}
\mathbb{P}[\mathcal{M}(x)=x^*]=  \frac{\alpha}{|T^+|} \cdot A + \frac{1-\alpha}{|T^-|}  \cdot (1-A).
\end{align}

Given $\alpha > 1/2$ and $|T^+| \leq |T^-|$, it follows that $\frac{\alpha}{|T^+|} > \frac{1-\alpha}{|T^+|} \geq \frac{1-\alpha}{|T^-|}$. Since $\left(  \frac{1}{2}+\frac{1}{2} x_j \cdot v_j \right) \geq 0$ for any $x_j \in [-1, 1]$, $v_j \in \{-1,1\}$, and $j \in [1,d]$,  then both $A$ and $1-A$ are \mbox{non-negative}. Then, it holds $0 \leq A \leq 1$.

Therefore, the maximum value of $A$ is $1$, and the minimum value of $A$ is $0$. Then, the Eq.~(\ref{eqn-prob-left-3}) will take the maximum value when $A=1$ and take the minimum value when $A=0$. Since $A=\sum_{ _{x^*\cdot v>0}^{v \in \{-1,1\}^d:}} \prod_{j=1}^d \left(  \frac{1}{2}+\frac{1}{2} x_j \cdot v_j \right)$ and $x\in[-1,1]^d$, we can easily know that $A=1$ if
% \begin{align}
    $x\in \{ v:v\in\{-1,1\}^d, x^*\cdot v>0 \}$.
% \end{align}
And, $A=0$ if
% \begin{align}
    $x\in \{ v:v\in\{-1,1\}^d, x^*\cdot v\leq 0 \}$.
% \end{align}
Hence, the maximum value of Eq.~(\ref{eqn-prob-left-3}) is $\frac{\alpha}{|T^+|}$ and the minimum value of Eq.~(\ref{eqn-prob-left-3}) is $\frac{1-\alpha}{|T^-|}$. Similarly, it can get the same results when given input $x'$. We omit the proof for brevity. \qeda

Therefore, based on Eq.~(\ref{eqn-eps-delta-ldp-single}) and Lemma~\ref{max-min-value}, to achieve $(\epsilon, \delta)$-local differential privacy, we only need to guarantee
\begin{align}\label{max-min-proof-ldp}
    \frac{\alpha}{|T^+|}\leq \frac{1-\alpha}{|T^-|}\cdot e^\epsilon + \delta.
\end{align}
By combining Eqs.~(\ref{eqn-T-odd}), (\ref{eqn-T-even}) and (\ref{max-min-proof-ldp}), we can obtain
\begin{align} \label{alpha-our}
    \alpha=
    \begin{cases}
    \frac{e^\epsilon+|T^+|\cdot\delta}{e^\epsilon+1},&\text{~if~}d\text{~is odd,} \\
    \frac{|T^+|\cdot e^\epsilon+|T^+|\cdot|T^-|\cdot\delta}{|T^+|\cdot e^\epsilon+|T^-|},&\text{~if~}d\text{~is even.}
    \end{cases}
\end{align}
By taking Eqs.~(\ref{eqn-T-odd-val}) and (\ref{eqn-T-even-val}) into Eq.~(\ref{alpha-our}), we can get
\begin{align}\label{alpha-val-our}
    \alpha=
    \begin{cases}
    \frac{e^\epsilon+C_d\cdot\delta}{e^\epsilon+1},&\text{~if~}d\text{~is odd,} \\
    \frac{e^\epsilon \cdot C_d + \delta \cdot C_d(2^d-C_d)}{(e^\epsilon-1)C_d+2^d},&\text{~if~}d\text{~is even.}
    \end{cases}
\end{align}

Additionally, it should be noted that it needs to ensure $C_d\cdot \delta<1$ in Eq.~(\ref{alpha-val-our}) in order to make $\alpha<1$.

\begin{lem}\label{algo-multi-unbiased} 
Algorithm~\ref{ldp-algo-our-multi} is an unbiased estimator of the input $x$ when $B$ is calculated by Eq.~(\ref{B-value}).
\end{lem}

\noindent\textbf{Proof.} We present the proof in Appendix~A.3
of the online full version~\cite{fullversion} due to space limitation. \qeda

%Appendix~\ref{appen-proof-unbiased} for the complete proofs.

\begin{thm}\label{err-algo-multi}
For any $j\in[1,d]$, let $Z_j=\frac{1}{N}\sum_{i=1}^N x_j^*(i)$ and $X_j=\frac{1}{N}\sum_{i=1}^N x_j(i)$. Then Algorithm~\ref{ldp-algo-our-multi} ensures that with at least $1-\beta$ probability,
\begin{align}
    \underset{j\in[1,d]}{\max}|Z_j-X_j|=O\left( \frac{\sqrt{d\log(d/\beta)}}{(\epsilon+2^d\cdot \delta)\sqrt{N}} \right).
\end{align}
\end{thm}

\noindent\textbf{Proof.} For any dimension $j\in[1,d]$ and user $i\in[1,N]$, it holds
% the variance of $x_j^*(i)-x_j(i)$ equals:
\begin{align}
    Var[x_j^*(i)-x_j(i)]&=Var[x_j^*(i)]
    =\mathbb{E}[(x_j^*(i))^2]-(\mathbb{E}[x_j^*(i)])^2 \nonumber\\
    &=\sum_{x_j^*(i)}(x_j^*(i))^2\bp{x_j^*(i)}-(x_j(i))^2 \nonumber\\
    &=\sum_{x_j^*(i)}B^2\bp{x_j^*(i)}-(x_j(i))^2 \nonumber\\
    &=B^2-(x_j(i))^2 \leq B^2
\end{align}
Since Algorithm~\ref{ldp-algo-our-multi} is an unbiased estimator of the input $x$, based on Lemma~\ref{algo-multi-unbiased}, it holds
% \begin{align}
    $|x_j^*(i)-x_j(i)|\leq B+1$.
% \end{align}
Then, by the Bernstein inequality (see Definition 4.1 of~\cite{cormode2018marginal}), we have
\begin{small}
\begin{align}
    &\bp{|Z_j-X_j|\geq \lambda} =\bp{\bigg|\frac{1}{N}\sum_{i=1}^{N}\{x_j^*(i)-x_j(i)\}\bigg|\geq \lambda} \nonumber\\
    &\leq 2\cdot \exp\bigg( -\frac{N\lambda^2}{\frac{2}{N}\sum_{i=1}^N Var[x_j^*(i)-x_j(i)]+\frac{2}{3}\lambda(B+1)} \bigg) \nonumber\\
    &=2\cdot \exp\bigg( -\frac{N\lambda^2}{2B^2+\frac{2}{3}\lambda(B+1)} \bigg).
\end{align}
\end{small}
Based on the union bound, it holds that
\begin{small}
\begin{align}
    \bp{\underset{j\in[1,d]}{\max}|Z_j-X_j|\geq \lambda}
    &=\bp{\{|Z_1-X_1|\geq \lambda\} \cup \cdots \cup \{|Z_d-X_d|\geq \lambda\}} \nonumber\\
    &\leq \sum_{j=1}^{d}\bp{|Z_j-X_j|\geq \lambda} \nonumber\\
    &\leq 2d\cdot \exp\bigg( -\frac{N\lambda^2}{2B^2+\frac{2}{3}\lambda(B+1)} \bigg)  .\nonumber
\end{align}
\end{small}
Then, to ensure that $\underset{j\in[1,d]}{\max}|Z_j-X_j|<\lambda$ holds with at least $1-\beta$ probability, it suffices to enforce 
\begin{small}
\begin{align}
    2d\cdot \exp\bigg( -\frac{N\lambda^2}{2B^2+\frac{2}{3}\lambda(B+1)} \bigg) = \beta. \label{eqn-less-beta-1}
\end{align}
\end{small}
% To ensure $\underset{j\in[1,d]}{\max}|Z_j-X_j|<\lambda$ holds with at least $1-\beta$ probability is equivalent to make
% \begin{small}
% \begin{align} 
    % &\bp{\underset{j\in[1,d]}{\max}|Z_j-X_j|\geq \lambda}
%     \nonumber\\
%     &=2d\cdot \exp\bigg( -\frac{N\lambda^2}{2B^2+\frac{2}{3}\lambda(B+1)} \bigg)\leq \beta. \label{eqn-less-beta-1}
% \end{align}
% \end{small}
By solving Eq.~(\ref{eqn-less-beta-1}), we get
% \begin{align}
    $\lambda = O\left(  B\cdot \sqrt{\log(d/\beta)}/\sqrt{N} \right)$.
% \end{align}

We now analyze $B$ in Eq.~(\ref{B-value}); i.e., $B:=
    \begin{cases}
    \frac{2^d+C_d\cdot(e^\epsilon-1)}{\binom{d-1}{(d-1)/2}\cdot(e^\epsilon +2^d\cdot \delta-1)},&\text{~if~}d\text{~is odd},\\ 
    \frac{2^d+C_d\cdot(e^\epsilon-1)}{\binom{d-1}{d/2}\cdot(e^\epsilon +2^d\cdot \delta -1)},&\text{~if~}d\text{~is even}.
    \end{cases}$ First, $C_d :=  2^{d-1}$ for odd $d$, and $C_d:=  2^{d-1}-\frac{1}{2}\binom{d}{d/2} = 2^{d-1}-o(2^{d-1})$ for even and large $d$, where $o(2^{d-1})$ represents a quantity $f(d)$ which satisfies $\frac{f(d)}{2^{d-1}} \to 0$ as $d\to \infty$.  Hence, we obtain $2^d+C_d\cdot(e^\epsilon-1) = O\left( 2^{d-1}(e^\epsilon+1) \right) = O\left( 2^{d} \right)$ for large $d$ and small $\epsilon$. We define the relation ``$\sim$'' such that two positive sequences $f_1(d)$ and $f_2(d)$ satisfy $f_1(d) \sim f_2(d)$ if and only if $ \frac{ f_1(d)}{f_2(d)} \to 1 $ as $d\to \infty$. Then for large and odd $d$, we obtain from Stirling's approximation~\cite{marsaglia1990new} that $(d-1)! \sim \sqrt{2\pi\cdot(d-1)}\cdot \left( \frac{d-1}{e} \right)^{d-1}  $ and $(\frac{d-1}{2})!  \sim \sqrt{2\pi\cdot \frac{d-1}{2}} \cdot\left( \frac{d-1}{2e} \right)^{\frac{d-1}{2}}  $, leading to $\binom{d-1}{(d-1)/2} =\frac{(d-1)!}{[(\frac{d-1}{2})!]^2}   \sim \frac{2^{d-1}}{\sqrt{\pi(d-1)/2}}    \sim \frac{2^{d}}{\sqrt{d}}   $. In a similar way, for large and even $d$, we obtain $\binom{d-1}{d/2}  \sim \frac{2^{d}}{\sqrt{d}}   $. For small $\epsilon$, we have $e^\epsilon-1 = \epsilon + o(\epsilon)$, where $o(\epsilon)$ represents a quantity $g(\epsilon)$ which satisfies $\frac{g(\epsilon)}{\epsilon} \to 0$ as $\epsilon\to 0$. Combining the above results, we finally derive $B = O\Big(  \frac{2^{d}}{ \frac{2^{d}}{\sqrt{d}} \cdot (\epsilon+2^d\cdot \delta) } \Big) =O\left(\frac{\sqrt{d}}{\epsilon+2^d\cdot \delta} \right)$. Hence, there exists $\lambda=O\left( \frac{\sqrt{d\log (d/\beta)}}{(\epsilon+2^d\cdot \delta)\sqrt{N}} \right)$ such that $\underset{j\in[1,d]}{\max}|Z_j-X_j|<\lambda$ holds with at least $1-\beta$ probability. \qeda

\begin{rem}
In Algorithm~\ref{ldp-algo-our-multi}, we select $T^+(v)$ (resp. $T^-(v)$) be the set of all tuples $x^*\in\{-B,B\}^d$ such that $x^*\cdot v > 0$ (resp. $x^*\cdot v\leq 0$). It should be noted that we can also select $T^+(v)$ (resp. $T^-(v)$) be the set of all tuples $x^*\in\{-B,B\}^d$ such that $x^*\cdot v \geq 0$ (resp. $x^*\cdot v < 0$). In this case, when $d$ is odd, the result is the same as Eq.~(\ref{alpha-val-our}) since $T^+(v)=T^-(v)$. But when $d$ is even, we have
\begin{align}\label{eqn-T-even-2}
    % \begin{cases}
    \left | T^+(v) \right |=\sum_{j\geq\frac{d}{2}}\binom{d}{j},~
    \left | T^-(v) \right |=\sum_{j\leq\frac{d}{2}-1}\binom{d}{j}.
    % \end{cases}
\end{align}
Thus, we can get
\begin{align}\label{eqn-T-even-val-2}
    \begin{cases}
    \left | T^+(v) \right |=2^{d-1}+\frac{1}{2}\binom{d}{d/2},\\
    \left | T^-(v) \right |=2^{d-1}-\frac{1}{2}\binom{d}{d/2}. 
    \end{cases}
\end{align}
By taking Eqs.~(\ref{eqn-T-odd-val}) and (\ref{eqn-T-even-val-2}) into Eq.~(\ref{alpha-our}), it can obtain
\begin{align}\label{alpha-val-our-2}
    \alpha=
    \begin{cases}
    \frac{e^\epsilon+C_d\cdot\delta}{e^\epsilon+1},&\text{~if~}d\text{~is odd,} \\
    \frac{e^\epsilon \cdot (2^d-C_d) + \delta \cdot C_d(2^d-C_d)}{e^\epsilon \cdot (2^d-C_d)+C_d},&\text{~if~}d\text{~is even.}
    \end{cases}
\end{align}
\end{rem}

\subsection{Our Second Solution for Multiple Numeric Attributes under \mbox{$(\epsilon,\delta)$-LDP}}

Before introducing our second mechanism for multiple numeric attributes, we first show the algorithm that preserves single numeric attributes under \mbox{$(\epsilon,\delta)$-LDP}. Based on Algorithm~\ref{ldp-algo-our-multi} in Section~\ref{sec-numeric-first}, we can easily deduce the \mbox{$(\epsilon,\delta)$-LDP} mechanism for one-dimensional numeric data. Algorithm~\ref{ldp-algo-our-one} presents the pseudo-code of the solution for one-dimensional numeric attribute under \mbox{$(\epsilon,\delta)$-LDP}. It can be seen that given a tuple $x \in [-1,1]$, the algorithm returns a perturbed tuple $x^*$ that equals either $\frac{e^\epsilon + 1}{e^\epsilon+2\delta-1}$ or $-\frac{e^\epsilon + 1}{e^\epsilon+2\delta-1}$, with the following probabilities:
\begin{align} \label{ldp-prob}
\mathbb{P}[x^* \mid x] = 
\begin{cases}
\frac{e^\epsilon + 2\delta -1}{2(e^\epsilon+1)}\cdot x +\frac{1}{2}, &\text{~if~}x^*=\frac{e^\epsilon +1}{e^\epsilon+2\delta-1},\\
-\frac{e^\epsilon  + 2\delta -1}{2(e^\epsilon+1)}\cdot x +\frac{1}{2}, &\text{~if~}x^*=-\frac{e^\epsilon +1}{e^\epsilon+2\delta-1}.
\end{cases}
\end{align}

\begin{algorithm}[tb]\label{ldp-algo-our-one}
\small
\setstretch{1} 
 \caption{Mechanism for One-dimensional Numeric Data under \mbox{$(\epsilon, \delta)$-LDP}} 
  \KwIn{tuple $x\in [-1, 1]$ and privacy parameters $\epsilon$ and $\delta$} 
  \KwOut{perturbed tuple $x^*\in \{-\frac{e^\epsilon +1}{e^\epsilon+2\delta-1}, \frac{e^\epsilon + 1}{e^\epsilon+2\delta-1}\}$} 
    Sample a Bernoulli variable $u$ such that $\mathbb{P}[u=1]=\frac{e^\epsilon + 2\delta -1}{2(e^\epsilon+1)}\cdot x +\frac{1}{2}$\; 
    \eIf{$u=1$}
    {$x^* = \frac{e^\epsilon + 1}{e^\epsilon+2\delta-1}$\;}
    {$x^* = -\frac{e^\epsilon + 1}{e^\epsilon+2\delta-1}$\;}
    \textbf{return} $x^*$\; 
\end{algorithm}

\begin{thm}\label{thm-algo-1-ldp} 
Algorithm~\ref{ldp-algo-our-one} satisfies \mbox{$(\epsilon, \delta)$-local differential privacy}.
\end{thm}

We omit the proof of Theorem~\ref{thm-algo-1-ldp} since Algorithm~\ref{ldp-algo-our-one} is the simple version of Algorithm~\ref{ldp-algo-our-multi} when $d=1$.

\begin{lem}\label{thm-algo-one-unbiased} 
Algorithm~\ref{ldp-algo-our-one} is an unbiased estimator of the input value $x$. And, the variance of the perturbed value $x^*$ in the worst-case is 
\begin{align}
    Var[x^*]=\left(\frac{e^\epsilon +1}{e^\epsilon+2\delta-1}\right)^2.
\end{align}
\end{lem}

\noindent\textbf{Proof.} Since $x^* \in \{-\frac{e^\epsilon+1}{e^\epsilon+2\delta-1}, \frac{e^\epsilon+1}{e^\epsilon+2\delta-1} \}$, the expectation of $x^*$ is computed as
\begin{small}
\begin{align}
\mathbb{E}[x^*]&=\frac{e^\epsilon +1}{e^\epsilon+2\delta-1}\cdot \frac{x \cdot(e^\epsilon+2\delta-1)+e^\epsilon+1}{2(e^\epsilon+1)} \nonumber \\
&\quad\quad + \Big(-\frac{e^\epsilon +1}{e^\epsilon+2\delta-1}\Big)\cdot\frac{-x \cdot(e^\epsilon + 2\delta -1)+e^\epsilon+1}{2(e^\epsilon+1)} \nonumber\\
%&=\frac{e^\epsilon+1}{e^\epsilon+2\delta-1}\cdot \nonumber\\ &\left[\frac{x_i(e^\epsilon+2\delta-1)+e^\epsilon+1+x_i(e^\epsilon+2\delta-1)-e^\epsilon-1}{2(e^\epsilon+1)}\right] \nonumber\\
&=\frac{e^\epsilon+1}{e^\epsilon+2\delta-1}\cdot\Big(\frac{2x(e^\epsilon+2\delta-1)}{2(e^\epsilon+1)}\Big) =x.\nonumber
\end{align}
\end{small}
Then, the variance is computed as:
\begin{small}
\begin{align}\label{ldp-variance}
    Var[x^*]&=\mathbb{E}[(x^*)^2]-(\mathbb{E}[x^*])^2 \nonumber\\
    &=\Big(\frac{e^\epsilon +1}{e^\epsilon+2\delta-1}\Big)^2\cdot\frac{x\cdot(e^\epsilon+2\delta-1)+e^\epsilon+1}{2(e^\epsilon+1)}\nonumber \\
    &\quad\quad +\Big(\frac{-(e^\epsilon+1)}{e^\epsilon+2\delta-1}\Big)^2\cdot\frac{-x\cdot(e^\epsilon+2\delta-1)+e^\epsilon+1}{2(e^\epsilon+1)} - x^2 \nonumber\\
    &=\Big(\frac{e^\epsilon +1}{e^\epsilon+2\delta-1}\Big)^2-x^2.
\end{align}
\end{small}
Therefore, the worst-case variance of $x^*$ equals to $\Big(\frac{e^\epsilon +1}{e^\epsilon+2\delta-1}\Big)^2$, and it occurs when $x=0$. \qeda

\begin{algorithm}[tb]\label{ldp-algo-our-one-d}
\small
\setstretch{1} 
 \caption{Mechanism for Multi-dimensional Numeric Data under \mbox{$(\epsilon, \delta)$-LDP} (\mbox{Mechanism-2})} 
  \KwIn{tuple $x\in [-1, 1]^d$ and privacy parameters $\epsilon$ and $\delta$} 
  \KwOut{perturbed tuple $x^*\in \{-\frac{e^\epsilon +1}{e^\epsilon+2\delta-1}\cdot \frac{d}{k}, 0, \frac{e^\epsilon + 1}{e^\epsilon+2\delta-1}\cdot \frac{d}{k}\}^d$}
    Initialize $x^*=\left \langle 0,0,\cdots,0 \right \rangle^d$\;
    Let $k=\max\{1,\min\{d,\left \lfloor \frac{\epsilon}{\tau} \right \rfloor\}\}$\; 
    Sample $k$ values uniformly without replacement from $\{1,2,\cdots,d\}$\;
    \For{each sampled value $j$ in $k$}
    {Take $x_j$, $\frac{\epsilon}{k}$ and $\frac{\delta}{k}$ as input to Algorithm~\ref{ldp-algo-our-one} and obtain a noisy value $\Bar{x}_j$\;
    $x_j^*=\frac{d}{k}\Bar{x}_j$\;}
    \textbf{return} $x^*$\; 
\end{algorithm}

\begin{thm}\label{err-algo-one}
Let $Z=\frac{1}{N}\sum_{i=1}^N x^*(i)$ and $X=\frac{1}{N}\sum_{i=1}^N x(i)$. The Algorithm~\ref{ldp-algo-our-one} holds that with at least $1-\beta$ probability,
\begin{align}
    |Z-X|=O\left( \frac{\sqrt{\log(1/\beta)}}{(\epsilon+2\delta)\sqrt{N}} \right).
\end{align}
\end{thm}

We omit the proof of Theorem~\ref{err-algo-one} since it's a special case of Theorem~\ref{err-algo-multi} when $d=1$ in Section~\ref{sec-numeric-first}.

When collecting multiple numeric attributes privately, a straightforward method is to use a one-dimensional numeric data perturbation algorithm (e.g., Algorithm~\ref{ldp-algo-our-one}), such that the privacy parameters of each attribute are given as $\epsilon/d$ and $\delta/d$. By composition theorem \cite{mcsherry2009privacy,kasiviswanathan2011can}, this method satisfies \mbox{$(\epsilon,\delta)$-LDP}. However, based on Theorem~\ref{err-algo-one}, the noise bound of each attribute will be $O\left( \frac{d\sqrt{\log d}}{(\epsilon+2\delta)\sqrt{N}} \right)$, which is super-linear to $d$. So, this solution leads to an inferior data utility especially when $d$ becomes large.

To address this problem, we follow the spirit of~\cite{wang2019collecting} to only perturb $k$ attributes instead of $d$ attributes, which will increase the privacy budget of each attribute from $\epsilon/d$ to $\epsilon/k$, thus reducing the noise variance in turn. 
Algorithm~\ref{ldp-algo-our-one-d} shows the pseudo-code of our extension of Algorithm~\ref{ldp-algo-our-one} for multi-dimensional numeric data. Given any $d$-dimensional tuple $x\in[-1,1]^d$, Algorithm~\ref{ldp-algo-our-one-d} will return a perturbed tuple $x^*$ with $k$ non-zero values. Specifically, it uniformly at random selects $k$ attributes from all $d$ attribute and perturbs these $k$-dimensional data instead, where $k$ is chosen by Lemma~\ref{choose-k}. Then, for each sampled dimension $j\in[1,k]$, Algorithm~\ref{ldp-algo-our-one-d} takes $x_j$, $\epsilon/k$ and $\delta/k$ as inputs to Algorithm~\ref{ldp-algo-our-one} and outputs a noisy value $\Bar{x}_j$. Thus, the finally returned value is $x_j^* = \frac{d}{k}\Bar{x}_j$.

% for some positive $\tau$ in the same spirit as~\cite{wang2019collecting}:
% \begin{align}\label{k-value}
%     k=\max\{1,\min\{d,\left \lfloor \frac{\epsilon}{\tau} \right \rfloor\}\}.
% \end{align}
% Generally, we take the value of $\tau$ such that $\tau>\epsilon/2$ for the consideration of achieving high utility. This can be seen from the proof of Theorem~\ref{optimal-bound} below.

% Additionally, from Eq.~(\ref{k-value}), we can observe that the value of $k$ is always $1$ by setting $\tau>\epsilon/2$. So in particular, each user only needs to transmit one bit to the aggregator by indicating its sign, then the aggregator can re-scale the reported value using parameters $\epsilon$ and $d$. Therefore, the communication overhead of Algorithm~\ref{ldp-algo-our-one-d} is exactly $1$ bit, which is clearly optimal in reality.

\begin{lem}\label{choose-k}
The optimal $k$ of Algorithm~\ref{ldp-algo-our-one-d} is chosen as
\begin{align}\label{k-value}
    k=\max\{1,\min\{d,\left \lfloor \frac{\epsilon}{2.17} \right \rfloor\}\}.
\end{align}
% (1) if $\frac{\epsilon}{2.177}\leq 1$, then $k=1$; (2) if $\frac{\epsilon}{2.177}\geq d$, then $k=d$; (3) $1<\frac{\epsilon}{2.177}<d$, then $k=\lfloor \frac{\epsilon}{2.177} \rfloor$ or $k=\lceil \frac{\epsilon}{2.177} \rceil$, which is determined by the experimental results.
\end{lem}

\noindent\textbf{Proof.} For each dimension $j\in[1,d]$, we can compute
\begin{align}
    \mathbb{E}[\Bar{x}_j^2]
    &=Var[\Bar{x}_j]+(\mathbb{E}[\Bar{x}_j])^2 =\Big(\frac{e^{\frac{\epsilon}{k}}+1}{e^{\frac{\epsilon}{k}}+\frac{2\delta}{k}-1}\Big)^2-\Bar{x}_j^2+\Bar{x}_j^2
    =\Big( \frac{e^{\frac{\epsilon}{k}}+1}{e^{\frac{\epsilon}{k}}+\frac{2\delta}{k}-1} \Big)^2. \nonumber
\end{align}
Then, the variance is computed as 
\begin{align}\label{var-compute-k}
    &Var[x_j^*]=\mathbb{E}[(x_j^*)^2]-(\mathbb{E}[x_j^*])^2
    =\frac{k}{d}\mathbb{E}[(\frac{d}{k}\Bar{x}_j)^2]-x_j^2
    =\frac{d}{k}\mathbb{E}[\Bar{x}_j^2]-x_j^2 \nonumber\\
    &=\frac{d}{k}\Big( \frac{e^{\frac{\epsilon}{k}}+1}{e^{\frac{\epsilon}{k}}+\frac{2\delta}{k}-1} \Big)^2 -x_j^2.
\end{align}

Hence, the worst-case variance is $\frac{d}{k}\Big( \frac{e^{\frac{\epsilon}{k}}+1}{e^{\frac{\epsilon}{k}}+\frac{2\delta}{k}-1} \Big)^2$. In order to compute the optimal $k$ that minimizes the worst-case variance, we define $a:=\epsilon/k$ and $b:=2\delta/\epsilon$ so that the worst-case variance equals $\frac{d}{\epsilon} \cdot   f(a) $, for
\begin{align}\label{var-reformulated}
    f(a) : =a\left( \frac{e^a+1}{e^a+ab-1} \right)^2.
\end{align}
Thus, computing the minimum worst-case variance is equivalent to compute the minimum value of Eq.~(\ref{var-reformulated}). We first show that Eq.~(\ref{var-reformulated}) has the minimum value since it is monotonically increasing first and then monotonically decreasing in interval $(0,+\infty)$. The derivative of Eq.~(\ref{var-reformulated}) with respect to $a$ is computed as
\begin{align}\label{derivative-a}
f'(a)=\frac{(e^a+1)[(e^a+1+2ae^a)(e^a+ab-1)-2a(e^a+1)(e^a+b)]}{(e^a+ab-1)^3}.
\end{align}
Making $f'(a)=0$ is equivalent to make $(e^a+1+2ae^a)(e^a+ab-1)-2a(e^a+1)(e^a+b)=0$. By reduction formula, it gets $-a(e^a+1-2ae^a)\cdot b + e^{2a}-4ae^a-1=0$. Defining $b:=g(a)$, $g_1(a):=a(e^a+1-2ae^a)$ and $g_2(a):=e^{2a}-4ae^a-1$, we have $-g_1(a)\cdot b + g_2(a)=0$ and $b=g(a)=\frac{g_2(a)}{g_1(a)}$. It can easily compute that $a=0,0.7388$ are the solutions of $g_1(a)=0$, and $a=0,2.177$ are the solutions of $g_2(a)=0$.

Since $a>0$, thus we discuss the sign of $f'(a)$ in interval $(0,+\infty)$ in two cases. (\textit{i}) When $a<0.7388$, it holds $g_1(a)>0$ and $g_2(a)<0$, so we have $f'(a)<0$. (\textit{ii}) When $a>0.7388$, we can observe by plotting $g(a)$ that $g(a)=\frac{e^{2a}-4ae^a-1}{a(e^a+1-2ae^a)}$ is monotonically decreasing from $+\infty$ to $-\infty$. Thus, there exists one and only one $a^*>0.7388$ that satisfies $b=\frac{g_2(a^*)}{g_1(a^*)}$. For $0.7388<a<a^*$, it holds $\frac{g_2(a)}{g_1(a)}>b$ and $g_1(a)<0$, so we have $f'(a)<0$. For $0.7388<a^*<a$, it holds $\frac{g_2(a)}{g_1(a)}<b$ and $g_1(a)<0$, thus we have $f'(a)>0$. Combining the above analyses, we finally derive that $f'(a)<0$ if $a\in(0,a^*)$ and $f'(a)>0$ if $a\in(a^*,+\infty)$. Therefore, it can conclude that $f(a)$ has the minimum value.

Based on above analysis, the solution $a^*$ that minimizes the value of Eq.~(\ref{var-reformulated}) can be computed by making the derivative of Eq.~(\ref{var-reformulated}) with respect to $a$ equal to 0. By solving this, we can obtain that $2.177<a^*<2.176$ when $0<b<10^{-3}$. Note that $b$ denoting $2\delta/\epsilon$ is an extremely small value (e.g., $b = 2 \times 10^{-6} $ for $\epsilon=1$ and $\delta=10^{-6}$). Hence, we can take 2.17 as an approximate value of $a^*$.

Therefore, it holds that the variance (i.e., Eq.~(\ref{var-compute-k})) will be smallest when $k=\frac{\epsilon}{a}=\frac{\epsilon}{2.17}$. Thus, it can be known easily that the optimal $k$ is determined by $\epsilon/2.17$. Specifically, we have (\textit{i}) if $\frac{\epsilon}{2.17}\leq 1$, then $k=1$; (\textit{ii}) if $\frac{\epsilon}{2.17}\geq d$, then $k=d$; (\textit{iii}) $1<\frac{\epsilon}{2.17}<d$, then $k=\lfloor \frac{\epsilon}{2.17} \rfloor$ (choosing $\lfloor \frac{\epsilon}{2.17} \rfloor$ is because it outperforms $\lceil \frac{\epsilon}{2.17} \rceil$ through experiments). This completes the proof.
\qeda

\begin{lem}\label{algo-our-one-d-unbiased}
Algorithm~\ref{ldp-algo-our-one-d} satisfies $(\epsilon, \delta)$-local differential privacy. In addition, for any $d$-dimensional input $x\in[-1,1]$, the perturbed output $x^*$ holds $\mathbb{E}[x_j^*]=x_j$  for all dimension $j\in[1,d]$.
\end{lem}

\noindent\textbf{Proof.} Because Algorithm~\ref{ldp-algo-our-one-d} composes $k$ number of $(\frac{\epsilon}{k}, \frac{\delta}{k})$-LDP perturbation algorithms, thus based on composition theorem \cite{mcsherry2009privacy,kasiviswanathan2011can}, the Algorithm~\ref{ldp-algo-our-one-d} satisfies \mbox{$(\epsilon, \delta)$-LDP}.

As we can seen from Algorithm~\ref{ldp-algo-our-one-d}, each perturbed output $x_j^*$ equals to $\frac{d}{k}\Bar{x}_j$ with probability $k/d$ or equals to 0 with probability $1-k/d$. Thus, based on Lemma~\ref{thm-algo-one-unbiased}, it holds $\mathbb{E}[x_j^*]=\frac{k}{d}\cdot\mathbb{E}[\frac{d}{k}\Bar{x}_j]=\mathbb{E}[\Bar{x}_j]=x_j.$ \qeda

\begin{thm}\label{optimal-bound}
For any $j\in[1,d]$, let $Z_j=\frac{1}{N}\sum_{i=1}^N x_j^*(i)$ and $X_j=\frac{1}{N}\sum_{i=1}^N x_j(i)$. The Algorithm~\ref{ldp-algo-our-one-d} holds that with at least $1-\beta$ probability,
\begin{align}
    \underset{j\in[1,d]}{\max}|Z_j-X_j|=O\left(\frac{\sqrt{d\log(d/\beta)}}{(\epsilon+2\delta)\sqrt{N}}\right).
\end{align}
\end{thm}

\noindent\textbf{Proof.} For each dimension $j\in[1,d]$, we can get $|x_j^*-x_j|\leq \frac{d}{k}\frac{e^{\epsilon/k}+1}{e^{\epsilon/k}+2\delta/k-1}=O(\frac{k}{\epsilon+2\delta})\cdot \frac{d}{k}=O(\frac{d}{\epsilon+2\delta})$ based on Lemma~\ref{algo-our-one-d-unbiased}. Besides, from Eq.~(\ref{var-compute-k}), we have $Var[x_j^*]=\frac{d}{k}\Big( \frac{e^{\frac{\epsilon}{k}}+1}{e^{\frac{\epsilon}{k}}+\frac{2\delta}{k}-1} \Big)^2 -x_j^2=O\left(\frac{dk}{(\epsilon+2\delta)^2}\right)$.

% Based on Lemma~\ref{algo-our-one-d-unbiased}, we know that
% \begin{align}
%     \mathbb{E}[\Bar{x}_j^2]
%     &=Var[\Bar{x}_j]+(\mathbb{E}[\Bar{x}_j])^2
%     =\left(\frac{e^{\frac{\epsilon}{k}}+1}{e^{\frac{\epsilon}{k}}+\frac{2\delta}{k}-1}\right)^2-\Bar{x}_j^2+\Bar{x}_j^2 \nonumber\\
%     &=\left( \frac{e^{\frac{\epsilon}{k}}+1}{e^{\frac{\epsilon}{k}}+\frac{2\delta}{k}-1} \right)^2 =O\left(\frac{k^2}{(\epsilon+2\delta)^2}\right).
% \end{align}
% Hence, we can get
% \begin{align}
%     Var[x_j^*]&=\frac{d}{k}O\left(\frac{k^2}{(\epsilon+\delta)^2}\right)-x_j^2 =O\left(\frac{dk}{(\epsilon+2\delta)^2}\right).
% \end{align} 
Then using the Bernstein inequality (see Definition 4.1 of~\cite{cormode2018marginal}), we have
\begin{small}
\begin{align}
    \bp{|Z_j-X_j|\geq \lambda} &=\bp{\bigg|\sum_{i=1}^{n}\{x_j^*(i)-x_j(i)\}\bigg|\geq n\lambda} \nonumber\\
    &=2\cdot \exp\left(\frac{-N\lambda^2}{\frac{2}{N}\sum_{i=1}^{N}Var[x_j^*(i)-x_j(i)]+\frac{2}{3}\lambda\frac{d}{k}\frac{e^{\frac{\epsilon}{k}}+1}{e^{\frac{\epsilon}{k}}+\frac{2\delta}{k}-1}}\right) \nonumber\\
    &=2\cdot \exp\left(\frac{-N\lambda^2}{O\left(\frac{dk}{(\epsilon+2\delta)^2}\right)+\lambda O\left(\frac{d}{\epsilon+2\delta}\right)}\right).
\end{align}
\end{small}

Based on the union bound, we have 
\begin{align}
    \bp{\underset{j\in[1,d]}{\max}|Z_j-X_j|\geq \lambda} &=\bp{\{|Z_1-X_1|\geq \lambda\} \cup \cdots \cup \{|Z_d-X_d|\geq \lambda\}} \nonumber\\
    &\leq \sum_{i=1}^{N}\bp{|Z_j-X_j|\geq \lambda} \nonumber\\
    &=2d\cdot \exp\left(\frac{-n\lambda^2}{O\left(\frac{dk}{(\epsilon+2\delta)^2}\right)+\lambda O\left(\frac{d}{\epsilon+2\delta}\right)}\right)  . \nonumber
\end{align} 
To ensure that $\underset{j\in[1,d]}{\max}|Z_j-X_j|<\lambda$ holds with at least $1-\beta$ probability, it suffices to enforce 
\begin{align}
   2d\cdot \exp\left(\frac{-n\lambda^2}{O\left(\frac{dk}{(\epsilon+2\delta)^2}\right)+\lambda O\left(\frac{d}{\epsilon+2\delta}\right)}\right) = \beta. \label{eqn-less-beta}
\end{align}
% To ensure $\underset{j\in[1,d]}{\max}|Z_j-X_j|<\lambda$ holds with at least $1-\beta$ probability is equivalent to make
% \begin{align} \label{eqn-less-beta}
%     &\bp{\underset{j\in[1,d]}{\max}|Z_j-X_j|\geq \lambda} \nonumber\\
%     &=d\cdot 2\cdot \exp\left(\frac{-N\lambda^2}{O\left(\frac{dk}{(\epsilon+2\delta)^2}\right)+\lambda O\left(\frac{d}{\epsilon+2\delta}\right)}\right)\leq \beta.
% \end{align}
Solving Eq.~(\ref{eqn-less-beta}), we obtain $\lambda=O\left( \frac{\sqrt{dk\log(d/\beta)}}{(\epsilon+2\delta)\sqrt{N}} \right)$, where $k$ is determined by Lemma~\ref{choose-k}. Since asymptotic expressions involving $\epsilon\rightarrow0$, $\lambda$ can also be written as $O\left( \frac{\sqrt{d\log(d/\beta)}}{(\epsilon+2\delta)\sqrt{N}} \right)$. \qeda

\subsection{Comparison with Related Work}
For collecting multi-dimensional numeric data, Duchi~\textit{et al.}~\cite{duchi2018minimax} propose to perturb multi-dimensional numeric data under \mbox{$\epsilon$-LDP}, which provides strong privacy guarantees and asymptotic error bound, but remains \mbox{$(\epsilon, \delta)$-LDP} unsolved. Inspired by Duchi~\textit{et al.}'s solution \cite{duchi2018minimax}, we firstly introduce Algorithm~\ref{ldp-algo-our-multi} which focuses on achieving \mbox{$(\epsilon, \delta)$-LDP} with high data utility when handling multi-dimensional numeric data. However, Duchi~\textit{et al.}'s solution is sophisticated when handling multi-dimensional data. Afterward, Nguy{\^e}n~\textit{et al.} \cite{nguyen2016collecting} proposed Harmony to only sample one dimensional data to perturb, which is simpler and achieves the optimal asymptotic error bound as \cite{duchi2018minimax}. Similar but differently, Wang~\textit{et al.}'s~\cite{wang2019collecting} propose to uniformly select $k$ dimensions from $d$, which also yields optimal asymptotic error bound. However, both \cite{nguyen2016collecting} and \cite{wang2019collecting} only achieve \mbox{$\epsilon$-LDP} and cannot handle the case of \mbox{$(\epsilon, \delta)$-LDP}. In this paper, our proposed Algorithm~\ref{ldp-algo-our-one-d} focuses on achieving \mbox{$(\epsilon, \delta)$-LDP} while ensuring high data utility. Particularly, following the idea of \cite{wang2019collecting}, our proposed Algorithm~\ref{ldp-algo-our-one-d} requires each user to randomized report only $k$ attributes that uniformly selected from $d$ attributes, which in turn reduces the total noise variance.

\section{Frequency Estimation for Categorical Attributes under \mbox{$(\epsilon, \delta)$-LDP}}\label{sec-categorical}

This section will investigate the mechanisms $\mathcal{M}$ to achieve $(\epsilon, \delta)$-local differential privacy for categorical attributes, supporting accurate frequency estimations of each possible value in each categorical attribute's domain. 

So far most existing algorithms \cite{erlingsson2014rappor,kairouz2014extremal,bassily2015local,wang2017locally,wangtt2017locally,wang2018locally} are designed for estimating the frequencies of categorical attributes while ensuring \mbox{$\epsilon$-LDP}. Wang~\textit{et~al.} \cite{wangtt2017locally} have introduced a framework for pure LDP which can be used to analyze and optimize different \mbox{$\epsilon$-LDP} protocols. They also proposed the optimized local hashing protocol to ensure better data utility under LDP. In this section, we firstly extend their framework for approximate LDP (e.g., \mbox{$(\epsilon,\delta)$-LDP}) and then analyze and optimize different \mbox{$(\epsilon,\delta)$-LDP} protocols for frequencies estimations of categorical attributes.

\begin{defn}[\mbox{$(\epsilon, \delta)$-LDP} Protocols]\label{local-protocol}
Consider two probabilities $p>q$. A local protocol given by $\mathcal{M}$ such that a user reports the true value with probability $p$ and reports each of other values with probability $q$, will satisfy \mbox{$(\epsilon,\delta)$-LDP} if and only if it holds $p\leq q\cdot e^\epsilon +\delta$.
\end{defn}

We now consider that each of $N$ users independently executes the mechanism in Definition~\ref{local-protocol}. In this context, from Theorem 2 of \cite{wangtt2017locally},   the variance for the number of times that the true value occurs among $N$ users' noisy values will be
\begin{align}\label{var-ldp}
    \text{Var}=\frac{Nq(1-q)}{(p-q)^2}+\frac{Nf_v(1-p-q)}{p-q},
\end{align}
where $f_v$ is the frequency of the value $v\in[1,k]$. Moreover, the variance of Eq.~(\ref{var-ldp}) will be dominated by the first term when $f_v$ is small. Hence, the approximation of the variance in Eq.~(\ref{var-ldp}) is denoted as
\begin{align}\label{var-ldp-approx}
    \text{Var}^*=\frac{Nq(1-q)}{(p-q)^2}.
\end{align}
In addition, it also holds Var$^*$=Var when $p+q=1$.

Recall the problem statement in Section~\ref{sec-preliminaries}, for a categorical attribute with domain $\{1,2,\cdots,k\}$, we use $[1,k]$ to denote the domain set $\{1,2,\cdots,k\}$. Then, based on Definition~\ref{local-protocol} and the existing \mbox{$\epsilon$-LDP} protocols, we will focus on proposing  local algorithms under \mbox{$(\epsilon, \delta)$-LDP} in the following.

\textbf{General Randomized Response under Approximate LDP (GRR-ALDP)}. General randomized response \cite{kairouz2014extremal} reports the true value with probability $p$, while reporting each incorrect value with probability $q=\frac{1-p}{k-1}$. Thus, in order to make GRR-ALDP meet \mbox{$(\epsilon, \delta)$-LDP} with $p\leq q\cdot e^\epsilon +\delta$, a general randomized local protocol $\mathcal{M}$ is required to output the perturbed value $y$ when given any input value $v\in[1,k]$ with the following distributions:
\begin{align}\label{grr-aldp}
    \bp{\mathcal{M}(v)=y}=
    \begin{cases}
    p=\frac{e^\epsilon+(k-1)\delta}{e^\epsilon+k-1},&\text{~if~}y=v, \\
    q=\frac{1-\delta}{e^\epsilon+k-1},&\text{~if~}y\neq v.
    \end{cases}
\end{align}

Then, by plugging $p$ and $q$ in Eq.~(\ref{grr-aldp}) into Eq.~(\ref{var-ldp-approx}), the variance of GRR-ALDP is
\begin{align}
    \text{Var}^*_{\text{GRR-ALDP}}=\frac{N(e^\epsilon+k-2+\delta)(1-\delta)}{(e^\epsilon+k\delta-1)^2}.
\end{align}

\textbf{Parallel Randomized Response \cite{erlingsson2014rappor} under Approximate LDP (PRR-ALDP)} first encodes the value $v \in \{1,2,\ldots,k\}$ into a length-$k$ binary vector $B$ where the $v$-th bit is 1, that is $B=[0,\cdots,0,1,0,\cdots,0]$. Then, PRR-ALDP perturbs each bit of $B$ with the following probability distribution
\begin{align}\label{prob-prr-aldp}
    \bp{B^*[i]=1}=
    \begin{cases}
    p,\text{~if~}B[i]=1, \\
    q,\text{~if~}B[i]=0,
    \end{cases}
\end{align}
where $p>q$.

Based on Eq.~(\ref{prob-prr-aldp}),
% \cite{erlingsson2014rappor} and \cite{wangtt2017locally}, 
for any inputs $v_1\in \{1,2,\ldots,k\}$ and $v_2\in \{1,2,\ldots,k\}$, and output $B^*$, it holds
\begin{align}
    &\bp{B^*|v_1}\leq \bp{B^*|v_2}+\delta \nonumber\\
    \Rightarrow\quad  &\prod_{i\in[k]}\bp{B^*[i]|v_1} \leq \prod_{i\in[k]}\bp{B^*[i]|v_2} +\delta \nonumber\\
    \Rightarrow\quad  & \begin{cases} \bp{B^*[v_1]=0|v_1}\cdot \bp{B^*[v_2]=0|v_1}  \leq \bp{B^*[v_1]=0|v_2}\cdot \bp{B^*[v_2]=0|v_2}+\delta, \\
    \bp{B^*[v_1]=0|v_1}\cdot \bp{B^*[v_2]=1|v_1}  \leq \bp{B^*[v_1]=0|v_2}\cdot \bp{B^*[v_2]=1|v_2}+\delta, \\ \bp{B^*[v_1]=1|v_1}\cdot \bp{B^*[v_2]=0|v_1}  \leq \bp{B^*[v_1]=1|v_2}\cdot \bp{B^*[v_2]=0|v_2}+\delta, \\ \bp{B^*[v_1]=1|v_1}\cdot \bp{B^*[v_2]=1|v_1}  \leq \bp{B^*[v_1]=1|v_2}\cdot \bp{B^*[v_2]=1|v_2}+\delta\end{cases} \nonumber\\
    \Rightarrow\quad  &p\cdot(1-q)\leq e^\epsilon\cdot q \cdot(1-p)+\delta~\text{(This last step uses $p>q$)}. \label{prr-aldp}
\end{align}
Therefore, PRR-ALDP will satisfy \mbox{$(\epsilon, \delta)$-LDP} if and only if Inequality~(\ref{prr-aldp}) holds. Letting the equal sign in~(\ref{prr-aldp}) hold, we set $p$ as follows:
\begin{align}\label{prr-aldp-jz}
 p = \frac{qe^\epsilon+\delta}{1-q+qe^\epsilon}.
\end{align}
Applying Eq.~(\ref{prr-aldp-jz})
% \begin{align}\label{prr-aldp}
%     p(1-p)\leq e^\epsilon\cdot (1-p)q+\delta.
% \end{align}
to Eq.~(\ref{var-ldp-approx}), we obtain
\begin{align}\label{var-prr}
    \text{Var}^*_{\text{PRR-ALDP}}=\frac{Nq(1-q)(1-q+qe^\epsilon)^2}{[q(1-q)(e^\epsilon-1)+\delta]^2}.
\end{align}

\textbf{Symmetric PRR-ALDP (SPRR-ALDP)}. In RAPPOR \cite{erlingsson2014rappor}, it chooses $p$ and $q$ such that $p+q=1$, leading to a symmetric perturbation on 1 and 0. Based on this observation and Eq.~(\ref{prr-aldp-jz}), we derive
\begin{align}
    p=\frac{e^\epsilon-\sqrt{e^\epsilon(1-\delta)+\delta}}{e^\epsilon-1},~
    q=\frac{\sqrt{e^\epsilon(1-\delta)+\delta}-1}{e^\epsilon-1}.
\end{align}
Then, the variance is
\begin{align}
    &\text{Var}^*_{\text{SPRR-ALDP}}=\frac{N(\sqrt{e^\epsilon(1-\delta)+\delta}-1)(e^\epsilon-\sqrt{e^\epsilon(1-\delta)+\delta})}{(e^\epsilon-2\sqrt{e^\epsilon(1-\delta)+\delta}+1)^2}.
\end{align}

% \textbf{Optimized PRR-ALDP (OPRR-ALDP)} is to choose $p$ and $q$ by minimizing (\ref{var-prr}) instead of making $p$ and $q$ symmetric. Specifically, we can solve $q$ by making the partial derivative of (\ref{var-prr}) with respect to $q$ equal to 0. \textcolor{red}{However, it's hard to solve the partial derivative of (\ref{var-prr}) with respect to $q$ ???
% \begin{align}
%     \frac{\partial \bigg[\frac{q(1-q)(1-q+qe^\epsilon)^2}{[q(1-q)(e^\epsilon-1)+\delta]^2}\bigg]}{\partial q}
% \end{align}
% }

\textbf{Local Hashing under Approximate LDP (LH-ALDP)} first hashes the input value into a domain $[g]$ such that $g<k$, and then perturbs the hashed value by the PRR-ALDP algorithm. Denote $\mathbb{H}$ as the universal hash function family such that each hash function $H\in\mathbb{H}$ hashes each input value into a value in $[g]$. Based on \cite{wangtt2017locally}, the universal property requires that
\begin{align}
    \forall v_1,v_2\in[k],v_1\neq v_2:\underset{H\in\mathbb{H}}P[H(v_1)=H(v_2)]\leq \frac{1}{g}.
\end{align}

Given any input value $v\in[k]$, LH-ALDP first outputs a value in $[g]$ by hashing, that is $x=H(v)$. Then, LH-ALDP perturbs $x$ with the following distribution
\begin{align}\label{dis-lh-aldp}
    \forall i\in[g], \bp{y=i}=
    \begin{cases}
    p=\frac{e^\epsilon+(g-1)\delta}{e^\epsilon+g-1},&\text{~if~}x=i,\\
    q=\frac{1-\delta}{e^\epsilon+g-1},&\text{~if~}x\neq i.
    \end{cases}
\end{align}
Based on Eq.~(\ref{dis-lh-aldp}), we can know that LH-ALDP satisfies \mbox{$(\epsilon, \delta)$-LDP} since it holds $p\leq qe^\epsilon+\delta$.

Then, while aggregating in the server, it holds that
\begin{align}
    p^*=p,~q^*=\frac{1}{g}p+\frac{g-1}{g}q=\frac{1}{g}.
\end{align}
Thus, by taking $p=p^*$ and $q=q^*$ into Eq.~(\ref{var-ldp}), the variance of LH-ALDP is
\begin{align}\label{var-lh-aldp}
    \text{Var}^*_{\text{LH-ALDP}}=\frac{N(e^\epsilon+g-1)^2}{(g-1)(e^\epsilon+g\delta-1)^2}.
\end{align}

\textbf{Optimized LH-ALDP (OLH-ALDP)}. As it can seen from Eq.~(\ref{var-lh-aldp}), we can minimize the variance of LH-ALDP by making the partial derivative of Eq.~(\ref{var-lh-aldp}) with respect to $g$ equal to 0.
% . That is
% \begin{align}
%     \frac{\partial \bigg[\frac{(e^\epsilon+g-1)^2}{(g-1)(e^\epsilon+g\delta-1)^2}\bigg]}{\partial g} \label{derivative-OLH}.
% \end{align}
% \textcolor{red}{However, it's hard to solve the partial derivative of (\ref{var-lh-aldp}) with respect to $g$ ???}
% To make the Eq.~(\ref{derivative-OLH}) with respect to $g$ equal to 0 is equivalent to solve the following equation:
That is, it needs to solve the following equation:
\begin{align}
    & -\delta^2\cdot g^3 - 3(e^\epsilon-1)\delta^2\cdot g^2 + \left[ (e^\epsilon-1)^2+2(e^\epsilon-1)\delta(\delta-2e^\epsilon+1) \right]\cdot g \nonumber\\
    & + (e^\epsilon-1)^2(2\delta -e^\epsilon -1) = 0. \label{derivative-equal-0}
\end{align}
Hence, the optimal $g$ is the solution to the cubic Eq.~(\ref{derivative-equal-0}), %Although its exact expression is complex and we omit the details here, it can be efficiently computed.
that is,
\begin{align}
    g=
    \frac{-3e^\epsilon\delta-\sqrt{e^\epsilon-1}\sqrt{(1-\delta)(e^\epsilon+\delta-9e^\epsilon\delta-1)}+e^\epsilon+3\delta-1}{2\delta}.
    \nonumber
\end{align}

\textbf{Optimal Gaussian Mechanism (\mbox{Opt-GM})}. When applying Gaussian mechanism on categorical attributes, an input value $v\in[1,k]$ is also encoded into a length-$k$ binary vector $B$ firstly. The vector $B$ has the same properties as described in PRR-ALDP. 

After encoding $v$ into a vector $B$, the \mbox{Opt-GM} will output the noisy vector $B^*$ such that each $B^*[i]$ is obtained by perturbing $B[i] \in \{0, 1\}$ via adding noise with a Gaussian distribution $\mathcal{N}(0,\sigma^2)$, where $\sigma$ is computed by Eq.~(\ref{eqn-DP-OPT}). And $\ell_2$-sensitivity is $\sqrt{2}$ since the binary vectors of two different input $v$ and $v'$ differ only in two bits. After collecting the noisy vectors from $N$ users ($B^*(j)$ for user $j \in \{1,2,\ldots,N\}$), the aggregator simply computes $ \sum_{j=1}^N B^*(j)[v]$ as the count for $v$ (if the result is not an integer, rounding can be applied; and a negative result can be considered as $0$). Although this method seems naive, its performance is not terrible due to large $N$ as shown in our experiments later. If we ignore the effect of rounding, the variance of Opt-GM is 
% \begin{align}
    $\text{Var}^*_{\text{Opt-GM}}=N\sigma^2$.
% \end{align}

\begin{figure}\centering
	\includegraphics[height=3.5cm]{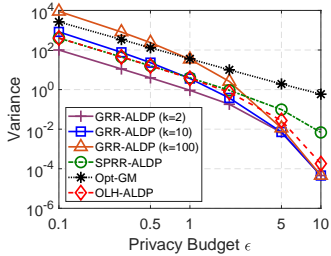}
	\vspace{-2mm}
	\caption{The noise variances of mechanisms under \mbox{$(\epsilon,\delta)$-LDP} for categorical attributes versus $\epsilon$ when $\delta=10^{-6}$}
	\label{Var-Cate}
	\vspace{-2mm}
\end{figure}

We compare the variance of the above mechanisms as shown in Fig.~\ref{Var-Cate}. For GRR-ALDP, the domain is set as $k=2, 10, 100$, respectively. It can be seen that the size of domain $k$ has a big impact on the variance of GRR-ALDP when the privacy protection level is high (i.e., the $\epsilon$ is small), that is a larger domain $k$ leads to a bigger variance. And this impact can be relatively eliminated under low privacy protection level (e.g., when $\epsilon=10$). In particular, GRR-ALDP has the smallest variance among all mechanisms when $k=2$. Overall, it shows that GRR-ALDP will be more appropriate when $k$ is small or when $\epsilon$ is extremely large.

In addition, we can observe that Opt-GM relatively has the biggest variance when compared with SPRR-ALDP and OLH-ALDP. Besides, the variance of SPRR-ALDP and OLH-ALDP are very close to each other when $\epsilon$ is small (i.e., when $\epsilon \leq 1$). And the OLH-ALDP will outperform SPRR-ALDP when $\epsilon$ becomes bigger (i.e., when $\epsilon > 1$). Therefore, to sum up, OLH-ALDP is always better than SPRR-ALDP and Opt-GM in a wide range of $\epsilon$. Moreover, OLH-ALDP is more applicable than GRR-ALDP in practical since the performance of the latter mechanism is overly dependent on the size of the domain.

\section{Experiments}\label{sec-experiments}

In this section, we evaluated the performance of our proposed mechanisms by using two public datasets (denoted as BR and MX) which contain census records from Brazil and Mexico, both extracted from the Integrated Public Use Microdata Series\footnote{\url{https://www.ipums.org}}. Both BR and MX have 4M tuples (e.g., users). Specifically, BR contains 16 attributes of which 6 are numeric attributes (e.g., income) and 10 are categorical attributes (e.g., gender); and MX contains 19 attributes of which 5 are numeric attributes and 14 are categorical attributes. Without loss of generality, we normalize the data domain of each numeric attribute into $[-1,1]$.

As mentioned before, we demonstrate the accuracy of our proposed mechanisms from two perspectives, that is, (i) the accuracy on mean/frequency estimation and (ii) the accuracy on building machine learning models. We implement all algorithms and experiments using Python 2.7, running on a Windows 10 PC with Intel Xeon E5-1650 3.20 GHz CPU and 16G RAM.

\begin{figure}
  \centering
  \begin{tabular}{cccc} 
    \hspace{-20mm}\includegraphics[height=3cm]{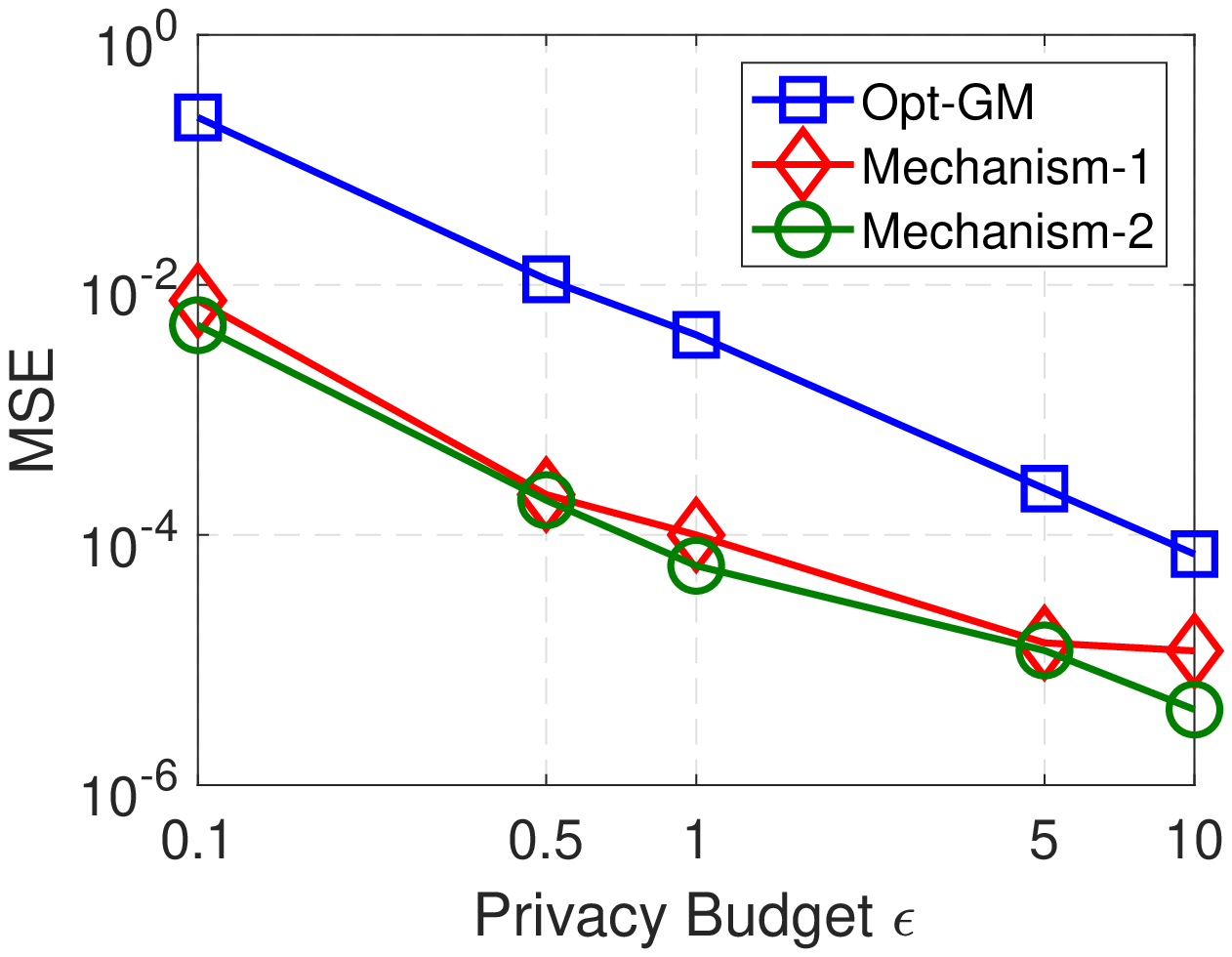}&
    \hspace{-3mm}\includegraphics[height=3cm]{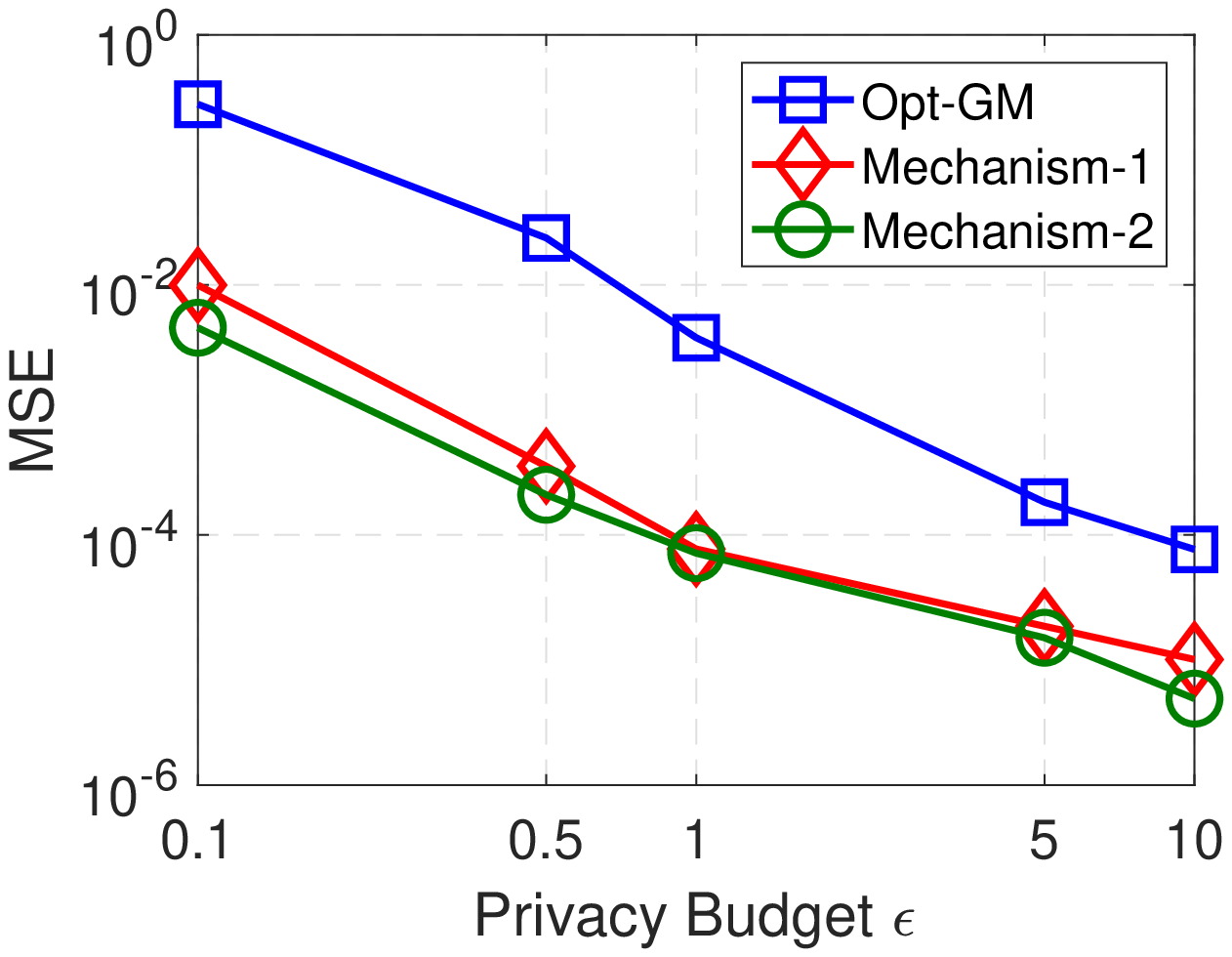}&
    \hspace{-3mm}\includegraphics[height=3cm]{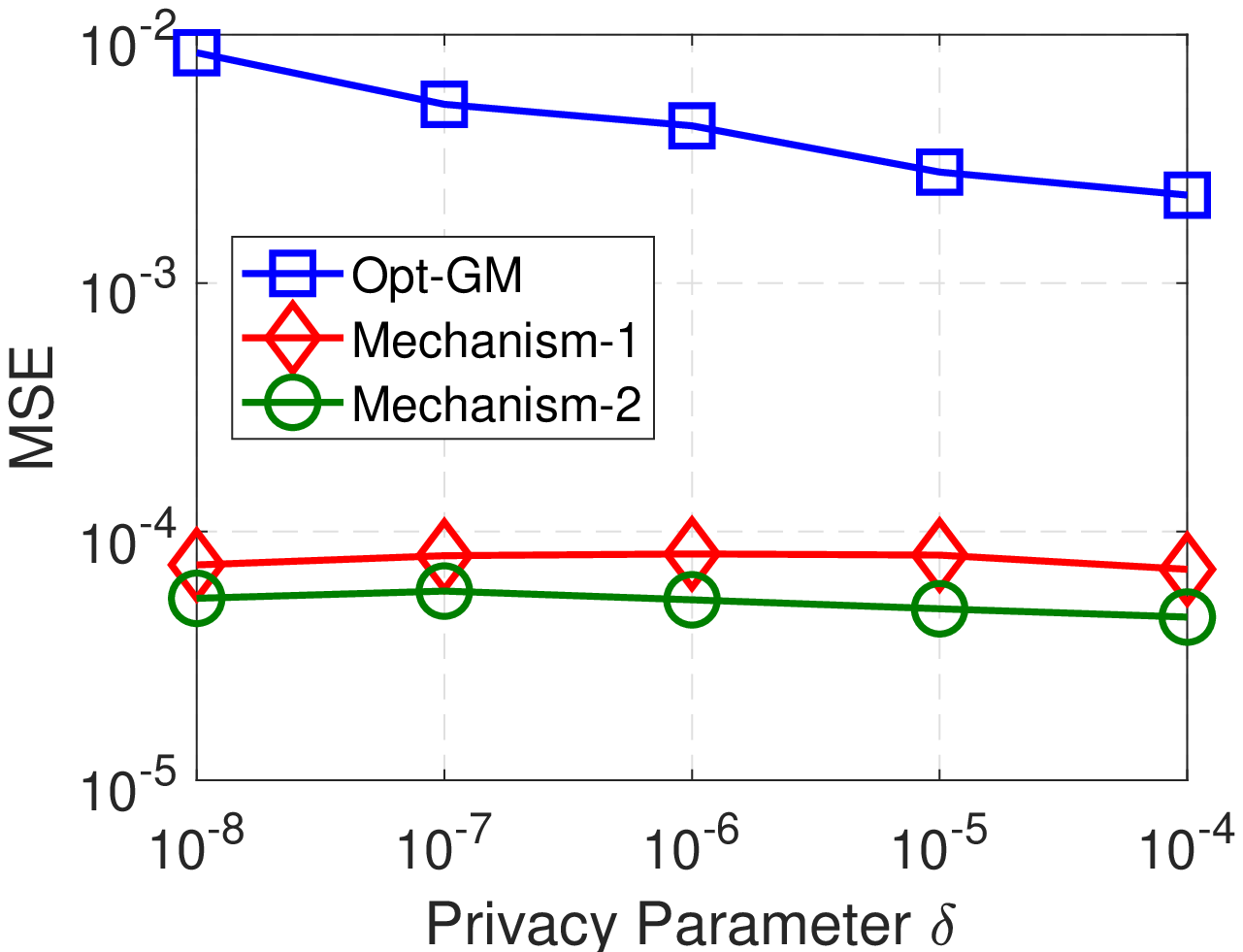}&
    \hspace{-3mm}\includegraphics[height=3cm]{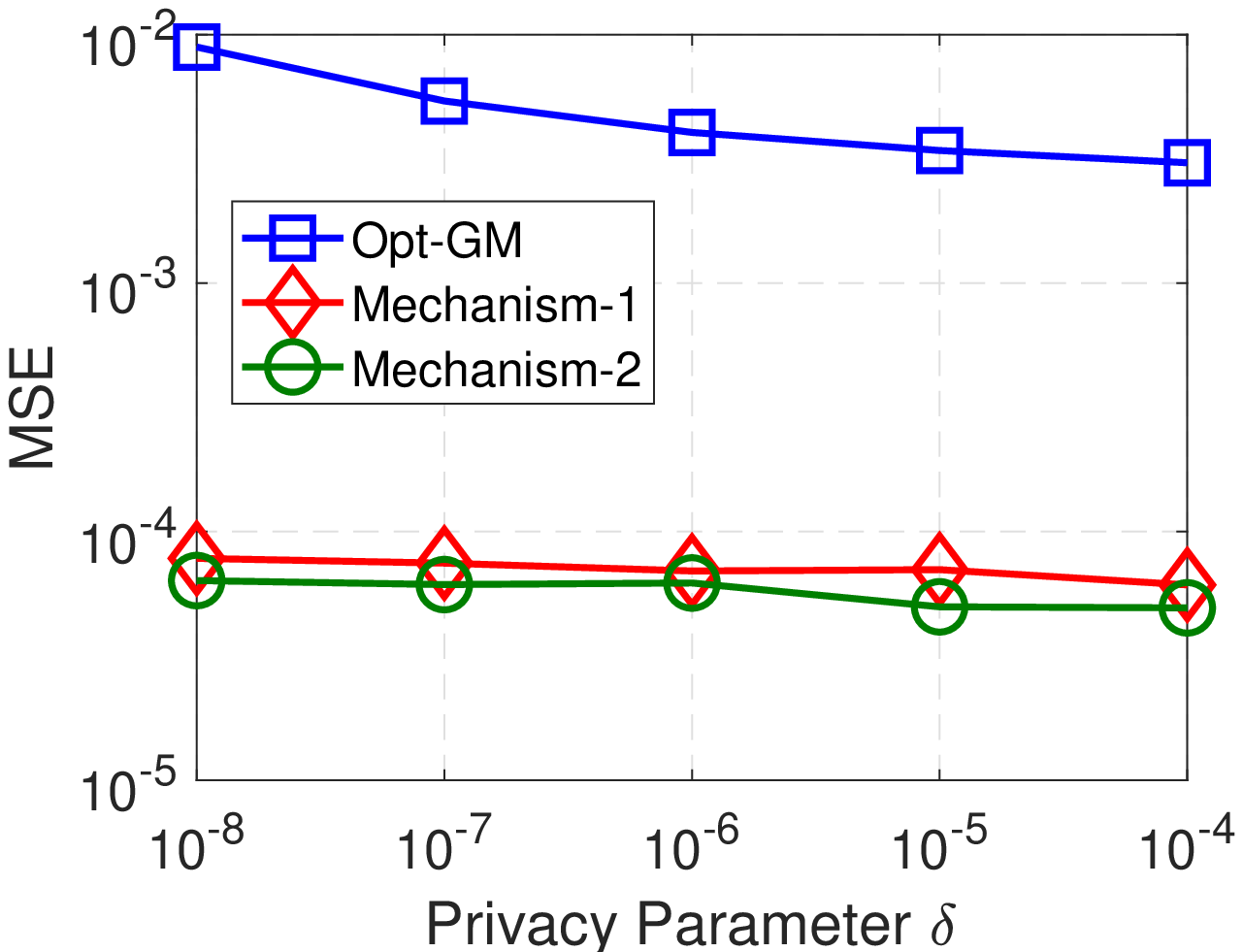}\\[-3mm]
   \hspace{-15mm}\scriptsize (a) MX-Numeric ($\delta=10^{-6}$) & 
   \hspace{-2mm}\scriptsize (b) BR-Numeric ($\delta=10^{-6}$) &
   \hspace{-2mm}\scriptsize (c) MX-Numeric ($\epsilon=1$) &
   \hspace{-2mm}\scriptsize (d) BR-Numeric ($\epsilon=1$)
   \end{tabular} 
 	\vspace{-5mm}
    \caption{Accuracy for mean estimation on numeric attributes.} \label{real-mean-vs-epsilon}
	\vspace{-2mm} 
\end{figure}

\subsection{Results on Mean/Frequency Estimation}

% \begin{figure}
%   \centering
%   \begin{tabular}{cc} 
%     \hspace{-10pt}\includegraphics[height=3.3cm]{figures/Nu_MX_eps_1.eps}&\hspace{-15pt}
%     \includegraphics[height=3.3cm]{figures/Nu_BR_eps_1.eps}\\[-1pt]
%   \scriptsize (a) MX-Numeric ($\epsilon=1$) & \scriptsize (b) BR-Numeric ($\epsilon=1$)
%   \end{tabular} 
%  	\vspace{-1mm}
%     \caption{Accuracy for mean estimation on numeric attributes.} \label{real-mean-vs-delta}
% 	\vspace{-5mm} 
% \end{figure}

In our first experimental settings, we consider the scenario that each user reports her/his multi-dimensional data tuple based on local differential privacy mechanisms and then the server collects and aggregates all the perturbed data and computes the estimations of the mean value for numeric attributes and the frequency value for categorical attributes. In particular, to show the accuracy of our proposed mechanisms, we evaluate the mean square error (MSE) of the estimated mean values for numeric attributes and frequencies for categorical attributes.

The accuracy for mean estimation on numeric attributes varying form different privacy parameters on both datasets MX and BR is shown in Fig.~\ref{real-mean-vs-epsilon}. On the whole, it can be seen that both our proposed two mechanisms significantly outperform the optimal Gaussian mechanism (i.e., Opt-GM) in all cases under \mbox{$(\epsilon,\delta)$-LDP}. And the MSE of \mbox{Mechanism-1} and \mbox{Mechanism-2} is close to each other and much smaller than Opt-GM. This is because \mbox{Mechanism-1} and \mbox{Mechanism-2} are unbiased estimations on mean values, thus holding much smaller variances than Opt-GM. Besides, Figs.~\ref{real-mean-vs-epsilon}(a) and (b) indicate the larger the privacy budget $\epsilon$ is, the lower MSE will be. In addition, it can be seen again from Figs.~\ref{real-mean-vs-epsilon}(c) and (d) that Opt-GM always has the biggest MSE among three mechanisms. With the increase of $\delta$, we can observe that the MSE of Opt-GM decreases gradually, while the MSEs of \mbox{Mechanism-1} and \mbox{Mechanism-2} are almost unchanged. This indicates the size of the privacy parameter $\delta$ has little impact on the accuracies of \mbox{Mechanism-1} and \mbox{Mechanism-2}.

Furthermore, we also conduct extensive experiments on synthetic datasets to compare the effects of different parameters, i.e., privacy parameters $\epsilon$ and $\delta$, dimension $d$. Specifically, each synthetic dataset contains $400,000$ tuples and is generated from a Gaussian distribution $\mathcal{N}(0,1/16)$ with mean value 0 and variance 1/16. We consider four synthetic datasets with different dimension in our experiments, i.e., $d=1,5,10,15$.

\begin{figure*}
  \centering
  \begin{tabular}{cccc} 
    \hspace{-20mm}\includegraphics[height=3cm]{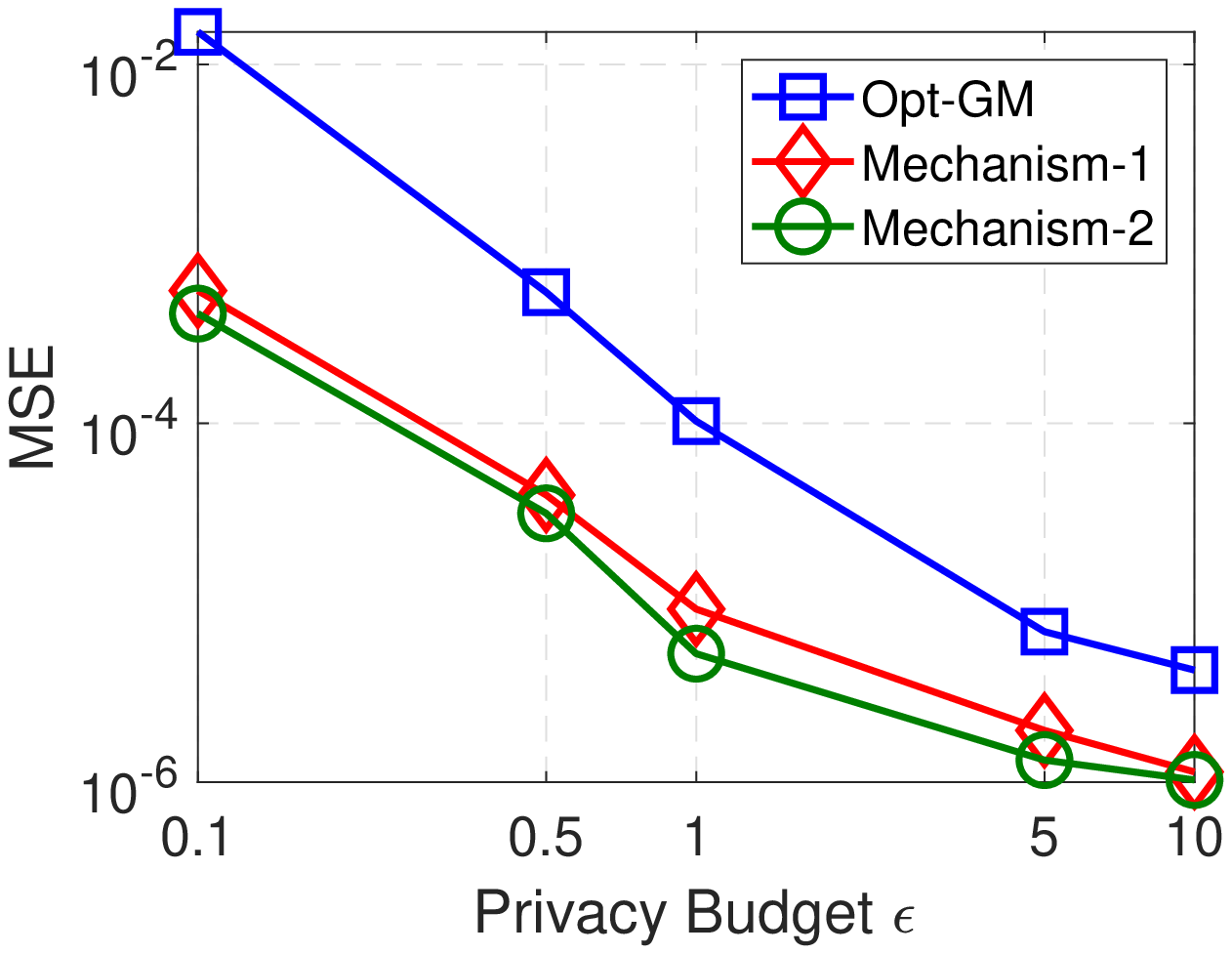}&
    \hspace{-3mm}\includegraphics[height=3cm]{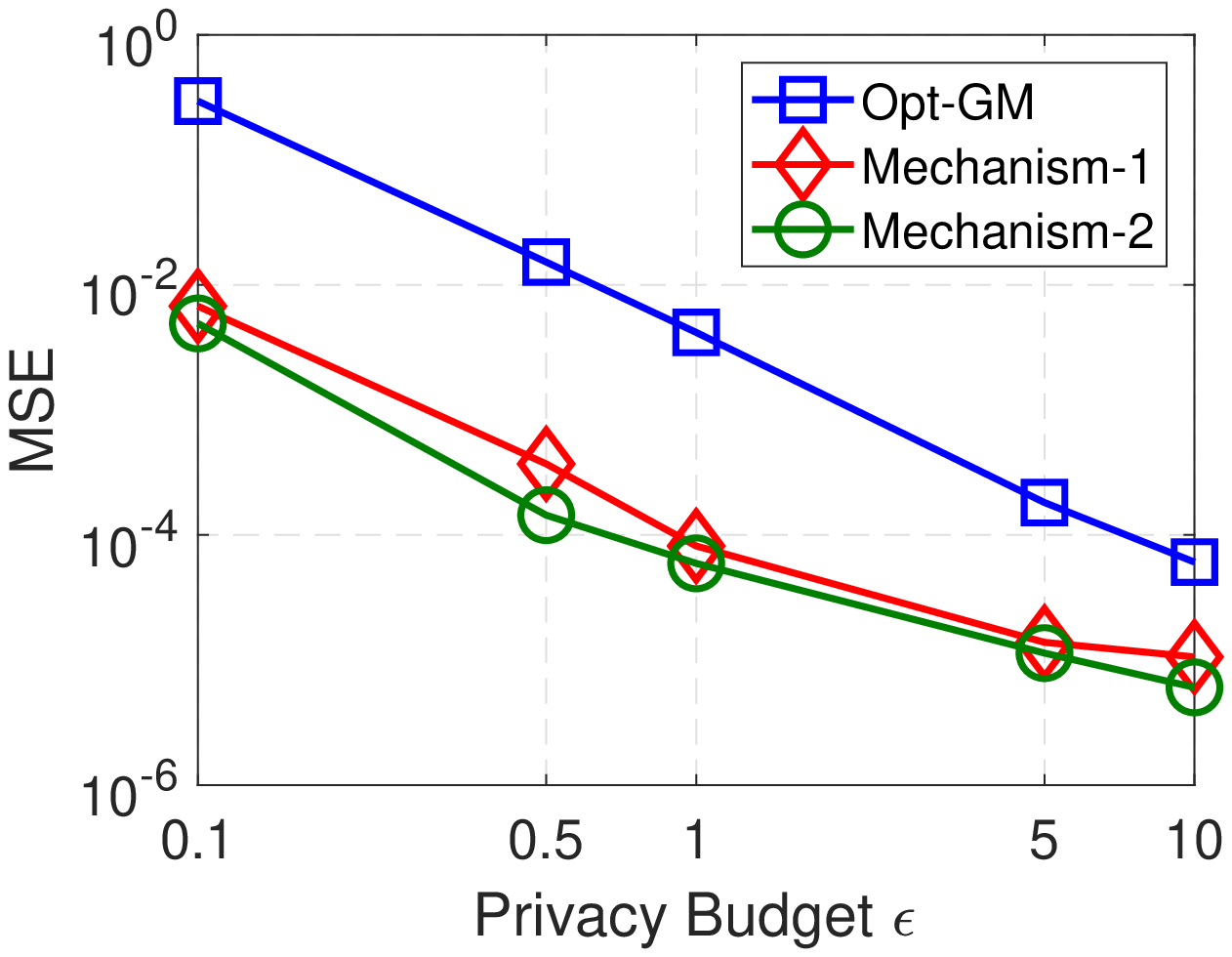}&
    \hspace{-3mm}\includegraphics[height=3cm]{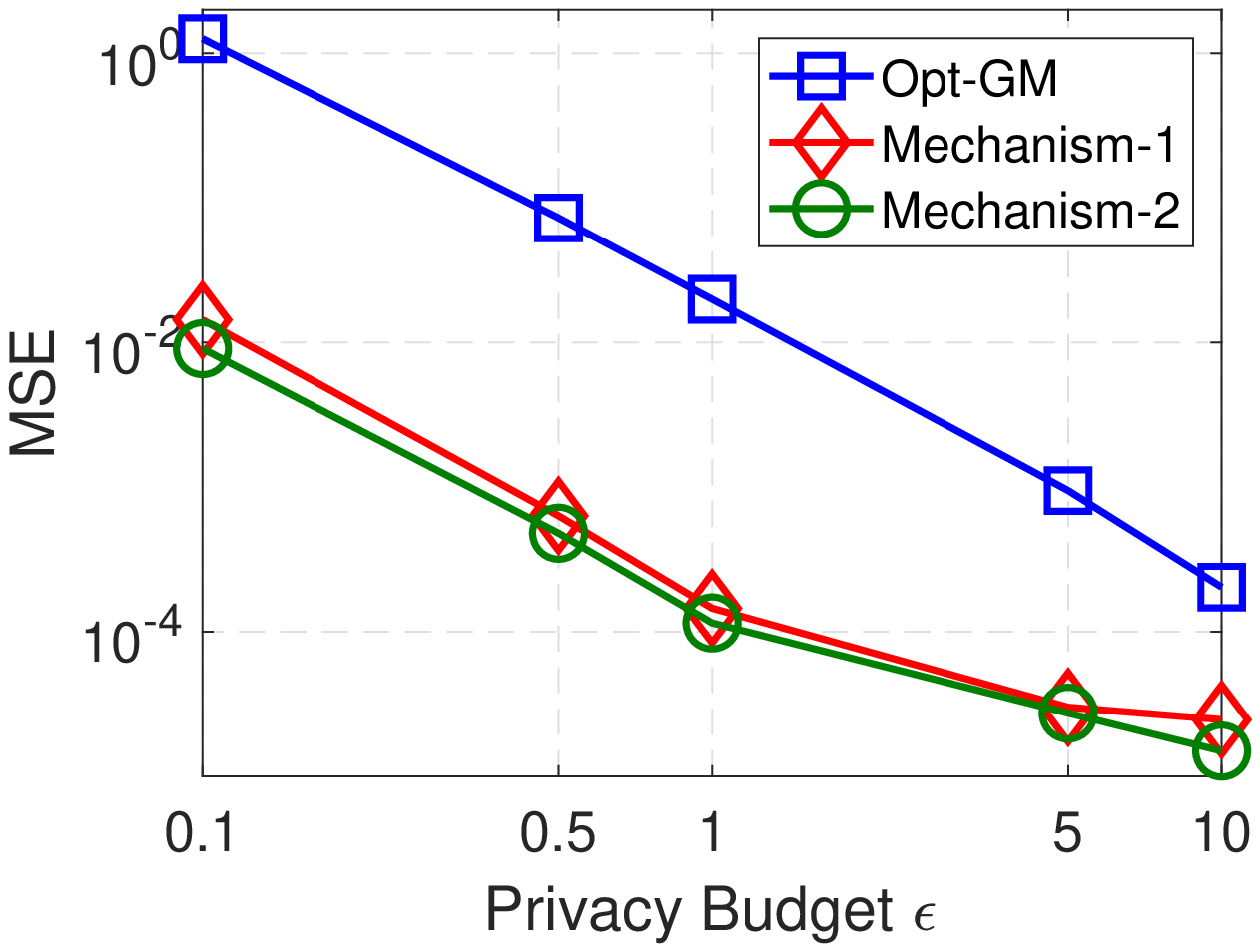}&
    \hspace{-3mm}\includegraphics[height=3cm]{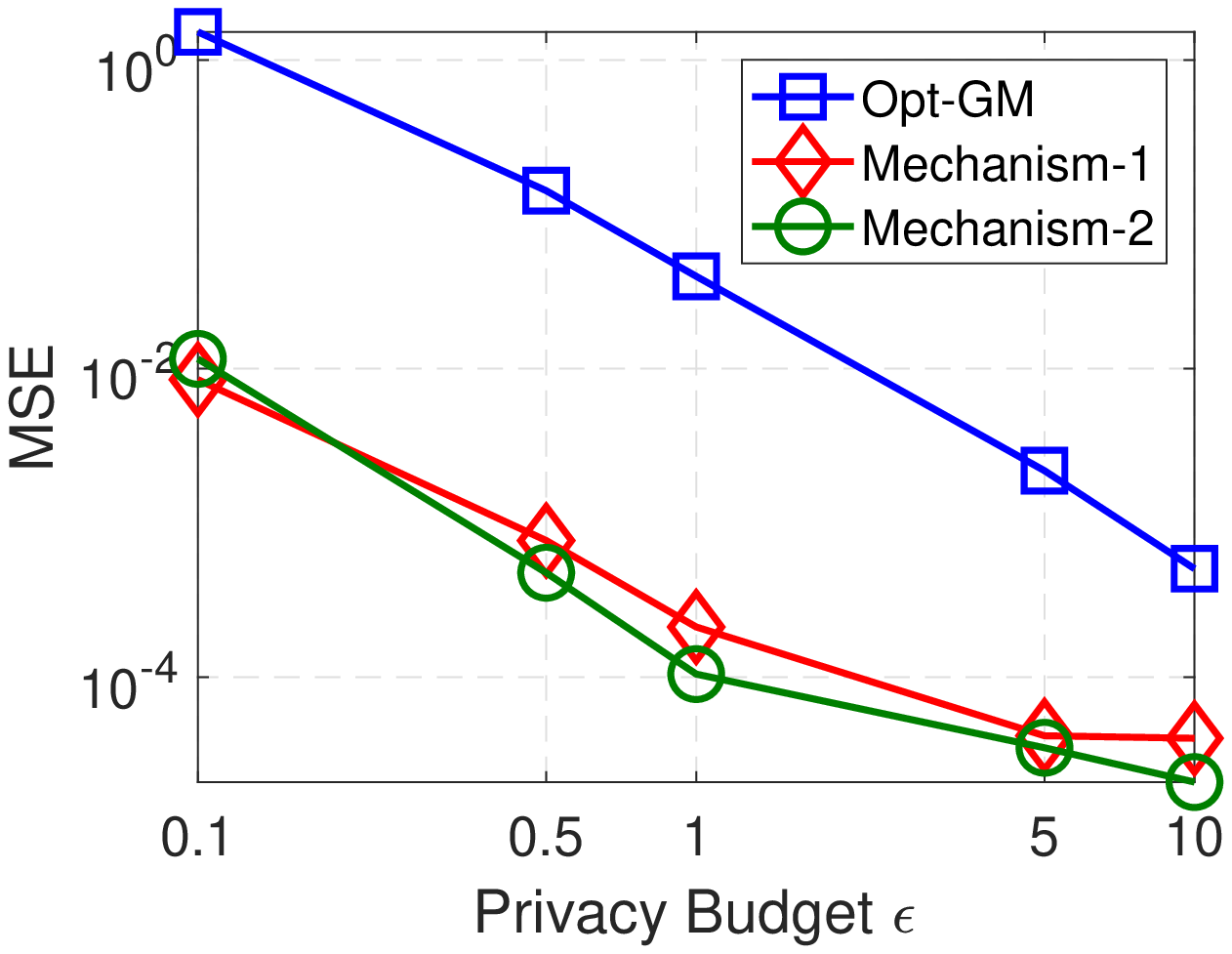}\\[-3mm]
   \hspace{-15mm}\scriptsize (a) $d=1$ & 
   \hspace{-2mm}\scriptsize (b) $d=5$ &
   \hspace{-2mm}\scriptsize (c) $d=10$ &
   \hspace{-2mm}\scriptsize (d) $d=15$
   \end{tabular} 
 	\vspace{-5mm}
    \caption{Accuracy for mean estimation on synthetic dataset under different dimensions ($\delta=10^{-6}$), each of which follows a Gaussian distribution $\mathcal{N}(0,1/16)$.} \label{syn-mean-vs-epsilon}
	\vspace{-1mm} 
\end{figure*}

Fig.~\ref{syn-mean-vs-epsilon} presents the accuracies of mean estimation on synthetic datasets varying from different privacy budgets $\epsilon$ and different dimensions $d$. It can be seen from all figures that \mbox{Mechanism-1} and \mbox{Mechanism-2} hold much higher accuracy than Opt-GM in all cases. By comparing four figures in Fig.~\ref{syn-mean-vs-epsilon}, the MSEs of all mechanisms will increase when the dimension $d$ increases from 1 to 15. Nonetheless, the MSEs of our proposed \mbox{Mechanism-1} and \mbox{Mechanism-2} increase much lower than that of Opt-GM. This demonstrates that our proposed \mbox{Mechanism-1} and \mbox{Mechanism-2} have better scalability for dimensions, which are more practical in reality.
% For example, under privacy parameters $\epsilon=0.1$ and $\delta=10^{-6}$, the MSE of Opt-GM has increased 100 times when $d$ increases from 1 to 15, while the MSE of \mbox{Mechanism-1} and \mbox{Mechanism-2} has increased 10 times under the same parameters.

\begin{figure*}
  \centering
  \begin{tabular}{cccc} 
    \hspace{-20mm}\includegraphics[height=3cm]{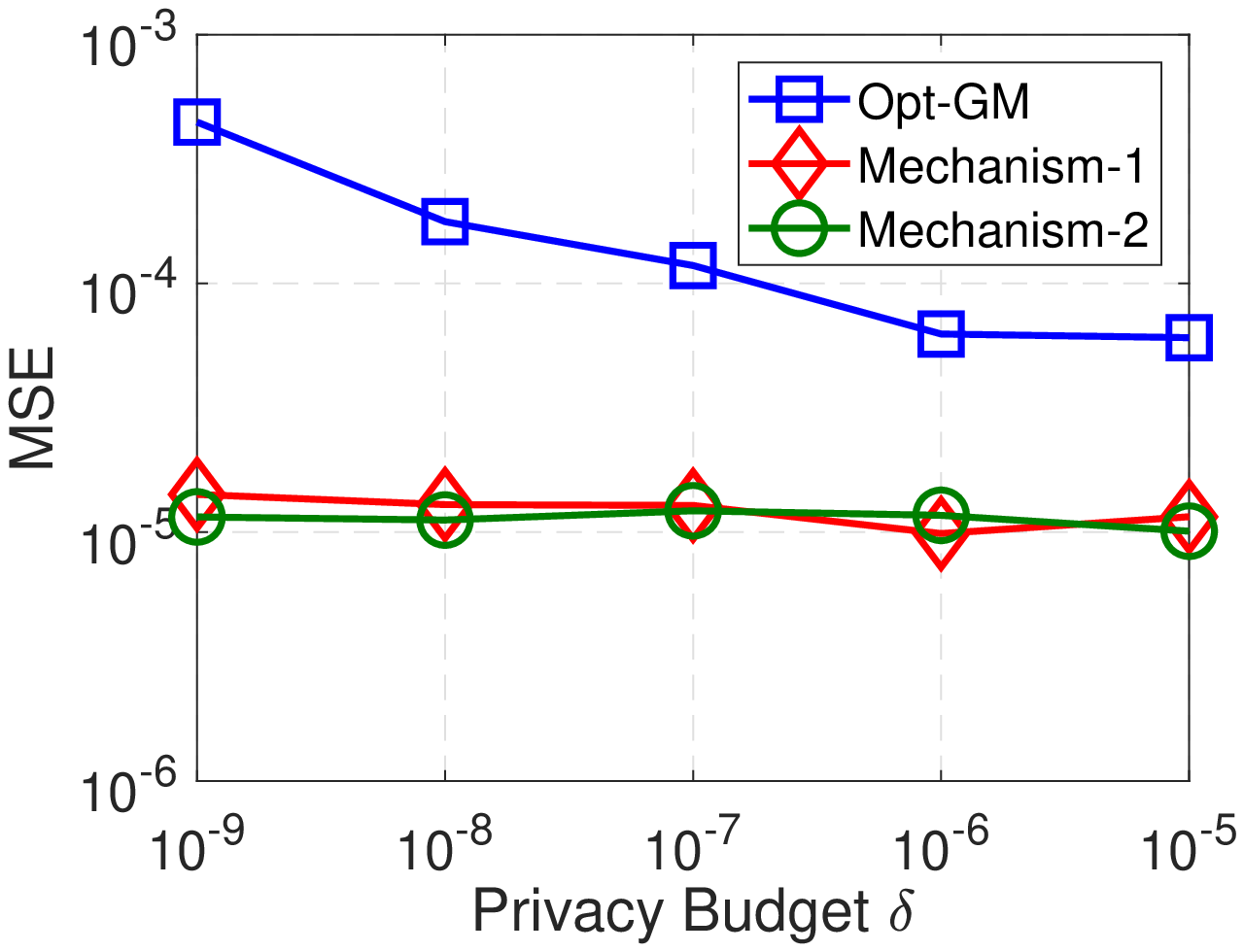}&
    \hspace{-3mm}\includegraphics[height=3cm]{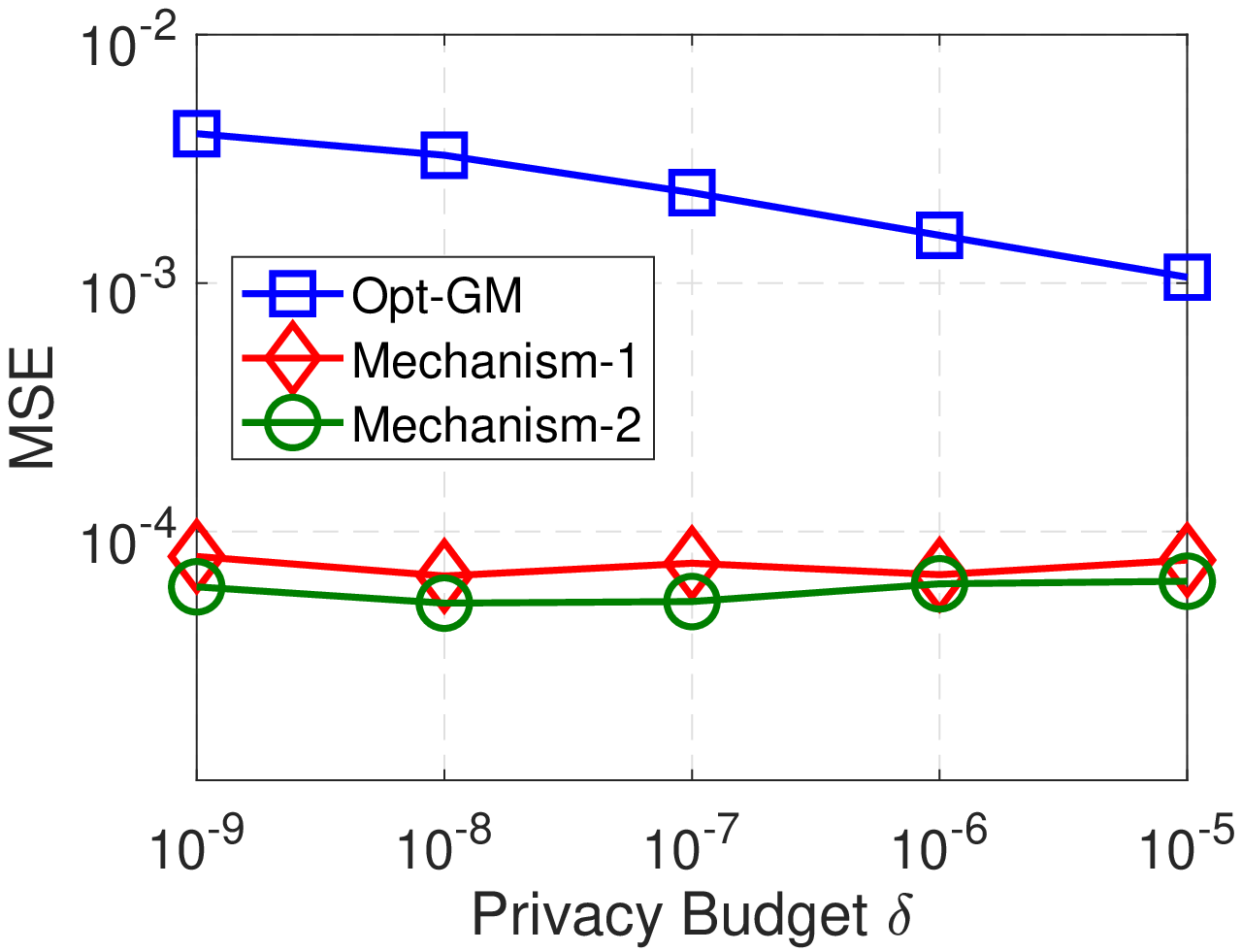}&
    \hspace{-3mm}\includegraphics[height=3cm]{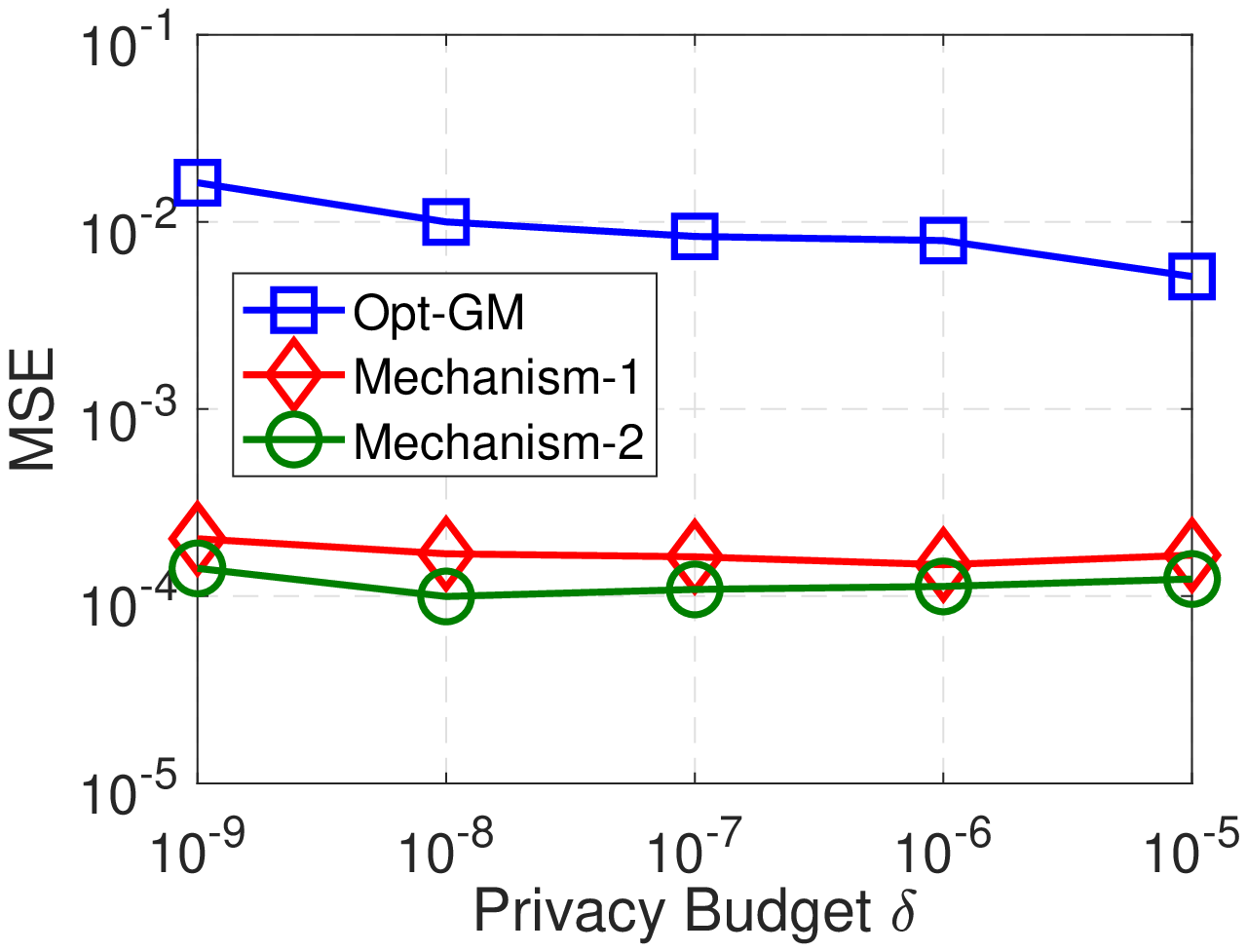}&
    \hspace{-3mm}\includegraphics[height=3cm]{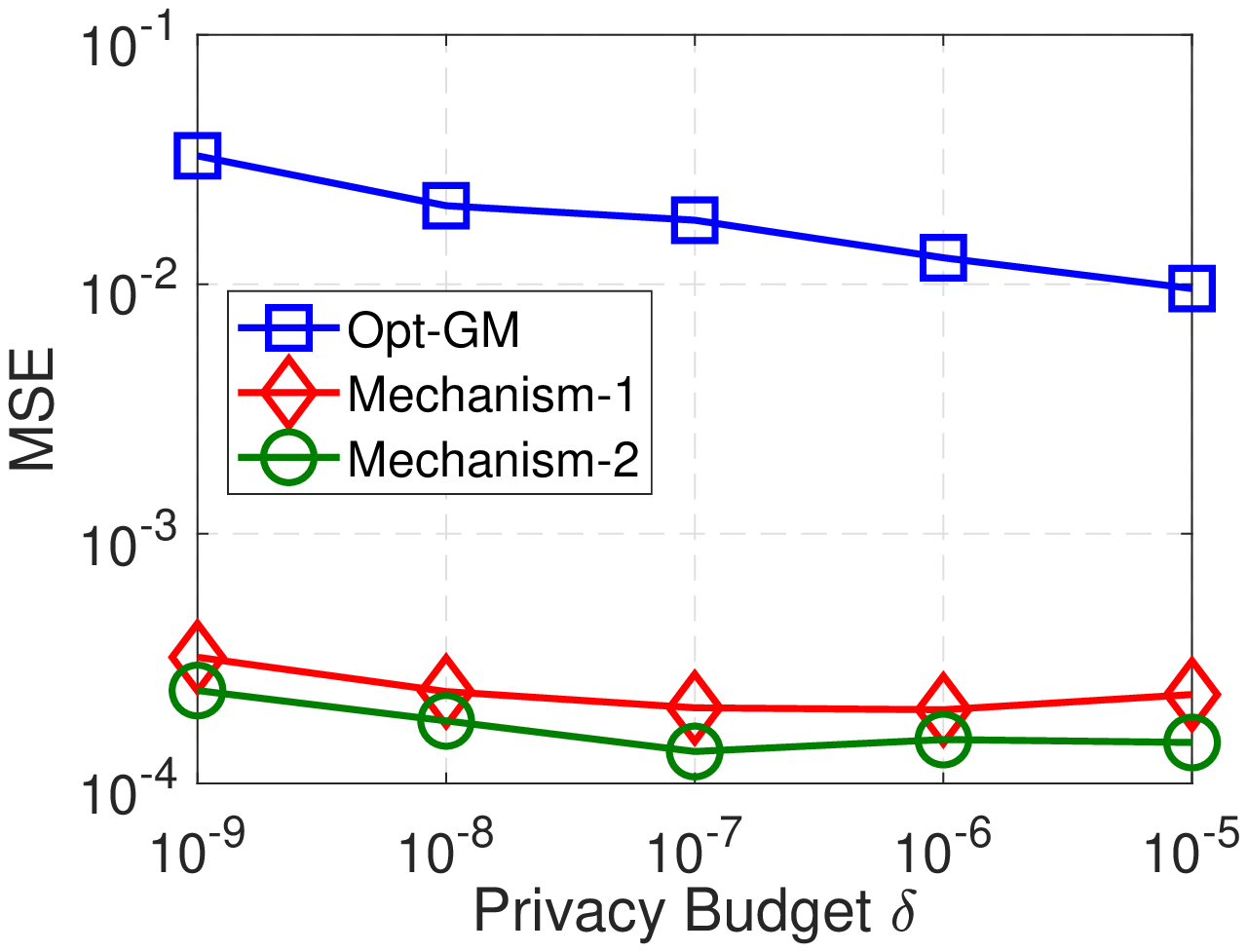}\\[-3mm]
   \hspace{-15mm}\scriptsize (a) $d=1$ & 
   \hspace{-2mm}\scriptsize (b) $d=5$ &
   \hspace{-2mm}\scriptsize (c) $d=10$ &
   \hspace{-2mm}\scriptsize (d) $d=15$
   \end{tabular} 
 	\vspace{-5mm}
    \caption{Accuracy for mean estimation on synthetic dataset under different dimensions ($\epsilon=1$), each of which follows a Gaussian distribution $\mathcal{N}(0,1/16)$.} \label{syn-mean-vs-delta}
	\vspace{-1mm} 
\end{figure*}

In addition, Fig.~\ref{syn-mean-vs-delta} shows the accuracy of mean estimation for numeric attributes on synthetic datasets varying from different privacy parameters $\delta$ and different dimensions $d$. It can be seen that the MSE of Opt-GM will decrease with the increasing of privacy parameter $\delta$. However, the MSEs of \mbox{Mechanism-1} and \mbox{Mechanism-2} are hardly affected by privacy parameter $\delta$. By comparing the four figures in Fig.~\ref{syn-mean-vs-delta}, we can find that the MSEs of our proposed \mbox{Mechanism-1} and \mbox{Mechanism-2} increase much slower than that of Opt-GM with the increasing of dimension $d$. Therefore, it demonstrates again that our proposed two mechanisms have better data utility with high scalability for dimensions.

\begin{figure*}
  \centering
  \begin{tabular}{cccc} 
    \hspace{-20mm}\includegraphics[height=3cm]{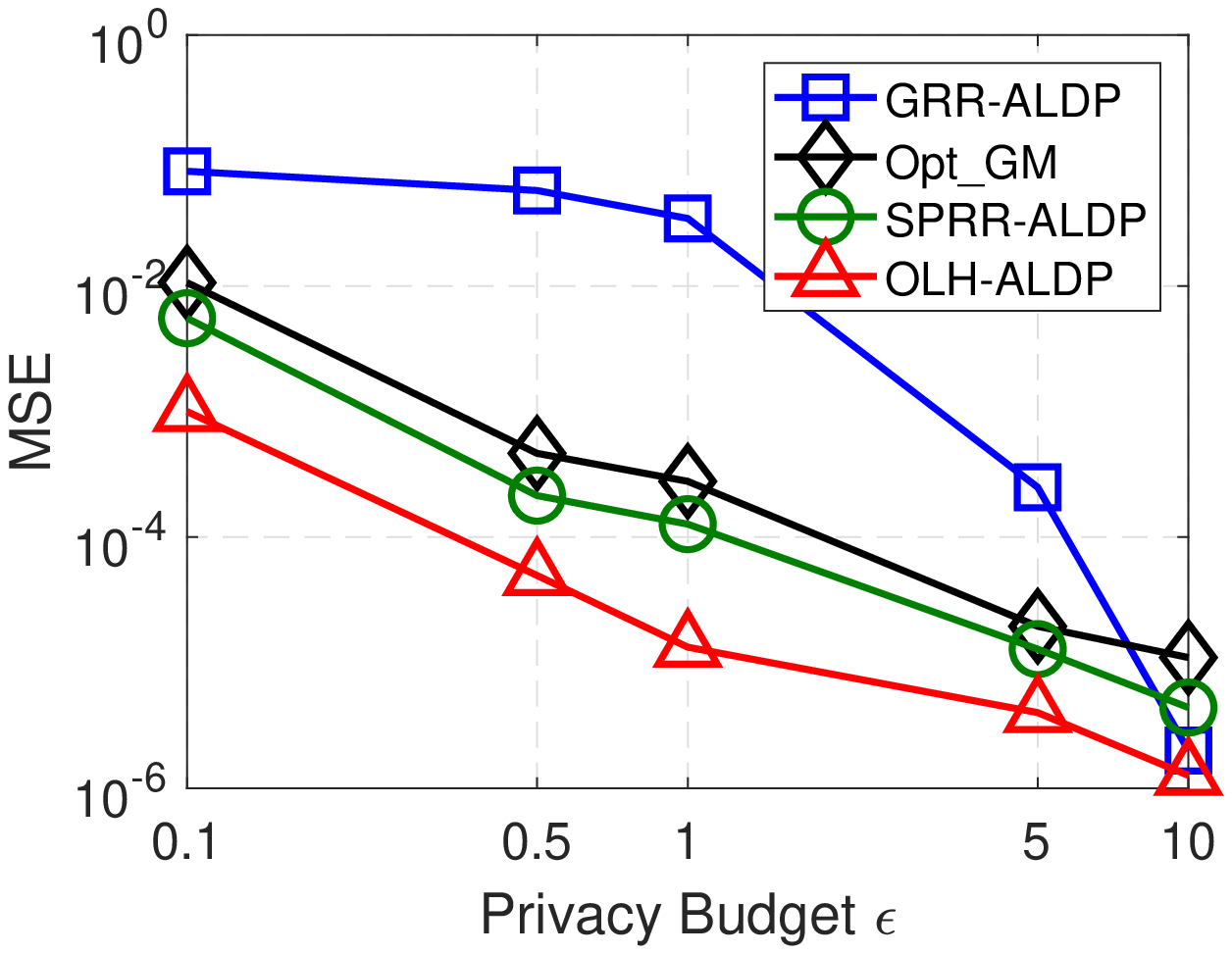}&
    \hspace{-3mm}\includegraphics[height=3cm]{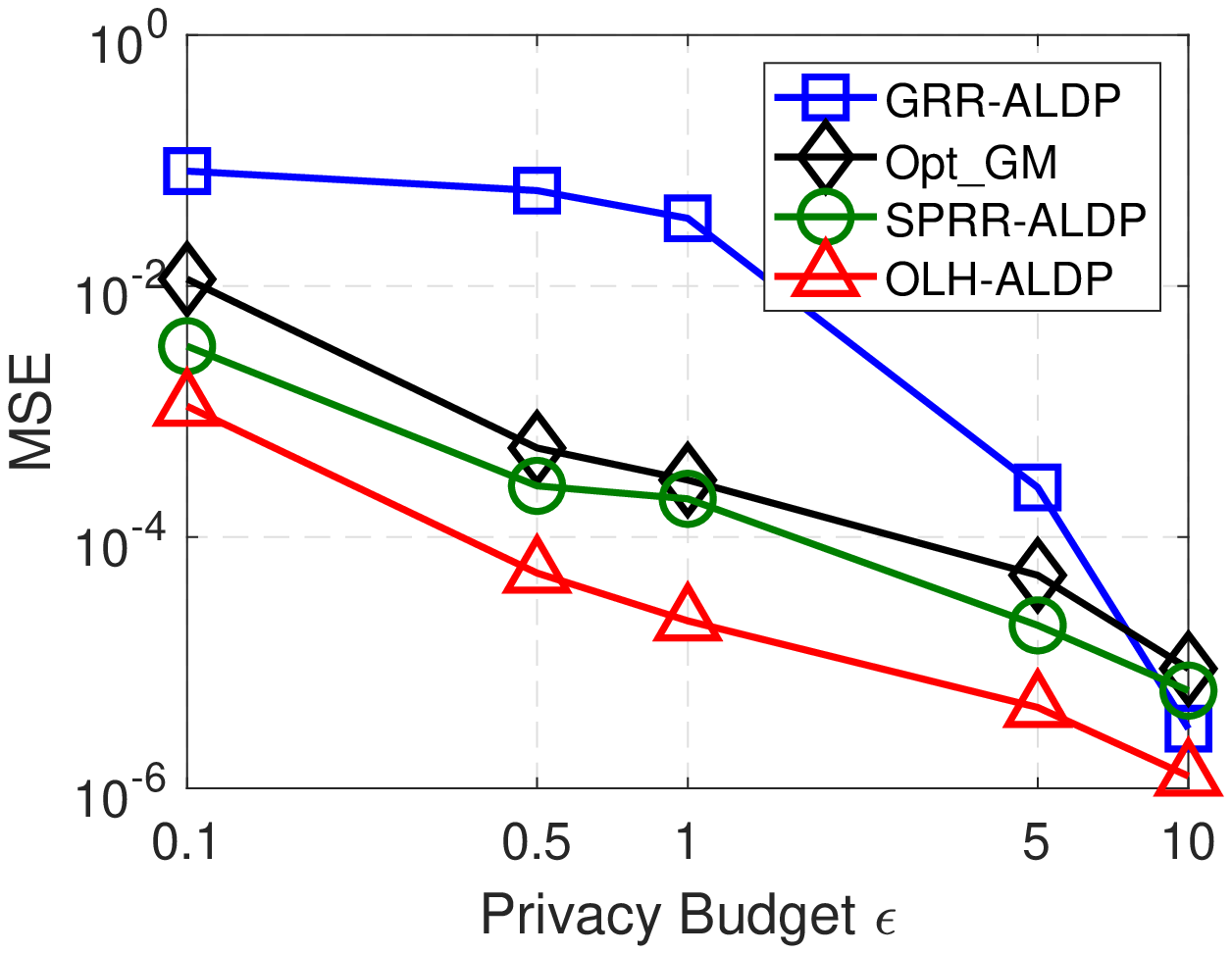}&
    \hspace{-3mm}\includegraphics[height=3cm]{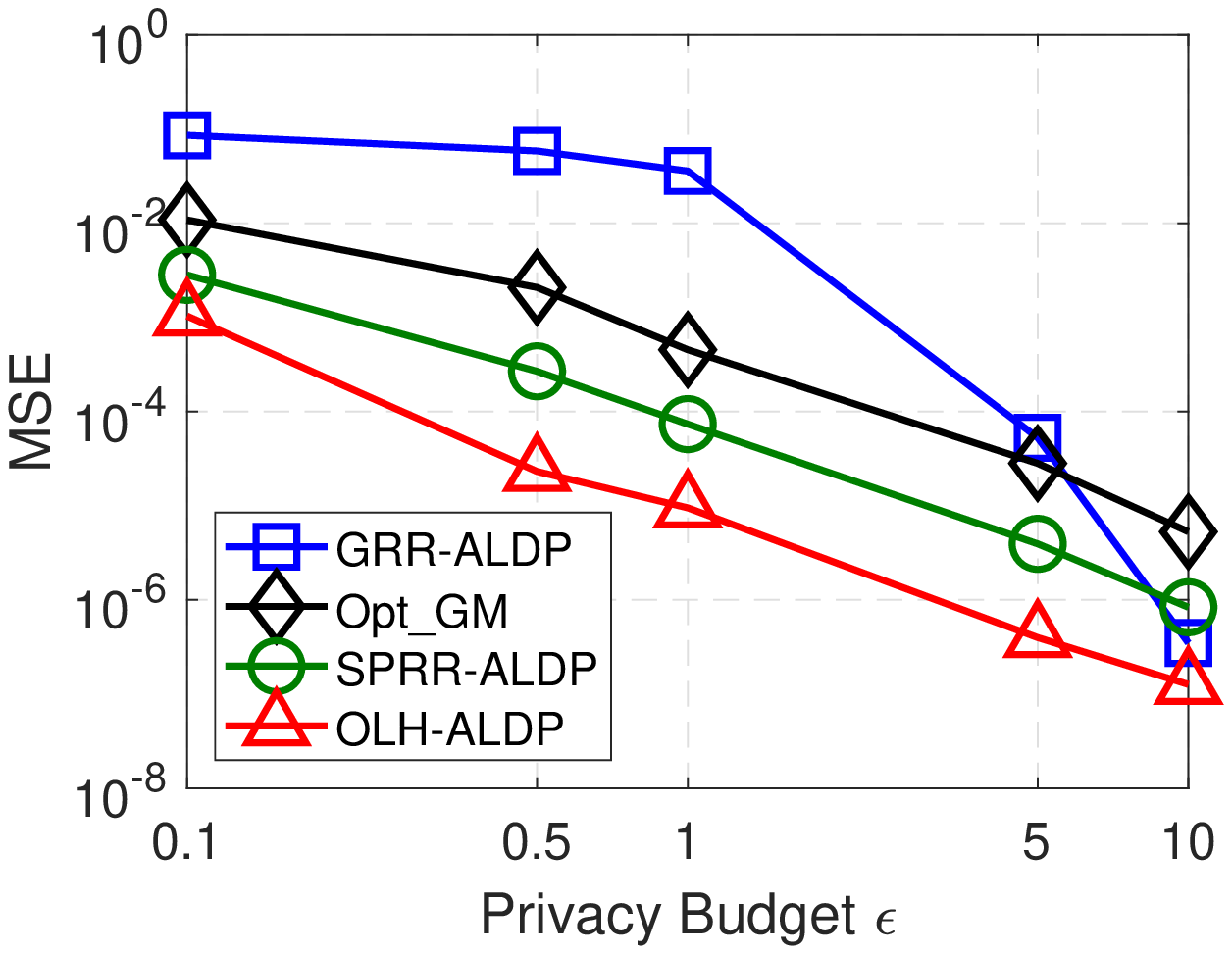}&
    \hspace{-3mm}\includegraphics[height=3cm]{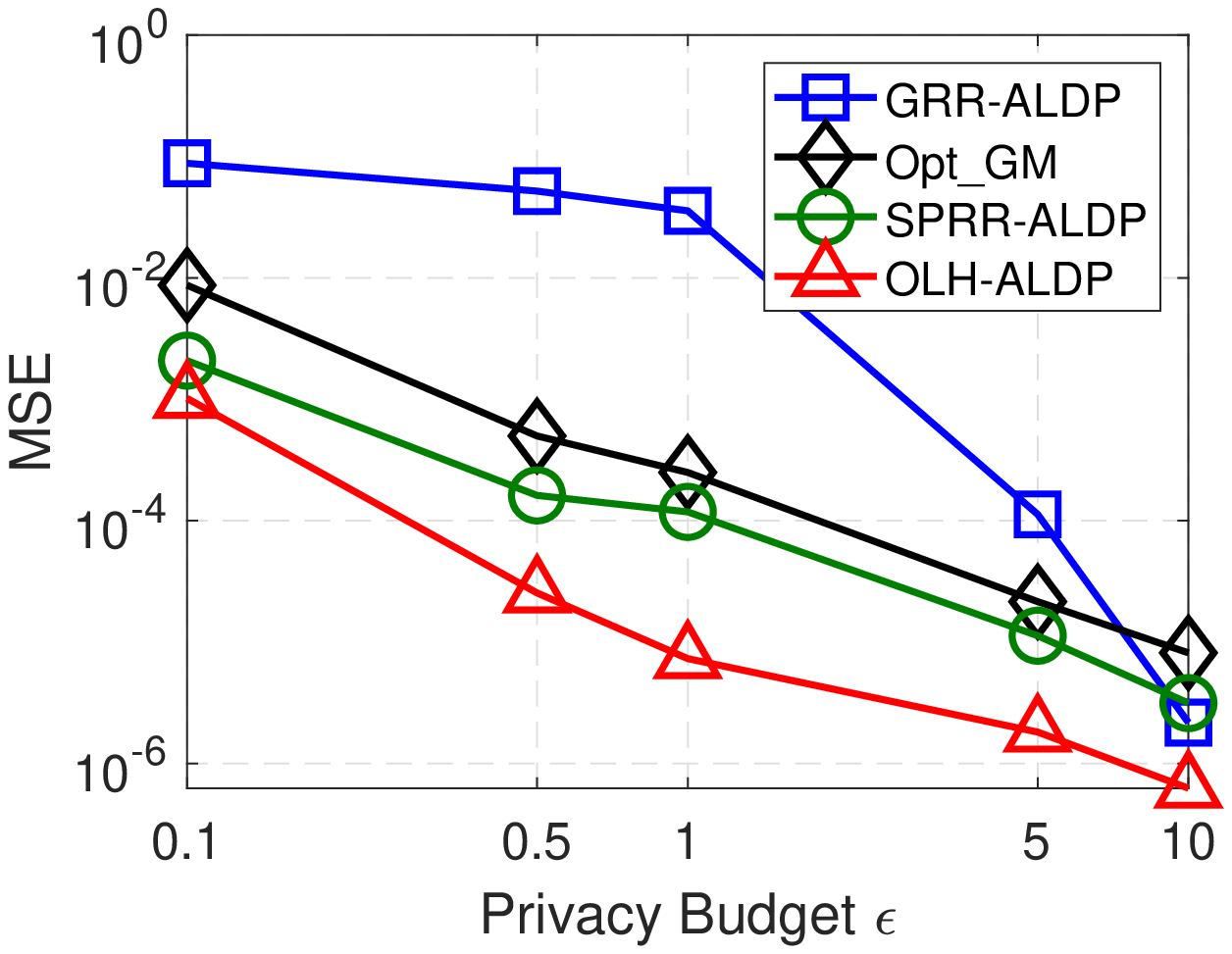}\\[-3mm]
   \hspace{-20mm}\scriptsize (a) MX-Categorical ($\delta=10^{-6}$) & \hspace{-2mm}\scriptsize (b) MX-Categorical ($\delta=10^{-7}$) &
   \hspace{-2mm}\scriptsize (c) BR-Categorical ($\delta=10^{-6}$) &
   \hspace{-2mm}\scriptsize (d) BR-Categorical ($\delta=10^{-7}$)
   \end{tabular} 
 	\vspace{-5mm}
    \caption{Accuracy for frequency estimation on categorical attributes.} \label{real-fre-vs-epsilon}
	\vspace{-1mm} 
\end{figure*}

As for categorical attributes, Fig.~\ref{real-fre-vs-epsilon} shows the accuracy of the frequency estimation of different mechanisms varying from different privacy parameters. On the whole, the MSEs of four
mechanisms decrease with an increase of $\epsilon$ from 0.1 to 10. Among four mechanisms ensuring \mbox{$(\epsilon,\delta)$-LDP} for categorical attributes, it can be seen that OLH-ALDP has the lowest MSE (e.g., the best data utility) in all cases, which corresponds to theoretical analysis. In addition, by comparing Fig.~\ref{real-fre-vs-epsilon}(a) and Fig.~\ref{real-fre-vs-epsilon}(b) (or, Fig.~\ref{real-fre-vs-epsilon}(c) and Fig.~\ref{real-fre-vs-epsilon}(d)), the MSEs of four mechanisms are almost unchanged when $\delta$ changes from $10^{-6}$ to $10^{-7}$. That is, the size of the privacy parameter $\epsilon$ primarily dominants the accuracy, while the size of the privacy parameter $\delta$ has a small effect on the accuracy.

In addition, we also implement different algorithms on synthetic datasets to compare the effects of the domain of categorical attributes. Each synthetic dataset is generated by following Zipf's distribution with an exponent parameter $s$=1.3 and each synthetic dataset contains 100,000 records.

\begin{figure*}
  \centering
  \begin{tabular}{cccc} 
    \hspace{-20mm}\includegraphics[height=3cm]{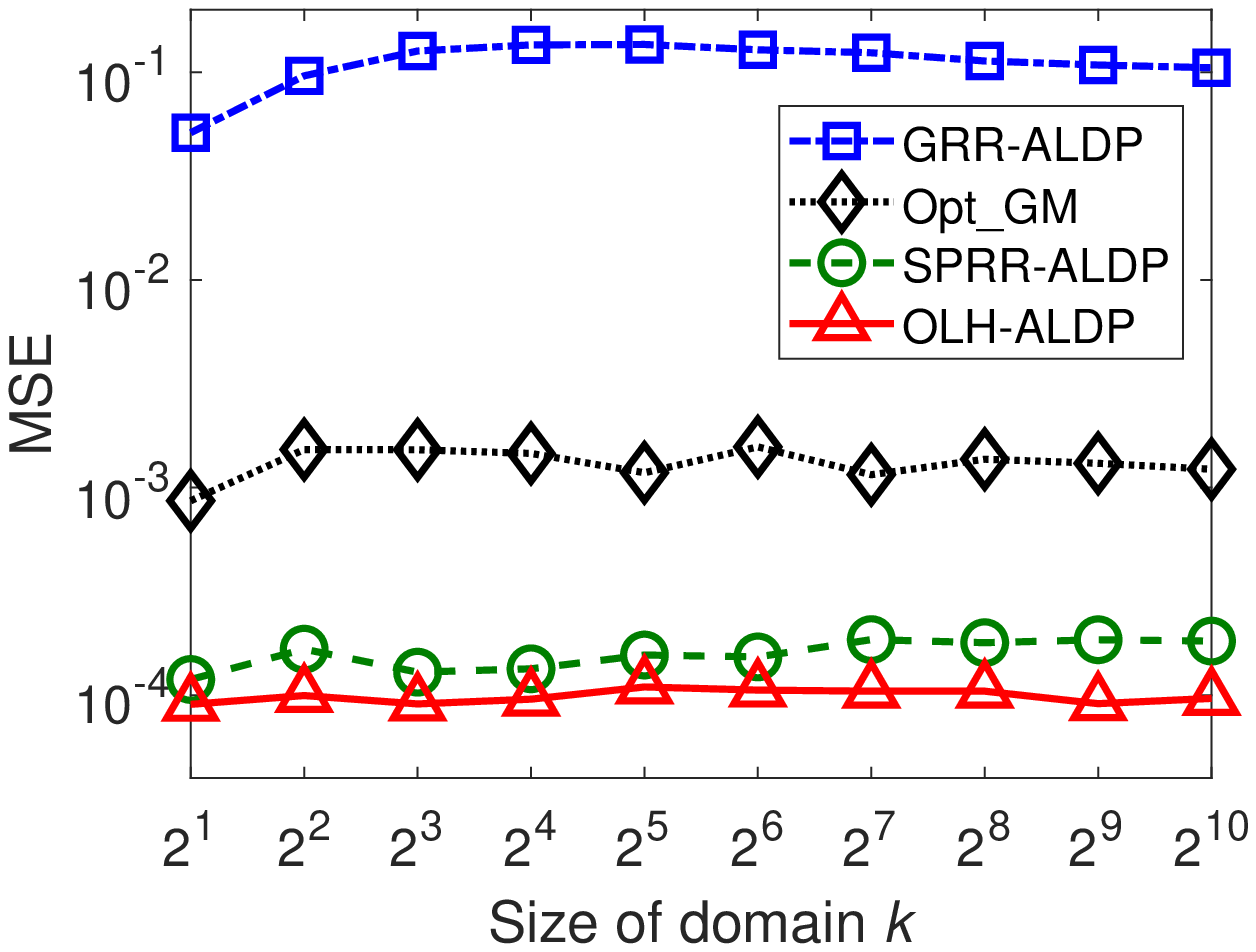}&
    \hspace{-3mm}\includegraphics[height=3cm]{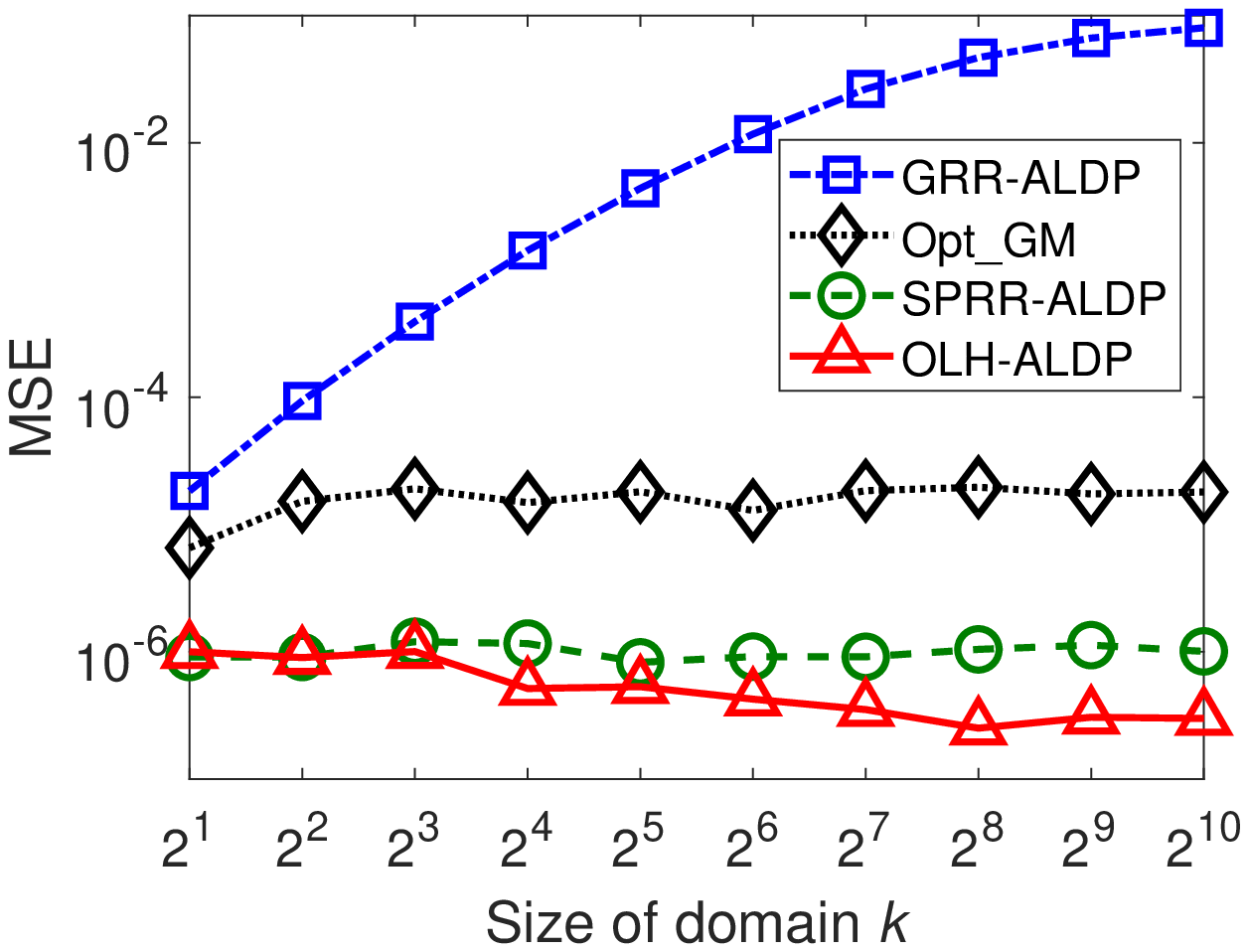}&
    \hspace{-3mm}\includegraphics[height=3cm]{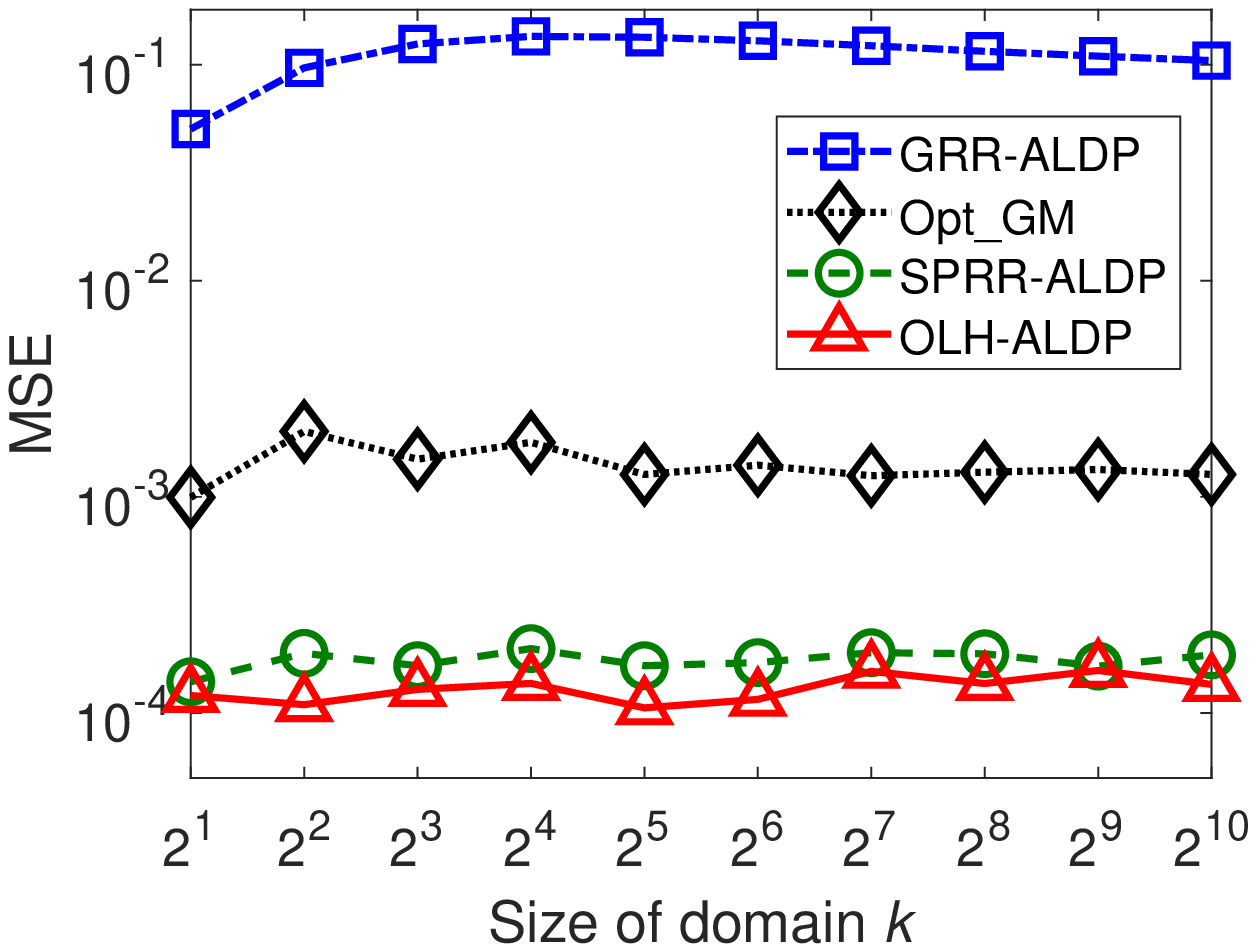}&
    \hspace{-3mm}\includegraphics[height=3cm]{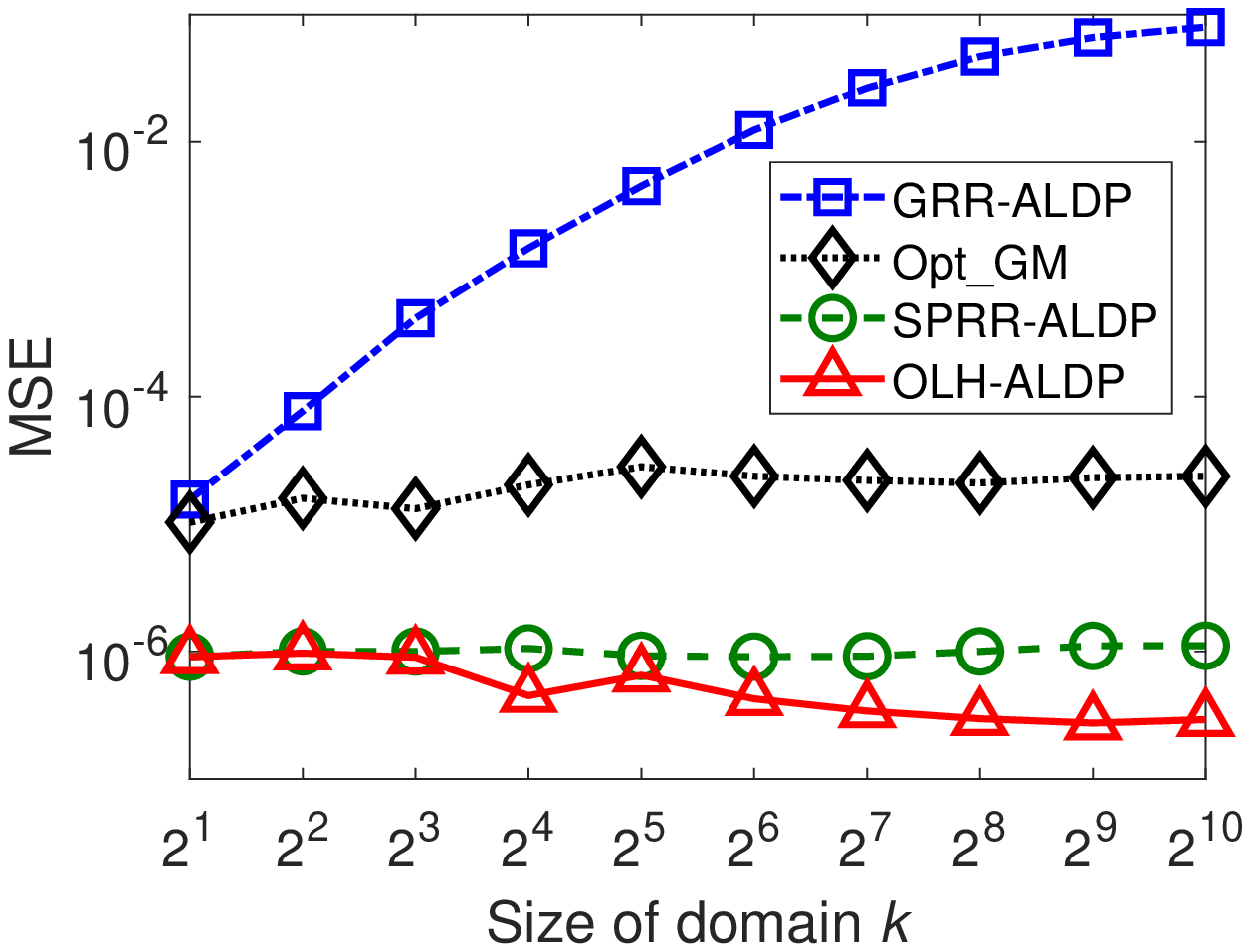}\\[-3mm]
   \hspace{-20mm}\scriptsize (a) Vary $k$ ($\epsilon = 0.5, \delta=10^{-6}$) &
   \hspace{-2mm}\scriptsize (b) Vary $k$ ($\epsilon = 5, \delta=10^{-6}$) &
   \hspace{-2mm}\scriptsize (c) Vary $k$ ($\epsilon = 0.5, \delta=10^{-7}$) &
   \hspace{-2mm}\scriptsize (d) Vary $k$ ($\epsilon = 5, \delta=10^{-7}$)
   \end{tabular} 
 	\vspace{-5mm}
    \caption{Accuracy for frequency estimation on synthetic data vary domain $k$, each of which follows Zipf's distribution with exponent $s$=1.3.} \label{syn-fre-vs-k}
	\vspace{-1mm} 
\end{figure*}

Fig.~\ref{syn-fre-vs-k} shows the accuracy of frequency estimation for categorical attributes on synthetic datasets varying from different domain size $k$. It can be seen that the MSEs of \mbox{Opt-GM}, SPRR-ALDP, and OLH-ALDP almost remain unchanged with the increasing of domain size $k$ in all cases. This is reasonable because these three methods are not relevant to the domain size in theory. In contrast, the domain size has a great impact on the MSE of GRR-ALDP. The larger the domain size is, the larger the MSE will be. This shows that GRR-ALDP leads to a low data utility for the categorical attributes that have large domain sizes, thus resulting in limited applications in reality.

Furthermore, we can see from Fig.~\ref{syn-fre-vs-k}(a) and Fig.~\ref{syn-fre-vs-k}(c) that the growth of MSE of GRR-ALDP are slower when privacy budget $\epsilon=0.5$. And the MSE of GRR-ALDP increases much quickly when privacy budget $\epsilon=5$, as shown in Fig.~\ref{syn-fre-vs-k}(b) and Fig.~\ref{syn-fre-vs-k}(d). This shows that the domain size will have a greater impact on the data utility when the privacy budget is relatively large. Thus, it demonstrates again that the data utility of GRR-ALDP suffers from both privacy parameters and domain size, leading a low data utility and poor availability. In contrast, the accuracies of both SPRR-ALDP and OLH-ALDP are much smaller in all cases and are minimally affected by the domain size.

\subsection{Results on Machine Learning Models}

% \begin{figure}
%   \centering
%   \begin{tabular}{cc} 
%     \hspace{-10pt}\includegraphics[height=3cm]{figures/SGD_Logistic_MX.eps}&
%     \hspace{-10pt}
%     \includegraphics[height=3cm]{figures/SGD_Logistic_BR.eps}\\[-1pt]
%   \scriptsize (a) MX ($\delta=10^{-26}$) & \scriptsize (b) BR ($\delta=10^{-13}$)
%   \end{tabular} 
%  	\vspace{-1mm}
%     \caption{Accuracy of logistic regression.} \label{logistic-vs-eps}
% 	\vspace{-3mm} 
% \end{figure}

In the second experimental setting, we build a class of machine learning models under \mbox{$(\epsilon,\delta)$-LDP} which are solved by stochastic gradient descent (SGD) \cite{wang2019collecting}. We focus on three common learning tasks: linear regression, logistic regression, and support vector machines (SVM) classification. We take the numeric attribute ``income'' as the label attribute in three tasks. In our experiments, each categorical attribute $A_j$ with $k$ values is transformed into $k-1$ bit binary values with domain $\{-1,1\}$ such that each new binary vector satisfies that, (\textit{i}) the $l$-th bit is set to 1 and the other $k-2$ bit are set to -1 for all $l$-th ($l<k$) value of $A_j$; (\textit{ii}) all $k-1$ bit are set to -1 for all $k$-th value of $A_j$. Then, the new datasets of BR and MX contain 42 and 85 dimensions, respectively. And for logistic regression and SVM classification, we process ``income'' into binary values such that the value larger than the mean is set to 1, and -1 otherwise.

Note that one tuple may be used in multiple iterations in the learning algorithms under non-private cases. However, the works \cite{nguyen2016collecting,wang2019collecting} have indicated that it will degrade the accuracy of the learning algorithms by iterating one tuple multiple times in the local private setting. Therefore, in the local private setting of SGD for machine learning, we assume each user (i.e., one tuple) only participates in at most one iteration. In each iteration, each user in one batch submits her noisy gradient to the aggregator. Then, the learning parameters will be updated by using Eq.~(\ref{eqn-sgd}).

\begin{figure*}
\begin{minipage}{.45\textwidth}
  \hspace{-20mm}
  \begin{tabular}{cc} 
    \hspace{-5mm}\includegraphics[height=3cm]{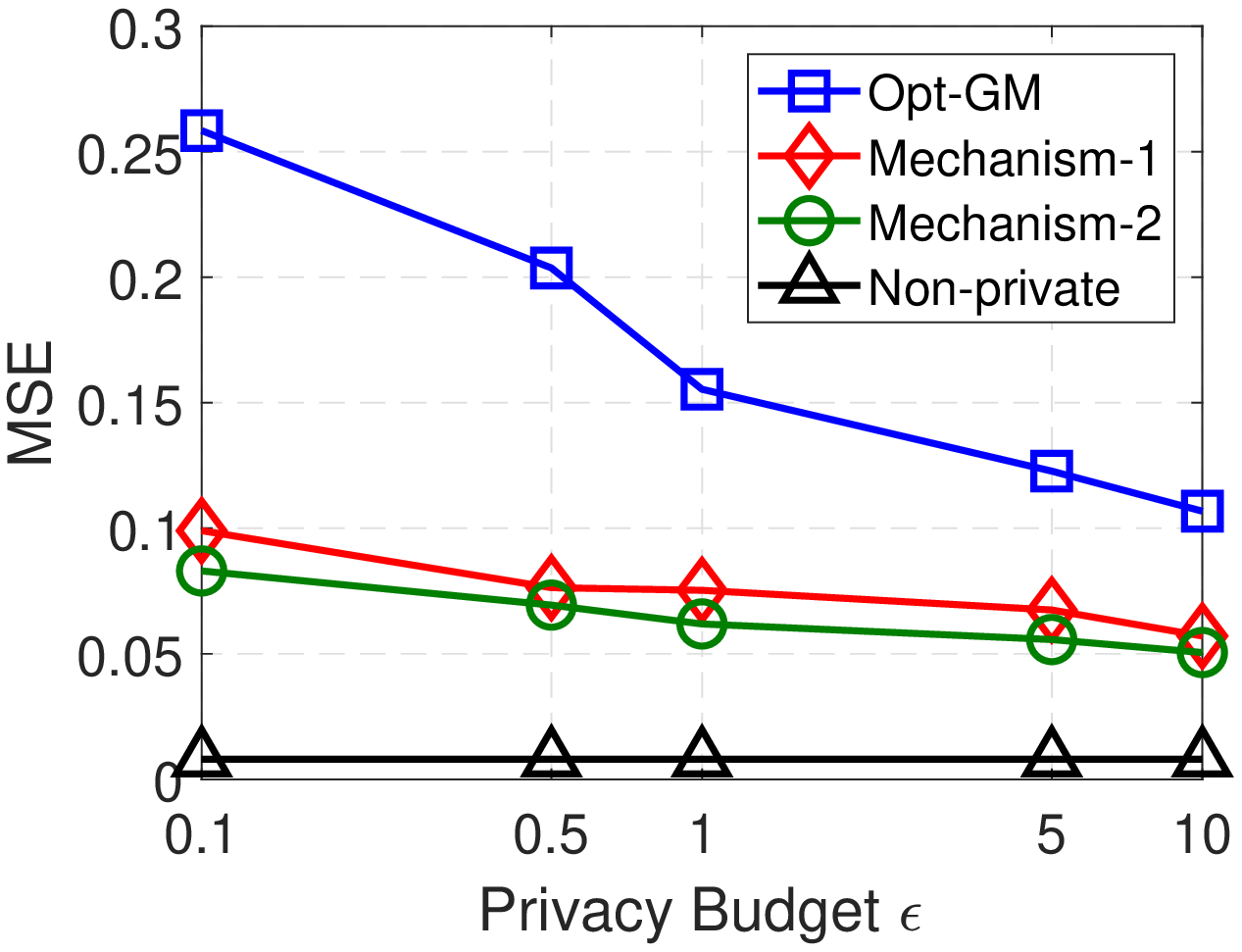}&
    \hspace{-4mm}
    \includegraphics[height=3cm]{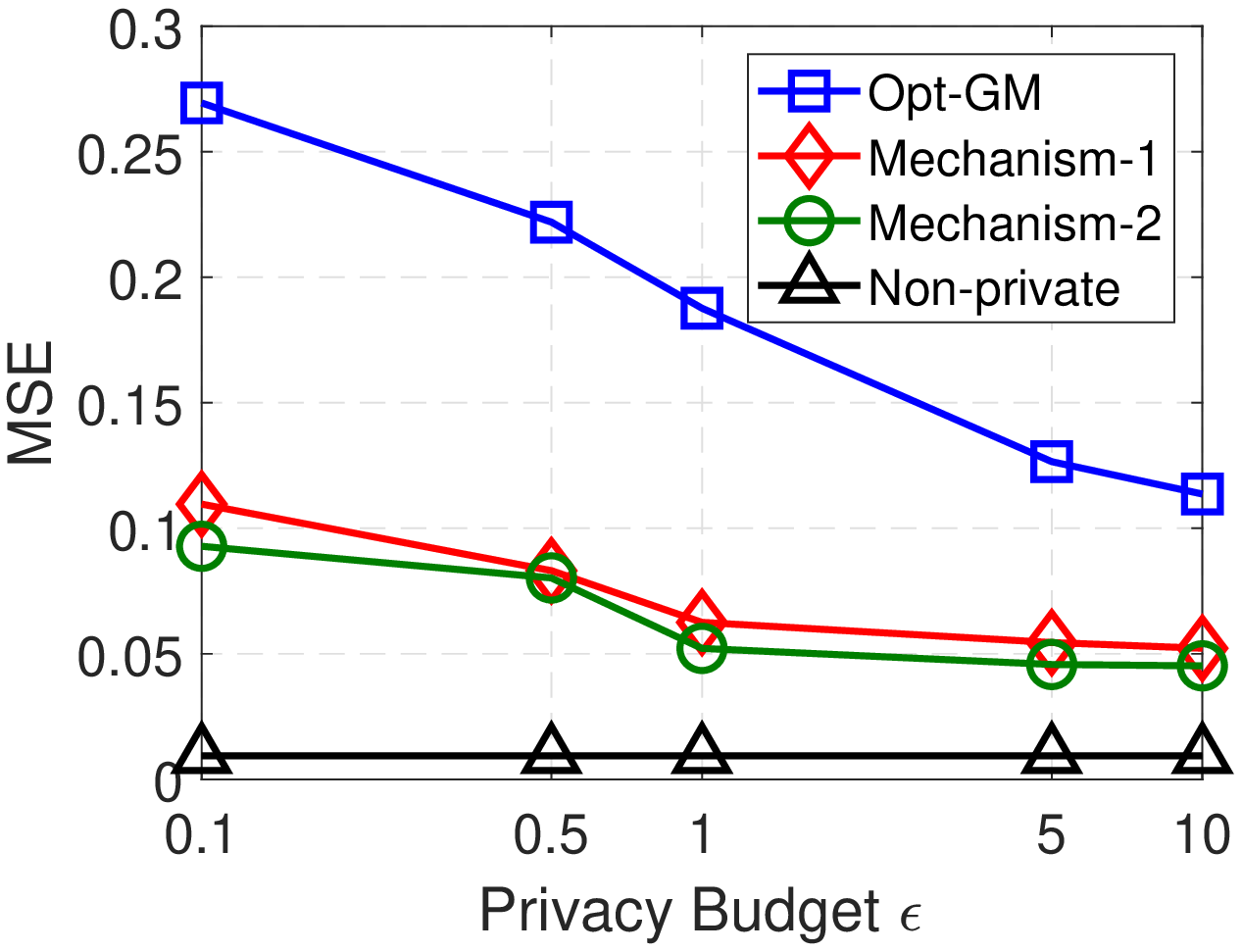}\\[-1pt]
   \scriptsize (a) MX ($\delta=10^{-26}$) & \scriptsize (b) BR ($\delta=10^{-13}$)
   \end{tabular} 
 	\vspace{-2mm}
    \caption{Accuracy of linear regression.} \label{linear-vs-eps}
	\vspace{-1mm} 
\end{minipage}
	~~~
	\hspace{3mm}
\begin{minipage}{.45\textwidth}
  \begin{tabular}{cc} 
    \hspace{-5mm}\includegraphics[height=3cm]{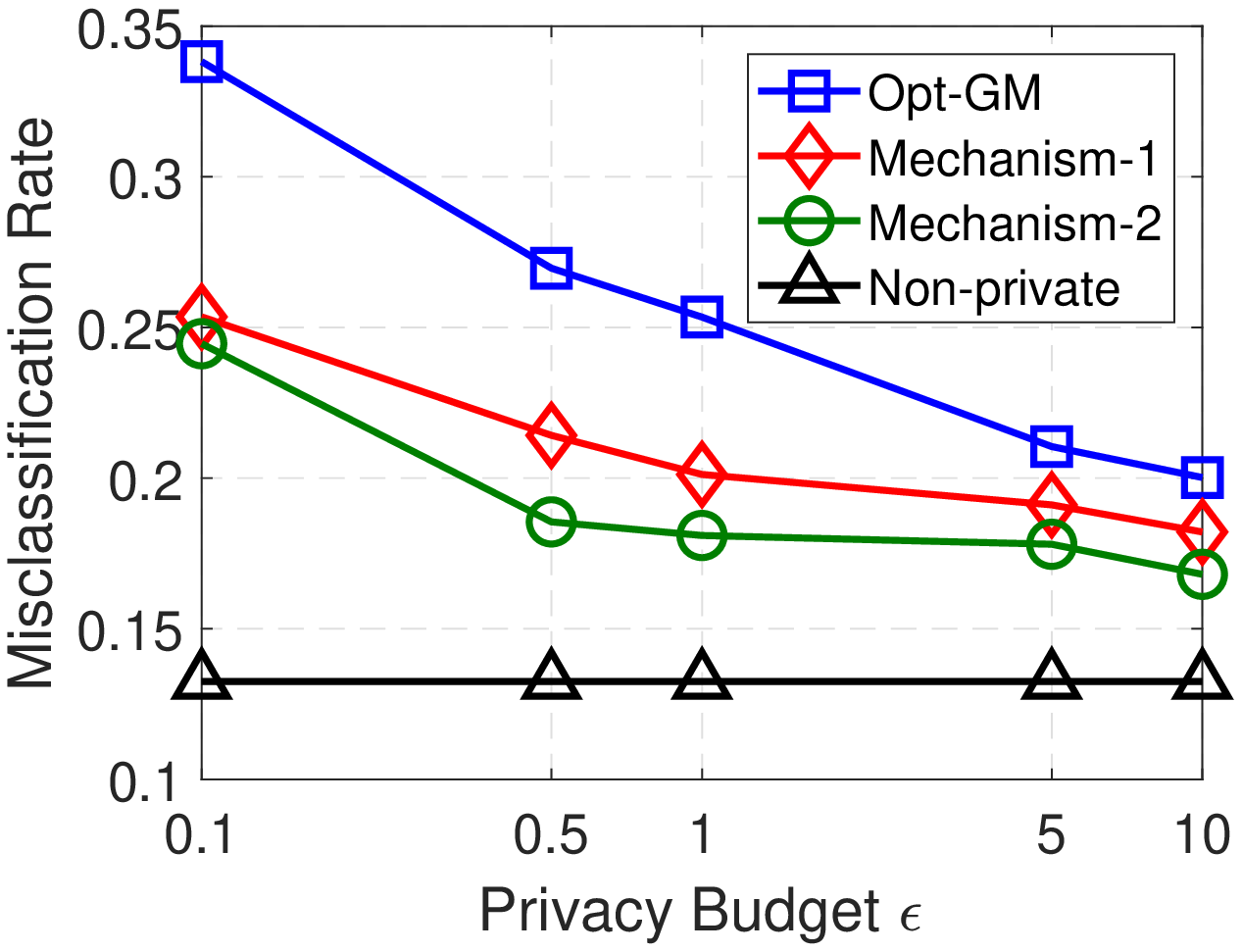}&
    \hspace{-4mm}
    \includegraphics[height=3cm]{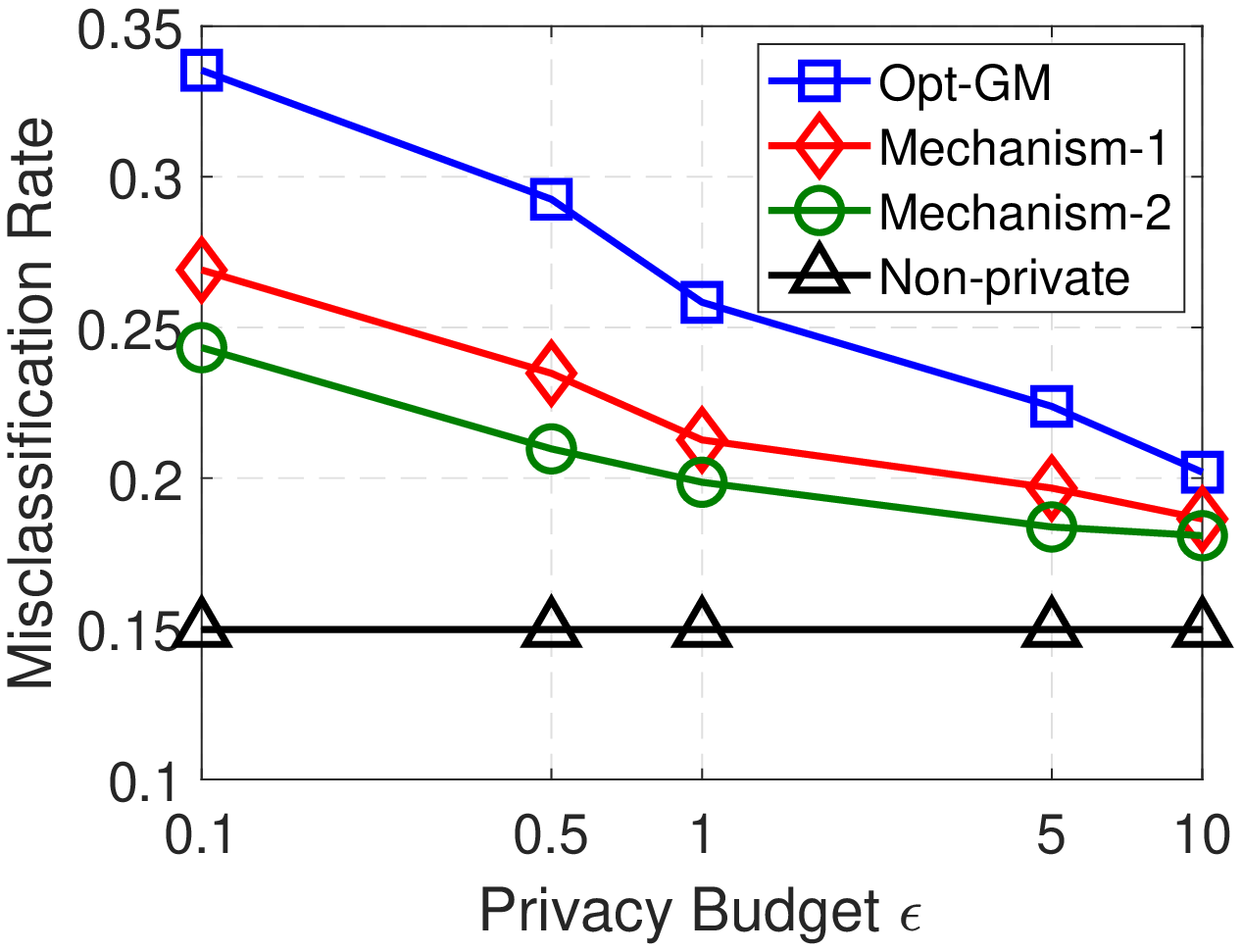}\\[-1pt]
   \scriptsize (a) MX ($\delta=10^{-26}$) & \scriptsize (b) BR ($\delta=10^{-13}$)
   \end{tabular} 
 	\vspace{-2mm}
    \caption{Accuracy of logistic regression.} \label{logistic-vs-eps}
	\vspace{-1mm} 
\end{minipage}
\end{figure*}

\begin{figure}
  \centering
  \begin{tabular}{cc} 
    \hspace{-10pt}\includegraphics[height=3cm]{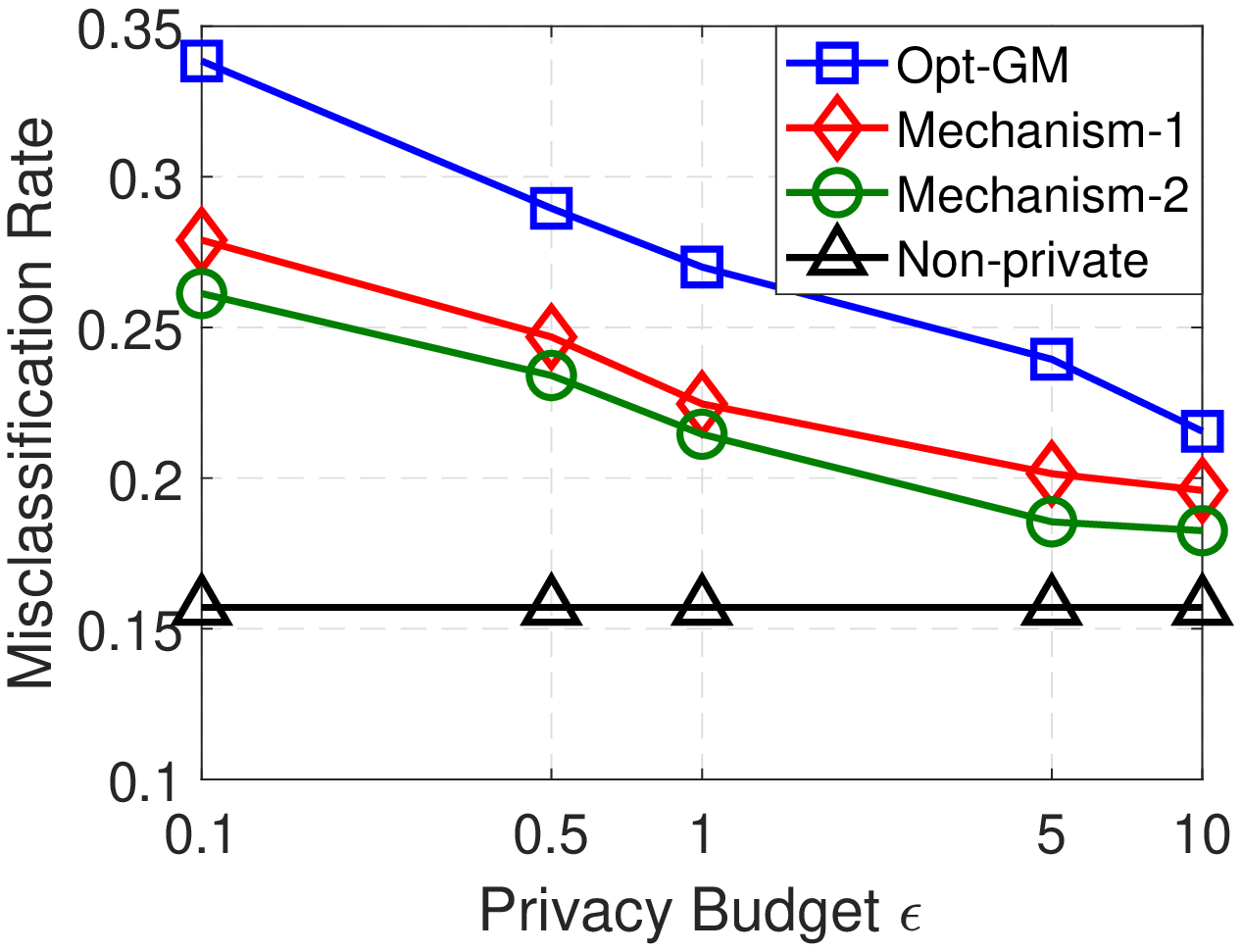}&
    \hspace{-10pt}
    \includegraphics[height=3cm]{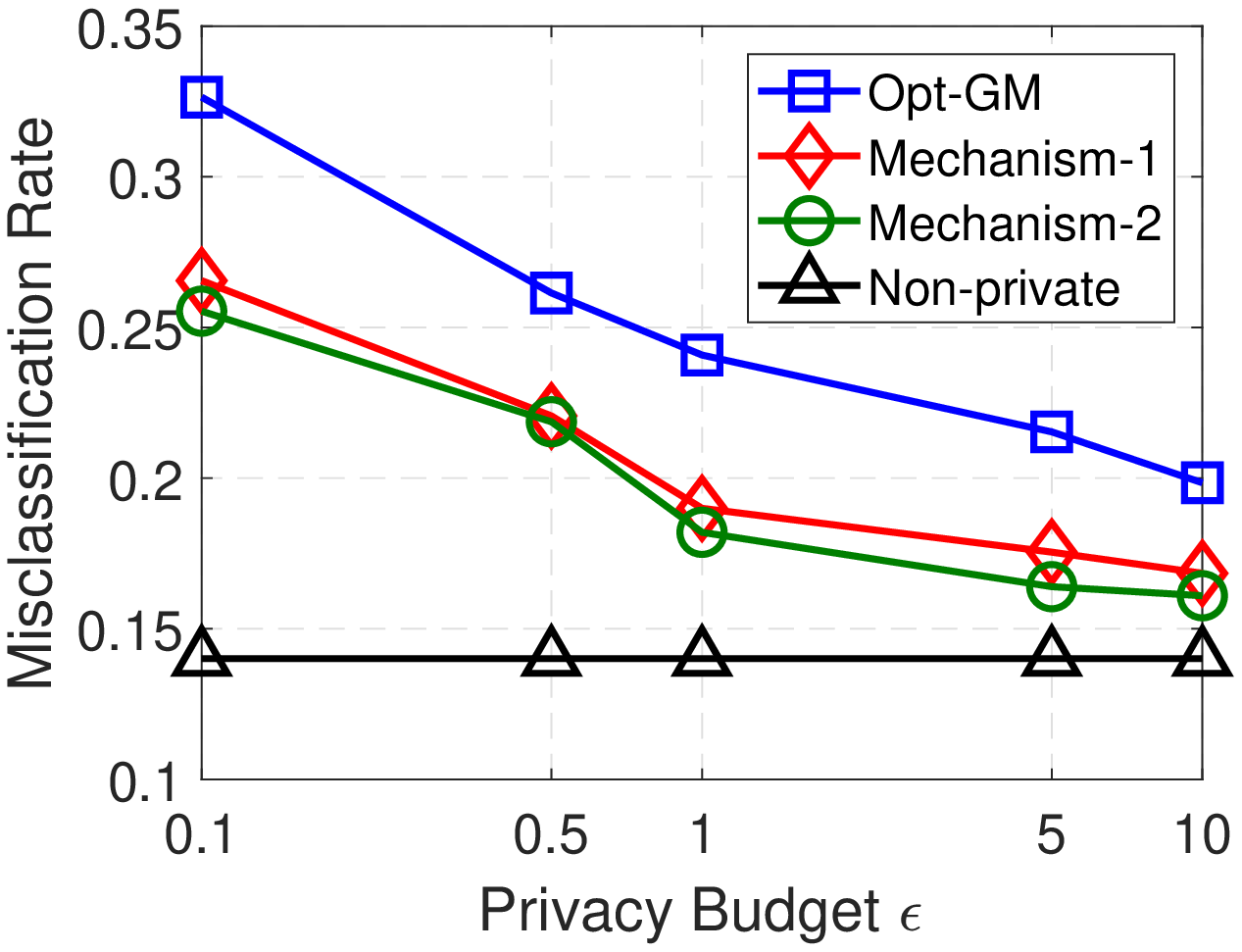}\\[-1pt]
   \scriptsize (a) MX ($\delta=10^{-26}$) & \scriptsize (b) BR ($\delta=10^{-13}$)
   \end{tabular} 
 	\vspace{-1mm}
    \caption{Accuracy of SVM classification.} \label{svm-vs-eps}
	\vspace{-3mm} 
\end{figure}

Fig.~\ref{linear-vs-eps} shows the mean squared error (MSE) of different mechanisms on the linear regression model varying values of privacy budget $\epsilon$ from 0.1 to 10. Note that we set $\delta=10^{-26}$ under MX dataset and $\delta=10^{-13}$ under BR dataset in order to ensure $\alpha <1$. It can been seen that our proposed \mbox{Mechanism-1} and \mbox{Mechanism-2} outperform \mbox{Opt-GM} in all case. This demonstrates that our proposed local differential privacy algorithms can ensure much lower errors than the optimal Gaussian mechanism when applying on the linear regression model.

Fig.~\ref{logistic-vs-eps} and Fig.~\ref{svm-vs-eps} present the misclassification rate of different mechanisms on logistic regression model and SVM classification model, respectively. We can observe from both figures that, with varying values of privacy budget $\epsilon$ from 0.1 to 10, our proposed two mechanisms always have a smaller misclassification rate than \mbox{Opt-GM}. Besides, the misclassification rates of our proposed mechanisms are close to that of the non-private method. In particular, when $\epsilon$ is large (i.e., $\epsilon \geq 5$), the accuracy of \mbox{Mechanism-1} and \mbox{Mechanism-2} approach to the non-private case, which demonstrates the high data utility of our proposed mechanism again.

\section{Conclusion}\label{sec-conclution}

This paper investigates the multi-dimensional data collection and analysis with $(\epsilon, \delta)$-local differential privacy under untrusted data curator. Aiming at both numeric data and categorical data, we have proposed novel solutions which can not only collect each user's data record in a randomized way to provide strong privacy guarantees, but also compute accurate statistics such that ensuring high accuracies on both mean/frequency estimation and machine learning models such as linear regression, logistic regression and SVM classification. Moreover, the theoretical analysis has shown that our solutions achieve low asymptotic error bound and the minimum variance. Extensive experimental results on real data and synthetic data have demonstrated the high accuracy of our proposed solutions on both simple data statistics and complex machine learning models.

\section*{Acknowledgement}
This work was supported in part by the Natural Science Foundation of China (NSFC) under grants: 61572398, 61772410 and 61802298. Any opinions, findings and conclusions or recommendations expressed in this material are those of the authors and do not necessarily reflect the views of the funding agencies.

\section*{References}
\small
\bibliographystyle{elsarticle-harv}
\bibliography{mybibfile}

\appendix
\section{Appendix}
\subsection{Proof of Duchi~\textit{et al.}'s Solution for Multi-dimensional Data}\label{appen-Duchi-solution}

Algorithm~\ref{algorithm-duchi} presents Duchi~\textit{et al.}'s mechanism for achieving \mbox{$\epsilon$-LDP} for multi-dimensional data. Actually, $B$ is the scaling factor that can ensure the expected value of noisy data is the same as the original data. Before choosing $B$, we first compute $C_d$ as
\begin{align}
    C_d=
    \begin{cases}
    2^{d-1},&\text{~if}~d~\text{is odd},\\
    2^{d-1}-\frac{1}{2}\binom{d}{d/2},&\text{~otherwise}.
    \end{cases}
\end{align}
Then, $B$ is calculated by
\begin{align}
    B=
    \begin{cases}
    \frac{2^d+C_d\cdot(e^\epsilon-1)}{\binom{d-1}{(d-1)/2}\cdot(e^\epsilon-1)},&\text{~if}~d~\text{is odd},\\
    \frac{2^d+C_d\cdot(e^\epsilon-1)}{\binom{d-1}{d/2}\cdot(e^\epsilon-1)},&\text{~otherwise}.
    \end{cases}
\end{align}

Nguy{\^e}n~\textit{et al.} \cite{nguyen2016collecting} have shown that Duchi~\textit{et al.}'s solution (i.e., Algorithm~\ref{algorithm-duchi}) doesn't guarantee local differential privacy when $d$ is even. However, they don't give the specific proofs. In the following, we will prove that Algorithm~\ref{algorithm-duchi} satisfies local differential privacy when $d$ is odd and Algorithm~\ref{algorithm-duchi} doesn't satisfy local differential privacy when $d$ is even.

% \noindent \textbf{Proof of Algorithm~\ref{algorithm-duchi} satisfies local differential privacy when $d$ is odd and doesn't satisfy local differential privacy when $d$ is even.} 

To achieve $\epsilon$-local differential privacy, it needs to guarantee $\mathbb{P}[\mathcal{M}(x) = x^*] \leq e^\epsilon \cdot \mathbb{P}[\mathcal{M}(x') = x^*]$. Based on Lemma~\ref{max-min-value}, we have
\begin{align}\label{max-min-proof-ldp-1}
    \frac{\alpha}{|T^+|}\leq \frac{1-\alpha}{|T^-|}\cdot e^\epsilon.
\end{align}
By combining (\ref{eqn-T-odd}), (\ref{eqn-T-even}) and (\ref{max-min-proof-ldp-1}), it will obtain
\begin{align}\label{alpha-duchi}
    \alpha\leq
    \begin{cases}
    \frac{e^\epsilon}{e^\epsilon+1},&\text{~if~}d\text{~is odd,} \\
    \frac{|T^+|\cdot e^\epsilon}{|T^+|\cdot e^\epsilon+|T^-|},&\text{~if~}d\text{~is even.}
    \end{cases}
\end{align}

Therefore, as we can see, when $d$ is odd, Algorithm~\ref{algorithm-duchi} satisfies $\epsilon$-local differential privacy. But when $d$ is even, Algorithm~\ref{algorithm-duchi} doesn't satisfy $\epsilon$-local differential privacy since the probability of Bernoulli variable $u=1$ is no longer equal to $\frac{e^\epsilon}{e^\epsilon+1}$.

\begin{algorithm}[tb]\label{algorithm-duchi}
\small
\setstretch{1} 
 \caption{Duchi~\textit{et al.}'s solution \cite{duchi2018minimax} for Multidimensional Data} 
  \KwIn{tuple $x\in [-1, 1]^d$ and privacy parameter $\epsilon$ and $\delta$} 
  \KwOut{tuple $x^*\in \{-B, B\}^d$}
    Generate a random vector $V : = [V_1, V_2, \ldots, V_d] \in \{-1,1\}^d$ by sampling each $V_j$ independently from the following distribution:
    \begin{align}
    \mathbb{P}[V_j=v_j]=\begin{cases}
    \frac{1}{2}+\frac{1}{2}x_j,~~\text{if}~~v_j=1\\
    \frac{1}{2}-\frac{1}{2}x_j,~~\text{if}~~v_j=-1
    \end{cases}\nonumber
    \end{align}\\
    {In the case of $V$ is sampled as $v$, let $T^+(v)$ (resp. $T^-(v)$) be the set of all tuples $x^*\in\{-B,B\}^d$ such that $x^*\cdot v > 0$ (resp. $x^*\cdot v\leq 0$)\;}
    {Sample a Bernoulli variable $u=1$ with probability $\frac{e^\epsilon}{e^\epsilon+1}$\;} 
    \eIf{$u=1$}
    {\textbf{return} a tuple $x^*$ uniformly from $T^+(v)$\;}
    {\textbf{return} a tuple $x^*$ uniformly from $T^-(v)$\;}
\end{algorithm}

% \end{document}

\subsection{Proof of Fixing Duchi~\textit{et al.}'s Mechanism to Satisfy LDP when $d$ is Even}\label{appen-fixing}

Nguy{\^e}n~\textit{et al.} \cite{nguyen2016collecting} have proposed one possible solution to fix the Algorithm~\ref{algorithm-duchi} to satisfy LDP while $d$ is even. Their method is to re-define a Bernoulli variable $u$ such that
\begin{align}\label{fixing-1}
    \bp{u=1}=\frac{e^\epsilon\cdot C_d}{(e^\epsilon-1)C_d+2^d}.
\end{align}

Note that Eq.~(\ref{fixing-1}) only fix the Algorithm~\ref{algorithm-duchi} with the situation that $T^+$ (resp, $T^-$) is the set of all tuples $x^*\in\{-B,B\}^d$ such that $x^*\cdot v > 0$ (resp. $x^*\cdot v\leq 0$). And the proof of Eq.~(\ref{fixing-1}) is not given. Thus, in the following, we will firstly show the proof of Eq.~(\ref{fixing-1}), and then propose the solution to fix the Algorithm~\ref{algorithm-duchi} with the situation that $T^+$ (resp, $T^-$) is the set of all tuples $x^*\in\{-B,B\}^d$ such that $x^*\cdot v \geq 0$ (resp. $x^*\cdot v < 0$). Note that only when $d$ is even, Algorithm~\ref{algorithm-duchi} violates LDP. Thus, without generality, $d$ is always even and we will no longer specify this in this subsection.

\noindent \textbf{Proof of Eq.~(\ref{fixing-1})}. Referring to Appendix~\ref{appen-Duchi-solution}, to achieve \mbox{$\epsilon$-LDP}, the probability of a Bernoulli variable $u=1$ should be
\begin{align}\label{fixing-3}
    \alpha \leq \frac{|T^+|\cdot e^\epsilon}{|T^+|\cdot e^\epsilon+|T^-|}.
\end{align}

%(\ref{eqn-a-7}) violates local differential privacy. Thus, we re-define a Bernoulli variable $u$ such that $\bp{u=1}=\alpha$ and $\bp{u=0}=1-\alpha$. Therefore, to achieve LDP, we only need to prove
% \begin{align}\label{fixing-2}
%     \frac{1-\alpha}{\alpha}\frac{\left | T_{x'}^+ \right |}{\left | T_{x}^- \right |}\geq e^{-\epsilon}.
% \end{align}

% By taking (\ref{eqn-a-6}) into (\ref{fixing-2}), we have
% \begin{align}
%     %&\frac{1-a}{a}\geq \frac{e^{-\epsilon}\cdot \sum_{j\leq\frac{d}{2}}\binom{d}{j}}{\sum_{j\geq\frac{d}{2}+1}\binom{d}{j}}, \\
%     %&a\leq \frac{\sum_{j\geq\frac{d}{2}+1}\binom{d}{j}}{e^{-\epsilon}\cdot \sum_{j\leq\frac{d}{2}}\binom{d}{j} + \sum_{j\geq\frac{d}{2}+1}\binom{d}{j}}.
%     &\frac{1-\alpha}{\alpha}\geq \frac{e^{-\epsilon}\cdot \left | T^- \right |}{\left | T^+ \right |}, \\
%     &\alpha\leq \frac{\left | T^+ \right |}{e^{-\epsilon}\cdot \left | T^- \right | + \left | T^+ \right |}. \label{fixing-3}
% \end{align}
From Eq.~(\ref{eqn-T-even-val}), it holds $\left | T^+ \right |=\left(2^d-\binom{d}{d/2}\right)/2$ and $\left | T^- \right |=\left(2^d+\binom{d}{d/2}\right)/2$. Thus, Eq.~(\ref{fixing-3}) can be re-written as
\begin{align}\label{fixing-4}
    \alpha\leq & \frac{\frac{2^d-\binom{d}{d/2}}{2}}{e^{-\epsilon}\cdot \frac{2^d+\binom{d}{d/2}}{2}+ \frac{2^d-\binom{d}{d/2}}{2}} \nonumber\\
    &=\frac{e^\epsilon [2^d-\binom{d}{d/2}]}{2^d+\binom{d}{d/2} + e^\epsilon[2^d - \binom{d}{d/2}]} 
    \nonumber\\ 
    &=\frac{e^\epsilon\cdot C_d}{(e^\epsilon-1)C_d+2^d}.
\end{align}
Therefore, we have proved that Algorithm~\ref{algorithm-duchi} can achieve \mbox{$\epsilon$-LDP} when $d$ is even as long as the probability of a Bernoulli variable $u=1$ is $\frac{e^\epsilon\cdot C_d}{(e^\epsilon-1)C_d+2^d}$. \qeda

As mentioned before, Eq.~(\ref{fixing-1}) can fix Algorithm~\ref{algorithm-duchi} only when $T^+$ (resp, $T^-$) is the set of all tuples $x^*\in\{-B,B\}^d$ such that $x^*\cdot v > 0$ (resp. $x^*\cdot v\leq 0$). In this paper, we have proposed the solution to fix Algorithm~\ref{algorithm-duchi} when $T^+$ (resp, $T^-$) is the set of all tuples $x^*\in\{-B,B\}^d$ such that $x^*\cdot v \geq 0$ (resp. $x^*\cdot v < 0$), that is re-defining a Bernoulli variable $u$ such that
\begin{align}\label{fixing-5}
    \bp{u=1}=\frac{e^\epsilon(2^d-C_d)}{e^\epsilon(2^d-C_d)+C_d}.
\end{align}

\noindent \textbf{Proof of Eq.~(\ref{fixing-5})}. When $d$ is even and $T^+$ (resp, $T^-$) is the set of all tuples $x^*\in\{-B,B\}^d$ such that $x^*\cdot v \geq 0$ (resp. $x^*\cdot v < 0$), it holds
\begin{align}\label{fixing-6}
    \begin{cases}
    \left | T^+ \right |=\sum_{j\geq d/2}\binom{d}{j},\\
    \left | T^- \right |=\sum_{j\leq d/2-1}\binom{d}{j}.
    \end{cases}
\end{align}
From Eq.~(\ref{fixing-6}), it holds $\left | T^+ \right | + \left | T^- \right |=2^d$ and $\left | T^+ \right | - \left | T^- \right |=\binom{d}{d/2}$. Then, we get $\left | T^+ \right |=\left(2^d+\binom{d}{d/2}\right)/2$ and $\left | T^- \right |=\left(2^d-\binom{d}{d/2}\right)/2$. Thus, Eq.~(\ref{fixing-3}) can be re-written as
\begin{align}\label{fixing-7}
    \alpha\leq & \frac{\frac{2^d+\binom{d}{d/2}}{2}}{e^{-\epsilon}\cdot \frac{2^d-\binom{d}{d/2}}{2}+ \frac{2^d+\binom{d}{d/2}}{2}} \nonumber\\
    &=\frac{e^\epsilon[2^d+\binom{d}{d/2}]}{2^d-\binom{d}{d/2} + e^\epsilon[2^d+\binom{d}{d/2}]}\nonumber\\ &=\frac{e^\epsilon(2^d-C_d)}{e^\epsilon(2^d-C_d)+C_d}.
\end{align}
This has completed the proof that Algorithm~\ref{algorithm-duchi} can achieve \mbox{$\epsilon$-LDP} when $d$ is even as long as the probability of a Bernoulli variable $u=1$ is $\frac{e^\epsilon(2^d-C_d)}{e^\epsilon(2^d-C_d)+C_d}$. \qeda

% \begin{align}  
%  &   \mathbb{P}[\mathcal{M}(x)=x^*] 
%  \nonumber\\
%     & =  \Bigg\{ \frac{\alpha}{|T^+|} \sum_{ _{x^*\cdot v>0}^{v \in \{-1,1\}^d:}} \prod_{j=1}^d \left(  \frac{1}{2}+\frac{1}{2} x_j \cdot v_j \right) \Bigg\} \nonumber\\
%     & \quad+\Bigg\{ \frac{1-\alpha}{|T^-|} \sum_{ _{x^*\cdot v \leq 0}^{v \in \{-1,1\}^d:}} \prod_{j=1}^d \left(  \frac{1}{2}+\frac{1}{2} x_j \cdot v_j \right) \Bigg\}.
% \end{align}

\subsection{Proof of Unbiased Estimation} \label{appen-proof-unbiased}
Based on Algorithm~\ref{ldp-algo-our-multi}, the expectation of $\mathcal{M}(x)$ is computed as
\begin{align}  
 & \mathbb{E}[\mathcal{M}(x)] =\sum_{x^*\in \{-B, B\}^d} \left\{ x^* \mathbb{P}[\mathcal{M}(x)=x^*] \right\}.
\end{align}
For $k \in \{1,2,\ldots, d\}$, the $k$-th dimension of $\mathbb{E}[\mathcal{M}(x)] $ is 
\begin{align}  
      \sum_{ _{x^*_k = B}^{x^*\in \{-B, B\}^d:}} \left\{B \cdot \mathbb{P}[\mathcal{M}(x)=x^*] \right\}+ \sum_{ _{x^*_k = - B}^{x^*\in \{-B, B\}^d:}} \left\{(-B) \cdot\mathbb{P}[\mathcal{M}(x)=x^*] \right\} .
\end{align}

To consider $k=1$, the first dimension of $\mathbb{E}[\mathcal{M}(x)] $ is
\begin{small}
\begin{align}  \label{eqn-109}
    & \sum_{ _{x^*_1 = B}^{x^*\in \{-B, B\}^d:}} \left\{B \cdot \mathbb{P}[\mathcal{M}(x)=x^*] \right\} + \sum_{ _{x^*_1 = - B}^{x^*\in \{-B, B\}^d:}} \left\{(-B) \cdot\mathbb{P}[\mathcal{M}(x)=x^*] \right\} \nonumber\\
    & = B \Bigg\{  \Bigg[ \sum_{x^*_{2:d}\in \{-B, B\}^{d-1}} \mathbb{P}[\mathcal{M}(x)=[B, x^*_{2:d}]] \Bigg] -  \Bigg[ \sum_{x^*_{2:d}\in \{-B, B\}^{d-1}}  \mathbb{P}[\mathcal{M}(x)=[-B, x^*_{2:d}]]   \Bigg]   \Bigg\}.
\end{align}
\end{small}
From Eq.~(\ref{eqn-109}), defining $J(x_{2:d}, v_{2:d})=\prod_{j=2}^d \left(  \frac{1}{2}+\frac{1}{2} x_j \cdot v_j \right)$, it will have
\begin{small}
\begin{align}  \label{eqn-110}
    & \sum_{x^*_{2:d}\in \{-B, B\}^{d-1}} \mathbb{P}[\mathcal{M}(x)=[B, x^*_{2:d}]] = \nonumber\\
    & \sum_{x^*_{2:d}\in \{-B, B\}^{d-1}} \Bigg\{ \Bigg[  \frac{\alpha}{|T^+|}  \times \sum_{ _{x^*\cdot v>0,x^*_{1}=B}^{v \in \{-1,1\}^d:}} \left(  \frac{1}{2}+\frac{1}{2} x_1\cdot v_1 \right)J(x_{2:d}, v_{2:d}) \Bigg] \nonumber\\
    & \quad+\Bigg[ \frac{1-\alpha}{|T^-|}  \times \sum_{ _{x^*\cdot v \leq 0,x^*_{1}=B}^{v \in \{-1,1\}^d:}} \left(  \frac{1}{2}+\frac{1}{2} x_1\cdot v_1 \right)  J(x_{2:d}, v_{2:d}) \Bigg]  \Bigg\},
\end{align}
\end{small}
where
\begin{small}
\begin{align}  
      & \sum_{ _{x^*\cdot v>0,x^*_{1}=B}^{v \in \{-1,1\}^d:}} \left(  \frac{1}{2}+\frac{1}{2} x_1\cdot v_1 \right)\prod_{j=2}^d \left(  \frac{1}{2}+\frac{1}{2} x_j \cdot v_j \right)   \nonumber\\
    & =  \sum_{ _{x^*\cdot v>0,x^*_{1}=B,v_{1}=1}^{v \in \{-1,1\}^d:}} \left(  \frac{1}{2}+\frac{1}{2} x_1\cdot v_1 \right)\prod_{j=2}^d \left(  \frac{1}{2}+\frac{1}{2} x_j \cdot v_j \right) \nonumber\\
    & \quad +  \sum_{ _{x^*\cdot v>0,x^*_{1}=B,v_{1}=-1}^{v \in \{-1,1\}^d:}} \left(  \frac{1}{2}+\frac{1}{2} x_1\cdot v_1 \right)\prod_{j=2}^d \left(  \frac{1}{2}+\frac{1}{2} x_j \cdot v_j \right) \nonumber\\
    & =  \left(  \frac{1}{2}+\frac{1}{2} x_1 \right)\sum_{ _{x^*_{2:d}\cdot v_{2:d}>-B}^{v_{2:d} \in \{-1,1\}^{d-1}:}} \prod_{j=2}^d \left(  \frac{1}{2}+\frac{1}{2} x_j \cdot v_j \right) \nonumber\\
    & \quad +  \left(  \frac{1}{2}-\frac{1}{2} x_1  \right) \sum_{ _{x^*_{2:d}\cdot v_{2:d}>B}^{v_{2:d} \in \{-1,1\}^{d-1}:}} \prod_{j=2}^d \left(  \frac{1}{2}+\frac{1}{2} x_j \cdot v_j \right),
\end{align}
\end{small}
and
\begin{small}
\begin{align}  
      & \sum_{ _{x^*\cdot v \leq 0,x^*_{1}=B}^{v \in \{-1,1\}^d:}} \left(  \frac{1}{2}+\frac{1}{2} x_1\cdot v_1 \right)\prod_{j=2}^d \left(  \frac{1}{2}+\frac{1}{2} x_j \cdot v_j \right)   \nonumber\\
    & =  \sum_{ _{x^*\cdot v \leq 0,x^*_{1}=B,v_{1}=1}^{v \in \{-1,1\}^d:}} \left(  \frac{1}{2}+\frac{1}{2} x_1\cdot v_1 \right)\prod_{j=2}^d \left(  \frac{1}{2}+\frac{1}{2} x_j \cdot v_j \right) \nonumber\\
    & \quad +  \sum_{ _{x^*\cdot v \leq 0,x^*_{1}=B,v_{1}=-1}^{v \in \{-1,1\}^d:}} \left(  \frac{1}{2}+\frac{1}{2} x_1\cdot v_1 \right)\prod_{j=2}^d \left(  \frac{1}{2}+\frac{1}{2} x_j \cdot v_j \right) \nonumber\\
    & =  \left(  \frac{1}{2}+\frac{1}{2} x_1 \right) \sum_{ _{x^*_{2:d}\cdot v_{2:d} \leq -B}^{v_{2:d} \in \{-1,1\}^{d-1}:}}\prod_{j=2}^d \left(  \frac{1}{2}+\frac{1}{2} x_j \cdot v_j \right) \nonumber\\
    & \quad +  \left(  \frac{1}{2}-\frac{1}{2} x_1  \right) \sum_{ _{x^*_{2:d}\cdot v_{2:d} \leq B}^{v_{2:d} \in \{-1,1\}^{d-1}:}} \prod_{j=2}^d \left(  \frac{1}{2}+\frac{1}{2} x_j \cdot v_j \right) .
\end{align}
\end{small}

Therefore, Eq.~(\ref{eqn-110}) can be deduced as
\begin{small}
\begin{align}  
      &  \sum_{x^*_{2:d}\in \{-B, B\}^{d-1}} \mathbb{P}[\mathcal{M}(x)=[B, x^*_{2:d}]]= \nonumber\\
    & \sum_{x^*_{2:d}\in \{-B, B\}^{d-1}} \Bigg\{ \Bigg[  \frac{\alpha}{|T^+|} \times  \Bigg( \left(  \frac{1}{2}+\frac{1}{2} x_1 \right) \sum_{ _{x^*_{2:d}\cdot v_{2:d} > -B}^{v_{2:d} \in \{-1,1\}^{d-1}:}}J(x_{2:d}, v_{2:d}) \nonumber\\
    & \quad \quad +  \left(  \frac{1}{2}-\frac{1}{2} x_1  \right) \sum_{ _{x^*_{2:d}\cdot v_{2:d} > B}^{v_{2:d} \in \{-1,1\}^{d-1}:}} J(x_{2:d}, v_{2:d})\Bigg)\Bigg] \nonumber\\
    & \quad+\Bigg[ \frac{1-\alpha}{|T^-|} \times \Bigg( \left(  \frac{1}{2}+\frac{1}{2} x_1 \right) \sum_{ _{x^*_{2:d}\cdot v_{2:d} \leq -B}^{v_{2:d} \in \{-1,1\}^{d-1}:}}J(x_{2:d}, v_{2:d}) \nonumber\\
    & \quad \quad +  \left(  \frac{1}{2}-\frac{1}{2} x_1  \right) \sum_{ _{x^*_{2:d}\cdot v_{2:d} \leq B}^{v_{2:d} \in \{-1,1\}^{d-1}:}} J(x_{2:d}, v_{2:d})\Bigg)\Bigg]  \Bigg\} .
\end{align}
\end{small}

Similarly, it also holds that
\begin{small}
\begin{align}  
    &  \sum_{x^*_{2:d}\in \{-B, B\}^{d-1}} \mathbb{P}[\mathcal{M}(x)=[-B, x^*_{2:d}]] = \nonumber\\
    &  \sum_{x^*_{2:d}\in \{-B, B\}^{d-1}} \Bigg\{ \Bigg[  \frac{\alpha}{|T^+|} \times  \Bigg( \left(  \frac{1}{2}+\frac{1}{2} x_1 \right) \sum_{ _{x^*_{2:d}\cdot v_{2:d} > B}^{v_{2:d} \in \{-1,1\}^{d-1}:}} J(x_{2:d}, v_{2:d}) \nonumber\\
    & \quad \quad +  \left(  \frac{1}{2}-\frac{1}{2} x_1  \right) \sum_{ _{x^*_{2:d}\cdot v_{2:d} > -B}^{v_{2:d} \in \{-1,1\}^{d-1}:}} J(x_{2:d}, v_{2:d})\Bigg)\Bigg] \nonumber\\
    & \quad+\Bigg[ \frac{1-\alpha}{|T^-|} \times \Bigg( \left(  \frac{1}{2}+\frac{1}{2} x_1 \right) \sum_{ _{x^*_{2:d}\cdot v_{2:d} \leq B}^{v_{2:d} \in \{-1,1\}^{d-1}:}} J(x_{2:d}, v_{2:d}) \nonumber\\
    & \quad \quad +  \left(  \frac{1}{2}-\frac{1}{2} x_1  \right) \sum_{ _{x^*_{2:d}\cdot v_{2:d} \leq -B}^{v_{2:d} \in \{-1,1\}^{d-1}:}} J(x_{2:d}, v_{2:d}) \Bigg)\Bigg]  \Bigg\}.
\end{align}
\end{small}

The first dimension of $\mathbb{E}[\mathcal{M}(x)] $ dividing $B$ is 
\begin{small}
\begin{align}  
    & \Bigg[ \sum_{x^*_{2:d}\in \{-B, B\}^{d-1}} \mathbb{P}[\mathcal{M}(x)=[B, x^*_{2:d}]] \Bigg] \nonumber\\
    & \quad -  \Bigg[ \sum_{x^*_{2:d}\in \{-B, B\}^{d-1}}  \mathbb{P}[\mathcal{M}(x)=[-B, x^*_{2:d}]]   \Bigg]  = \nonumber\\
    & \sum_{x^*_{2:d}\in \{-B, B\}^{d-1}} \Bigg\{ \Bigg[  \frac{\alpha}{|T^+|}  \times  \Bigg( \left(  \frac{1}{2}+\frac{1}{2} x_1 \right) \sum_{ _{x^*_{2:d}\cdot v_{2:d} > -B}^{v_{2:d} \in \{-1,1\}^{d-1}:}} J(x_{2:d}, v_{2:d}) \nonumber\\
    & \quad \quad +  \left(  \frac{1}{2}-\frac{1}{2} x_1  \right) \sum_{ _{x^*_{2:d}\cdot v_{2:d} > B}^{v_{2:d} \in \{-1,1\}^{d-1}:}} J(x_{2:d}, v_{2:d}) \Bigg)\Bigg] \nonumber\\
    & \quad+\Bigg[ \frac{1-\alpha}{|T^-|}  \times \Bigg( \left(  \frac{1}{2}+\frac{1}{2} x_1 \right) \sum_{ _{x^*_{2:d}\cdot v_{2:d} \leq -B}^{v_{2:d} \in \{-1,1\}^{d-1}:}} J(x_{2:d}, v_{2:d}) \nonumber\\
    & \quad \quad +  \left(  \frac{1}{2}-\frac{1}{2} x_1  \right) \sum_{ _{x^*_{2:d}\cdot v_{2:d} \leq B}^{v_{2:d} \in \{-1,1\}^{d-1}:}}  J(x_{2:d}, v_{2:d}) \Bigg)\Bigg]  \Bigg\}-    \nonumber\\    & \sum_{x^*_{2:d}\in \{-B, B\}^{d-1}} \Bigg\{ \Bigg[  \frac{\alpha}{|T^+|}  \times  \Bigg( \left(  \frac{1}{2}+\frac{1}{2} x_1 \right) \sum_{ _{x^*_{2:d}\cdot v_{2:d} > B}^{v_{2:d} \in \{-1,1\}^{d-1}:}} J(x_{2:d}, v_{2:d}) \nonumber\\
    & \quad \quad +  \left(  \frac{1}{2}-\frac{1}{2} x_1  \right) \sum_{ _{x^*_{2:d}\cdot v_{2:d} > -B}^{v_{2:d} \in \{-1,1\}^{d-1}:}} J(x_{2:d}, v_{2:d}) \Bigg)\Bigg] \nonumber\\
    & \quad+\Bigg[ \frac{1-\alpha}{|T^-|} \times \Bigg( \left(  \frac{1}{2}+\frac{1}{2} x_1 \right) \sum_{ _{x^*_{2:d}\cdot v_{2:d} \leq B}^{v_{2:d} \in \{-1,1\}^{d-1}:}} J(x_{2:d}, v_{2:d}) \nonumber\\
    & \quad \quad +  \left(  \frac{1}{2}-\frac{1}{2} x_1  \right) \sum_{ _{x^*_{2:d}\cdot v_{2:d} \leq -B}^{v_{2:d} \in \{-1,1\}^{d-1}:}}  J(x_{2:d}, v_{2:d}) \Bigg)\Bigg]  \Bigg\} \nonumber\\
    &  = \sum_{x^*_{2:d}\in \{-B, B\}^{d-1}} \Bigg\{ \Bigg[  \frac{\alpha}{|T^+|} \times  \Bigg( \left(  \frac{1}{2}+\frac{1}{2} x_1 \right) \sum_{ _{-B<x^*_{2:d}\cdot v_{2:d} \leq B}^{v_{2:d} \in \{-1,1\}^{d-1}:}} J(x_{2:d}, v_{2:d}) \nonumber\\
    & \quad \quad -  \left(  \frac{1}{2}-\frac{1}{2} x_1  \right) \sum_{ _{-B<x^*_{2:d}\cdot v_{2:d}\leq B}^{v_{2:d} \in \{-1,1\}^{d-1}:}}  J(x_{2:d}, v_{2:d}) \Bigg)\Bigg] \nonumber\\
    & \quad+\Bigg[ \frac{1-\alpha}{|T^-|} \times \Bigg( -\left(  \frac{1}{2}+\frac{1}{2} x_1 \right) \sum_{ _{-B < x^*_{2:d}\cdot v_{2:d} \leq B}^{v_{2:d} \in \{-1,1\}^{d-1}:}} J(x_{2:d}, v_{2:d}) \nonumber\\
    & \quad \quad +  \left(  \frac{1}{2}-\frac{1}{2} x_1  \right) \sum_{ _{-B < x^*_{2:d}\cdot v_{2:d} \leq B}^{v_{2:d} \in \{-1,1\}^{d-1}:}} J(x_{2:d}, v_{2:d}) \Bigg)\Bigg]  \Bigg\} \nonumber\\
    &  = \sum_{x^*_{2:d}\in \{-B, B\}^{d-1}} \Bigg\{ \Bigg[  \frac{\alpha}{|T^+|}  \times    x_1 \sum_{ _{-B<x^*_{2:d}\cdot v_{2:d} \leq B}^{v_{2:d} \in \{-1,1\}^{d-1}:}} J(x_{2:d}, v_{2:d})  \Bigg] \nonumber\\
    & \quad+\Bigg[ \frac{1-\alpha}{|T^-|} \times \Bigg( - x_1  \sum_{ _{-B < x^*_{2:d}\cdot v_{2:d} \leq B}^{v_{2:d} \in \{-1,1\}^{d-1}:}} J(x_{2:d}, v_{2:d})   \Bigg\}  \nonumber\\
    &  =   x_1 \bigg( \frac{\alpha}{|T^+|} - \frac{1-\alpha}{|T^-|}   \bigg) \times \sum_{x^*_{2:d}\in \{-B, B\}^{d-1}}   \sum_{ _{-B<x^*_{2:d}\cdot v_{2:d} \leq B}^{v_{2:d} \in \{-1,1\}^{d-1}:}} J(x_{2:d}, v_{2:d})  .
\end{align}
\end{small}
To ensure that the first dimension of $\mathbb{E}[\mathcal{M}(x)] $ equals $x_1$, we set $B$ as
\begin{align}  
      & \left[\bigg( \frac{\alpha}{|T^+|} - \frac{1-\alpha}{|T^-|}   \bigg) \times H \right]^{-1},
\end{align}
where
\begin{small}
\begin{align}  
    & H=\sum_{x^*_{2:d}\in \{-B, B\}^{d-1}}   \sum_{ _{-B<x^*_{2:d}\cdot v_{2:d} \leq B}^{v_{2:d} \in \{-1,1\}^{d-1}:}}\prod_{j=2}^d \left(  \frac{1}{2}+\frac{1}{2} x_j \cdot v_j \right) \nonumber\\
  & = \begin{cases}
  \mathlarger{  \sum_{x^*_{2:d}\in \{-B, B\}^{d-1}}   \sum_{ _{x^*_{2:d}\cdot v_{2:d} = 0}^{v_{2:d} \in \{-1,1\}^{d-1}:}}\prod_{j=2}^d \left(  \frac{1}{2}+\frac{1}{2} x_j \cdot v_j \right)} , \\ \quad\quad\quad\quad\quad\quad\quad\quad\quad\quad\text{ if $d$ is odd}, \\ \mathlarger{\sum_{x^*_{2:d}\in \{-B, B\}^{d-1}}   \sum_{ _{x^*_{2:d}\cdot v_{2:d} = B}^{v_{2:d} \in \{-1,1\}^{d-1}:}}\prod_{j=2}^d \left(  \frac{1}{2}+\frac{1}{2} x_j \cdot v_j \right), } \\ \quad\quad\quad\quad\quad\quad\quad\quad\quad\quad\text{ if $d$ is even}.
  \end{cases} \label{sum-117}
\end{align}
\end{small}

If $d$ is odd, then we have
\begin{small}
\begin{align}  
    &\sum_{x^*_{2:d}\in \{-B, B\}^{d-1}}   \sum_{ _{x^*_{2:d}\cdot v_{2:d} = 0}^{v_{2:d} \in \{-1,1\}^{d-1}:}}\prod_{j=2}^d \left(  \frac{1}{2}+\frac{1}{2} x_j \cdot v_j \right)\nonumber\\
    & =  \sum_{v_{2:d} \in \{-1,1\}^{d-1}}   \sum_{ _{x^*_{2:d}\cdot v_{2:d} = 0}^{x^*_{2:d}\in \{-B, B\}^{d-1}:}}\prod_{j=2}^d \left(  \frac{1}{2}+\frac{1}{2} x_j \cdot v_j \right) \nonumber\\
    & =  \sum_{v_{2:d} \in \{-1,1\}^{d-1}}  \left[ \binom{d-1}{\frac{d-1}{2}} \prod_{j=2}^d \left(  \frac{1}{2}+\frac{1}{2} x_j \cdot v_j \right) \right] \nonumber\\
    & =  \binom{d-1}{\frac{d-1}{2}} \sum_{v_{2:d} \in \{-1,1\}^{d-1}} \prod_{j=2}^d \left(  \frac{1}{2}+\frac{1}{2} x_j \cdot v_j \right) \nonumber\\
    &  =  \binom{d-1}{\frac{d-1}{2}}  \prod_{j=2}^d  \left[ \left(  \frac{1}{2}+\frac{1}{2} x_j   \right) + \left(  \frac{1}{2} - \frac{1}{2} x_j   \right) \right]  \nonumber\\
    &  = \binom{d-1}{\frac{d-1}{2}}  \prod_{j=2}^d 1
    = \binom{d-1}{\frac{d-1}{2}}.
\end{align}
\end{small}

If $d$ is even, then we have
\begin{small}
\begin{align}  
      &\sum_{x^*_{2:d}\in \{-B, B\}^{d-1}}   \sum_{ _{x^*_{2:d}\cdot v_{2:d} = B}^{v_{2:d} \in \{-1,1\}^{d-1}:}}\prod_{j=2}^d \left(  \frac{1}{2}+\frac{1}{2} x_j \cdot v_j \right)\nonumber\\
    & =  \sum_{v_{2:d} \in \{-1,1\}^{d-1}}   \sum_{ _{x^*_{2:d}\cdot v_{2:d} = B}^{x^*_{2:d}\in \{-B, B\}^{d-1}:}}\prod_{j=2}^d \left(  \frac{1}{2}+\frac{1}{2} x_j \cdot v_j \right) \nonumber\\
    & =  \sum_{v_{2:d} \in \{-1,1\}^{d-1}}  \left[ \binom{d-1}{\frac{d}{2}} \prod_{j=2}^d \left(  \frac{1}{2}+\frac{1}{2} x_j \cdot v_j \right) \right] \nonumber\\
    & =  \binom{d-1}{\frac{d}{2}} \sum_{v_{2:d} \in \{-1,1\}^{d-1}} \prod_{j=2}^d \left(  \frac{1}{2}+\frac{1}{2} x_j \cdot v_j \right) \nonumber\\
    &  =  \binom{d-1}{\frac{d}{2}}  \prod_{j=2}^d  \left[ \left(  \frac{1}{2}+\frac{1}{2} x_j   \right) + \left(  \frac{1}{2} - \frac{1}{2} x_j   \right) \right]  \nonumber\\
    &  = \binom{d-1}{\frac{d}{2}}  \prod_{j=2}^d 1
      = \binom{d-1}{\frac{d}{2}}.
\end{align}
\end{small}

Therefore, the above result in Eq.~(\ref{sum-117}) equals     
\begin{align}  
    H= \begin{cases}
    \binom{d-1}{(d-1)/2} , &\text{ if $d$ is odd}, \\ \binom{d-1}{d/2}, &\text{ if $d$ is even}.
    \end{cases}  
\end{align} 

Thus, the $B$ can be calculated as
\begin{align}  
    B= \begin{cases}
    \left[ \left( \frac{\alpha}{|T^+|} - \frac{1-\alpha}{|T^-|} \right) \binom{d-1}{\frac{d-1}{2}} \right]^{-1} , &\text{ if $d$ is odd}, \\  
    \left[\left( \frac{\alpha}{|T^+|} - \frac{1-\alpha}{|T^-|}   \right) \binom{d-1}{\frac{d}{2}}\right]^{-1}, &\text{ if $d$ is even}.
    \end{cases}
\end{align}
Since
\begin{align}
\begin{cases}
    \begin{cases}
    |T^+|=2^{d-1},\\ 
    |T^-|=2^{d-1},
    \end{cases}
    &\text{ if $d$ is odd,}\\
    \begin{cases}
    |T^+|=2^{d-1}-\frac{1}{2}\binom{d}{d/2},\\
    |T^-|=2^{d-1}+\frac{1}{2}\binom{d}{d/2},
    \end{cases}
    &\text{ if $d$ is even,}
\end{cases}
\end{align}
based on Eq.~(\ref{alpha-our}), we can obtain
\begin{align}
    B=
    \begin{cases}
    \frac{2^d+C_d\cdot(e^\epsilon-1)}{\binom{d-1}{(d-1)/2}\cdot(e^\epsilon +2^d\cdot \delta-1)},&\text{~if}~d~\text{is odd},\\
    \frac{2^d+C_d\cdot(e^\epsilon-1)}{\binom{d-1}{d/2}\cdot(e^\epsilon +2^d\cdot \delta -1)},&\text{~if}~d~\text{is even}.
    \end{cases}
\end{align}


% \bibliographystyle{cas-model2-names}

% \bibliography{cas-refs}


% Author biography without author photo.

% Author biography with author photo.

% add figure path in $\{\}$
% Author biography with author photo.

\end{document}